\documentclass{jfm}
\usepackage{booktabs}
\usepackage{subcaption}
\usepackage{graphicx,color}
\usepackage{float}
\usepackage{epstopdf, epsfig}
\usepackage{caption,subcaption}
\usepackage[intlimits]{amsmath}
\usepackage{lscape}
\usepackage{rotating}
\usepackage{amssymb, amsmath}
\usepackage{amsmath}
\usepackage[mathscr]{euscript}
\usepackage{soul}

\newcommand{\CRCA}[1]{{\textcolor{black}{#1}}}

\newcommand\Web{\mbox{\textit{We}}}
\newcommand\Mar{\mbox{\textit{Ma}}}

\shorttitle{Effects of insoluble surfactants on breaking waves}
\shortauthor{B. Wang and others}

\title{
Effects of insoluble surfactants on breaking waves: regular and spilling regimes
}

\author{B. Wang$^{1}$, \ns J. Chergui$^2$, \ns S. Shin$^3$,  \ns D. Juric$^{2,4}$, C. R Constante-Amores$^{1}$\corresp{\email{crconsta@illinois.edu} }}

\affiliation{
$^1$Department of Mechanical Science and Engineering, University of Illinois, Urbana Champaign, IL 61801, USA
\\[\affilskip]
$^2$ Universit\'e Paris Saclay, Centre National de la Recherche Scientifique (CNRS), Laboratoire Interdisciplinaire des Sciences du Num\'erique (LISN), 91400 Orsay, France
\\[\affilskip]
$^3$Department of Mechanical and System Design Engineering, Hongik University, Seoul 04066, Republic of Korea
\\[\affilskip]
$^4$Department of Applied Mathematics and Theoretical Physics, University of Cambridge,  Cambridge CB3 0WA, UK
\\[\affilskip]
}

\begin{document}

\maketitle

\begin{abstract}

We investigate the influence of insoluble surfactants on the spatio-temporal evolution of breaking waves, focusing on both regular and spilling regimes. Three-dimensional direct numerical simulations are conducted using an interface-tracking/level-set method that incorporates surfactant-induced Marangoni stresses. The simulations reveal that surfactant gradients, through Marangoni stresses, markedly alter the wave dynamics. While regular breakers exhibit only minor modifications in the presence of surfactants, increasing surfactant-induced Marangoni stresses in spilling breakers leads to  changes in the crest evolution and vorticity generation. \textcolor{black}{Surfactants enhance spilling breakers, primarily driven by Marangoni stresses rather than surface tension reduction.}
To quantify these effects, we also extend circulation-based theoretical frameworks to account for surfactant contributions. 

\end{abstract}

\section{Introduction}

Surface wave breaking plays a significant role in the exchange of mass, momentum, and energy across the air-sea boundary \citep{melville1996role,garbe2014transfer,iafrati2009numerical,deike2022mass}. When waves break, they generate turbulence, entrain air, and enhance mixing, processes that significantly affect oceanic and atmospheric dynamics \citep{melville1996role,deike2022mass}. The underlying physics of wave breaking, particularly spilling  and plunging breakers, has been extensively studied using both experiments and numerical simulations \citep{rapp1990laboratory,Chen,duncan1999gentle,duncan2001spilling,qiao2001gentle,liu2003effects,liu2006experimental,iafrati2009numerical,deike2015capillary,deike2016air}. 

Early studies by \cite{Chen} showed that vorticity generation and energy dissipation are strongly localized near the wave crest, corroborating experimental observations of wave-induced turbulence. These findings, obtained through two-dimensional simulations, captured  plunging and spilling breaker dynamics. Building on this, \cite{iafrati2009numerical} provided a comprehensive numerical framework for analyzing the mechanics of both plunging and spilling breakers. 
Their work classified different wave regimes as a function of the initial wave steepness, $\epsilon$, for fixed  Reynolds and Weber numbers  ($Re=10^4$, $We=100$).   Waves remained regular for $\epsilon<0.33$; transition to spilling breaker when $0.33\leq\epsilon<0.37$ and become plunging breakers for  $\epsilon\geq0.37$. These results agree with the theoretical analysis by \cite{grue2010model}.  Building on earlier studies, \citet{deike2015capillary,deike2016air} used direct numerical simulations (DNS) to investigate the mechanisms of energy dissipation in breaking waves, highlighting differences between plunging and spilling regimes. Their work emphasized the transition to turbulence, vorticity generation, and air entrainment as central features of breaking dynamics. \textcolor{black}{
It is important to note that the relationship between initial wave steepness and breaking type is not universal. Experimental and numerical studies show that the limiting steepness for  breaking spans a wide range, depending on the generation mechanism, spectral bandwidth, and  conditions \citep{duncan1981experimental,duncan1983breaking,ramberg1987laboratory,rapp1990laboratory,wu2004laboratory,babanin2010numerical}. As reviewed comprehensively by \cite{perlin2013breaking}, gravity-capillary wave has been observed over $\epsilon \approx 0.15$–$0.48$, and in some configurations spilling breakers may exhibit equal or even greater steepness than plunging ones. 
}

Breaking waves undergo a cascade of processes, from jet impact and vortex formation to bubble entrainment and droplet ejection, that enhance mixing and momentum transfer across the interface \citep{liu2006experimental,deike2022mass}. High-fidelity simulations by \citet{wang2016high} captured the formation of bubbles, droplets, and spray, offering insight into the multiphase dynamics following wave impact. 
To model energy dissipation, \citet{mostert2020inertial} developed a theoretical framework relating dissipation rates to wave height, water depth, and beach slope, which was validated numerically and extended empirically to offshore conditions. \CRCA{For deep-water plunging breakers, however, inertial dissipation is more appropriately described by the crest-length scaling of \citet{drazen2008inertial}, which links dissipation to wave kinematics and breaker geometry.} 
In a follow-up work,  \citet{mostert2022high} examined the transition from laminar to turbulent flow, showing that the energy dissipation rates and the size distributions of bubbles and droplets become independent of the Reynolds number once it exceeds a sufficiently high threshold. Recent studies have further examined the dynamics of plunging breakers, such as the work by \citet{di2022coherent}, who performed both two- and three-dimensional simulations across different Reynolds numbers, resolving the evolution of the air–water interface and associated bubble structures. Similarly, \citet{hu2024numerical} investigated the influence of wave slope and Reynolds number on spray formation and air entrainment during breaking.

Most of the recent studies on breaking waves have focused on clean, uncontaminated interfaces. However, natural  surface-active agents (surfactants), which are  ubiquitous   in marine environments due to biological activity and pollution, are often present in the ocean surface \citep{lapham1998wave,laxague2024suppression}. Surfactants may be either insoluble (residing primarily at the air-water interface) or soluble (dissolved in the bulk fluid and can adsorb to and desorb from the interface). In both cases, spatial variations in concentration in the interface result in  surface tension gradients, leading to Marangoni stresses driven flow   from regions of low to high surface tension \cite{manikantan_squires_2020}.  Surfactant-driven Marangoni stresses can modify  interfacial phenomena. For example, in  capillary singularities,  surfactant-induced flows can retard thread thinning and promote the formation of  microthreads \citep{ambravaneswaran1999effects,craster2002pinchoff,liao2006deformation,wee2020effects,Kamat_prf_2018,Constante-Amores_prf_2020}.
In small-scale systems related to wave breaking, such as bubbles and droplets, surfactants have been found to resist jet breakup, slow down the droplet ejection after the bursting of a bubble, and alter aerosol production \citep{takagi2011surfactant,constanteamores2020bb,ma2023effects}.

Surfactants alter every stage of wave breaking, such as capillary wave formation, 
and the structure of the turbulent flow beneath the crest \citep{ceniceros2003effects,liu2006experimental,liu2007weakly}. Two‑dimensional simulations of capillary waves with insoluble surfactant by \citet{ceniceros2003effects} showed that Marangoni stresses reduce both the bulge size and the wave amplitude while sharpening the toe curvature.  \citet{liu2003effects} and \citet{liu2006experimental}  revealed similar trends with a  suppression of capillary ripples, smoother crest evolution and a smaller, blunter bulge even at modest surfactant concentrations in wave‑tank experiments on weak spilling breakers with the soluble surfactant (sodium dodecyl sulphate). By tracking the time varying bulge geometry, they linked these macroscopic changes directly to surface‐rheological properties, concluding that surfactant‑induced surface tension gradients delay toe shedding and reduce energy dissipation.
Plunging breakers respond to the presence of surfactants even more dramatically. \citet{erinin2023effects} did  experiments with Triton X‑100 and surfactant‑depleted water, and  showed that surfactants deform the jet tip, fragment the entrained air cavity, and weaken the downward plunge.
They also presented numerical simulations that modeled surfactants as tracer particles along the interface. While these confirmed the qualitative trends, they could not capture the fully coupled surfactant-induced Marangoni stresses. Extending this work, \citet{Liu_Erinin_Liu_Duncan_2025} 
demonstrated that soluble surfactants reduce both droplet count and ejection velocity while shifting the size distribution towards larger diameters. Together, these studies establish that even trace amounts of surfactants suppress small scale capillary activity, delay the onset of turbulence beneath the crest and reshape the air entrainment and spray signature of breaking waves.

As revealed by the foregoing review, considerable progress has been made in understanding the role of surfactants in breaking wave dynamics.  However, to the best of our knowledge, the specific influence of surfactant-induced Marangoni stresses remains poorly understood.
In this study, we  investigate the influence of insoluble surfactants on breaking wave dynamics, with a particular focus on regular and  spilling regimes,  using direct numerical simulations.
The rest of this paper is structured as follows: in Section \ref{Numerical}, the numerical method, governing dimensionless parameters, problem configuration, and validation, are introduced. Section \ref{Results} presents the results, and concluding remarks are given in Section \ref{Con}.

\section{ Problem formulation and numerical method \label{Numerical}}

\textcolor{black}{High-resolution} simulations were performed by solving the two-phase incompressible Navier-Stokes equations with surface tension in a three-dimensional Cartesian domain $\mathbf{x} = \left(x, y, z \right)$ (see Figure \ref{configuration}). 
The interface between the gas and liquid is described by
a hybrid front-tracking/level-set method, where (insoluble) surfactant transport is explicitly resolved at the interface 
\citep{Shin_jcp_2018}. Here, and in what follows, all variables are made dimensionless (represented by tildes) using the following characteristic scales 
\begin{equation}
\quad \tilde{\mathbf{x}}=\frac{\mathbf{x}}{\lambda},
\quad \tilde{t}=\frac{t}{t_{r}}, 
\quad \tilde{\textbf{u}}=\frac{\textbf{u}} {c},
\quad \tilde{p}=\frac{p}{\rho_{\textcolor{black}{l}} c^2}, 
\quad \tilde{\sigma}=\frac{\sigma}{\sigma_s},
\quad \tilde{\Gamma}=\frac{\Gamma}{\Gamma_\infty},
\end{equation}

\noindent	
where, $t$, $\textbf{u}$, and $p$ stand for time, velocity, and pressure, respectively. The physical parameters correspond to the liquid density, $\rho_l$, liquid viscosity, $\mu_l$, and surfactant-free surface tension, $\sigma_s$.  The  characteristic velocity is  $c=\sqrt{g\lambda}$ with $\lambda$  the  wavelength of the wave; hence, the characteristic time scale is $t_{r}= \lambda/c$. 
The interfacial surfactant concentration, $\Gamma$, is scaled by the saturation interfacial concentration, $\Gamma_{\infty}$. 
As a result of this scaling, the dimensionless equations read 
\begin{equation}\label{div}
 \nabla \cdot \tilde{\textbf{u}}=0,
\end{equation}
\begin{equation*}
\tilde{\rho} (\frac{\partial \tilde{\textbf{u}}}{\partial \tilde{t}}+\tilde{\textbf{u}} \cdot\nabla \tilde{\textbf{u}}) + \nabla \tilde{p}  =  -\frac{\textbf{i}_z ~}{Fr^2} + \frac{1}{Re}~ \nabla\cdot  \left [ \tilde{\mu} (\nabla \tilde{\textbf{u}} +\nabla \tilde{\textbf{u}}^T) \right ] +
\end{equation*}

\begin{equation}\label{NS_Eq}
+\frac{1}{\Web}~
\int_{\tilde{A}\tilde{(t)}} 
(\tilde{\sigma} \tilde{\kappa} \textbf{n} +   \nabla_s  \tilde{\sigma})  \delta \left(\tilde{\textbf{x}} - \tilde{\textbf{x}}_{_f}  \right)\mbox{d}\tilde{A},
\end{equation}

 \begin{equation} 
 \frac{\partial \tilde{\Gamma}}{\partial \tilde{t}}+\nabla_s \cdot (\tilde{\Gamma}\tilde{\textbf{u}}_{\text{t}})=\frac{1}{Pe_s} \nabla^2_s \tilde{\Gamma}, 
 \label{gamma_equation}
 \end{equation}

\noindent
where the density and viscosity are given by $\tilde{\rho}=\rho_g/\rho_{\textcolor{black}{l}} + \left(1 -\rho_g/\rho_{\textcolor{black}{l}}\right) \mathcal{H}\left(\tilde{\textbf{x}},\tilde{t}\right)$ and $\tilde{\mu}=\mu_g/\mu_{\textcolor{black}{l}}+ \left(1 -\mu_g/\mu_{\textcolor{black}{l}}\right) \mathcal{H}\left( \tilde{\textbf{x}},\tilde{t}\right)$
wherein $\mathcal{H}\left( \tilde{\textbf{x}},\tilde{t}\right)$ represents a Heaviside function, which is zero in the gas phase and unity in the liquid phase, while the subscript `$g$' and `$l$' designate the gas, and liquid phases, respectively,
and
$\tilde{\textbf{u}}_{\text{t}}= \left ( \tilde{\textbf{u}}_{\text{s}} \cdot \textbf{t} \right ) \textbf{t}$ is the tangential velocity at the interface in which $\tilde{\textbf{u}}_{\text{s}}$ represents the interfacial velocity, and $\kappa$ is the curvature, and  $\mathbf{t}$ is the unit tangent vector to the interface.  The interfacial gradient is given by $\nabla_s=\left({\mathbf{I}}-\mathbf{n}\mathbf{n}\right)\cdot \nabla$ wherein $\mathbf{I}$ is the identity tensor and $\mathbf{n}$ is the outward-pointing unit normal. In addition, $\delta$ is a Dirac delta function, equal to unity at the interface and zero otherwise, and $\tilde{A} (\tilde{t})$ is the time-dependent interface area. 
\textcolor{black}{As justified by the final paragraph of \citet{Stone}, in our frame of reference $\textbf{u} \cdot \textbf{n}=0$, which gives rise to equation \ref{gamma_equation}}. 

The dimensionless groups that appear in the governing equations are defined as

\begin{equation}
Re =\frac{\rho_l  \textcolor{black}{c} \lambda}{\mu_l}, ~~~
We =g^{1/2}\lambda\sqrt{\frac{\rho_l}{\sigma_s}},  ~~~
Fr =\frac{ c }{\sqrt{g \lambda}},  ~~~
Pe_s=\frac{ c \lambda}{\mathcal{D}_s},~~~
\beta_s= \frac{\Re \mathcal{T} \Gamma_\infty}{\sigma_s}, 
\end{equation}
where $Re$, $We$, $Fr$, and $Pe_s$ stand for the Reynolds, Weber, Froude, and (interfacial) Peclet numbers.  \textcolor{black}{As a consequence of our nondimensionalisation, the phase speed is $c=\sqrt{g\lambda}$, so the Froude number becomes
$\mathrm{Fr}=c/\sqrt{g\lambda}=1$, consistent with the formulation of \citet{iafrati2009numerical}. 
We note that a depth-based definition would
yield $\mathrm{Fr}=\sqrt{2}$ for the present configuration, as described by \citep{treske1994undular,chanson2009current,gavrilyuk2016spilling,richard2019new}.
} 
The parameter $\beta_s$ is the surfactant elasticity number that is a measure of the sensitivity of the surface tension, $\sigma$, to the interface surfactant concentration, $\Gamma$. \textcolor{black}{Here, $\Re$ is the ideal gas constant value ($\Re = 8.314$ J K$^{-1}$ mol$^{-1}$),  $\mathcal{T}$ denotes temperature,  and $\mathcal{D}_s$ stands for the interfacial diffusion coefficient.}

The non-linear Langmuir equation
is used to describe $\sigma$ in terms of $\Gamma$, this is  
\begin{equation}
    \tilde{\sigma}=\text{max} [0.05, 1 + \beta_s \ln{ (1 -\tilde{\Gamma})]},
    \label{eq:Langmuir}
\end{equation}
where a lower limit of $\sigma$ has been set to $0.05$, below which the Langmuir equation of state may diverge. This formulation is common as previous studies \citep{constanteamores2020bb,atasi2018influence,muradoglu2008front,teigen2011diffuse,Shin_jcp_2018}. Finally, the Marangoni stress, $\tilde{\tau}$, is expressed as a function of $\tilde{\Gamma}$ as
\begin{equation}
\frac{1}{\Web} \nabla_s \tilde{\sigma}\cdot {\mathbf{t}}  \equiv
\frac{\tilde{\tau}}{\Web} = -
\frac{\Mar }{(1-\tilde{\Gamma})}  \nabla_s\tilde{\Gamma}\cdot {\mathbf{t}},
\end{equation}
\noindent
where $Ma=\beta_s/We=Re \mathcal{T} \Gamma_\infty/\rho_l \textcolor{black}{c}^2\lambda$ is the Marangoni parameter.
Tildes are dropped henceforth.

The incompressible Navier-Stokes equation  is integrated numerically using a second-order finite difference formulation on a staggered grid \citep{Harlow_1965,temam1968stabilite}. The flow field is represented on a fixed, uniform Eulerian mesh. All spatial derivatives are discretized with second-order centered differences, with the exception of the nonlinear advective terms  for which a second-order essentially non-oscillatory (ENO) scheme is employed to ensure numerical stability and limit spurious oscillations \citep{sussman1994level}. Viscous stresses are likewise evaluated using a second-order centered scheme. The incompressibility constraint is enforced through a fractional-step projection method, in which the pressure is obtained by solving a Poisson equation using a multigrid iterative solver \citep{Chorin_1968,kwak2004multigrid}. To accurately resolve interfacial dynamics, the numerical framework incorporates an adaptive Lagrangian mesh within a hybrid front-tracking/level-set methodology \citep{Shin_jmst_2007,Shin_jmst_2017}. Coupling between the Lagrangian interface and the Eulerian flow solver is achieved via an immersed boundary formulation \citep{peskin1977numerical}.

The Lagrangian interface is advanced in time according to $dx/dt=V$, 
where the interface velocity $V$ 
\textcolor{black}{ is obtained by second-order bilinear interpolation of the Eulerian velocity field.} Time integration of the interface position is performed using a second-order Runge-Kutta scheme. Time integration of the interface position is performed using a second-order Runge-Kutta scheme. Parallel computation is implemented through a domain-decomposition approach, with data communication across subdomains handled using the Message Passing Interface (MPI) \citep{Shin_jmst_2017}. 
The surfactant conservation equation on the evolving interface is discretised within the hybrid Lagrangian-Eulerian LCRM framework, following \citet{muradoglu2008front} for the transient and convective terms, which are integrated conservatively over each triangular interface element using the Leibniz transport theorem.  The surface tension coefficient $\sigma$ is evaluated locally on the Lagrangian interface from the surfactant concentration. We refer the reader to \citet{Shin_jcp_2018} for more details of the surfactant implementation, including validation against standard benchmark test cases.




\subsection{Numerical Setup and parameters}

 \begin{figure}
 \centering
 \begin{tabular}{c}
 \includegraphics[width=0.4\textwidth]{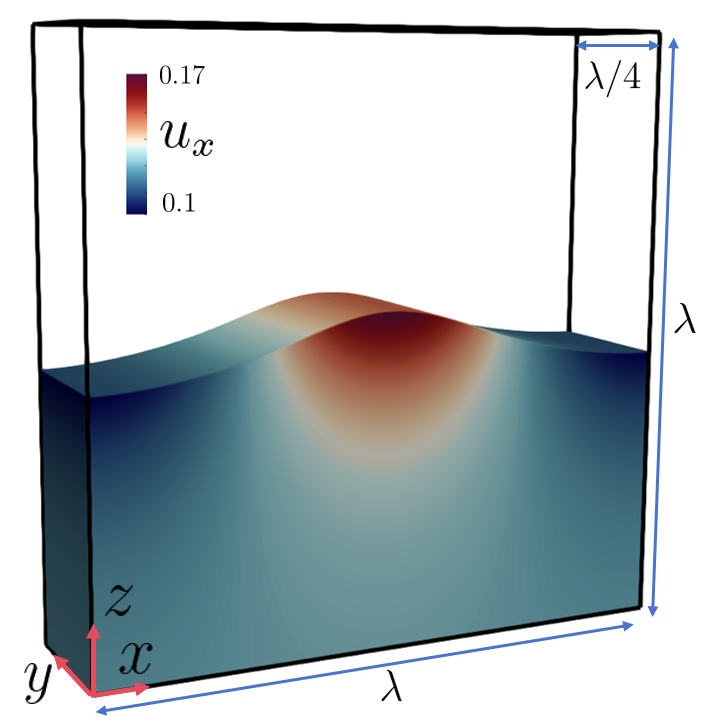}
 \end{tabular}
 \caption{Schematic representation of the flow configuration. Initial configuration showing the computational domain in a 3D Cartesian coordinate system, ${\bf x} = (x, y, z)$. }\label{configuration}
 \end{figure}

In this work, we study the evolution of a single breaking wave with wavelength $\lambda$. The computational domain is defined as $(\lambda \times \lambda/4 \times \lambda)$, as shown in  figure~\ref{configuration}. While previous studies have commonly employed cubic domains of size $\lambda^3$, the reduced spanwise extent adopted here is intentional, as we are not focusing on the strongly 3D dynamics associated with the plunging regime. By reducing the spanwise dimension, we limit the computational cost while retaining 3D dynamics. This approach has  also been adopted by \citet{di2024air}. \CRCA{No variations in surfactant concentration are observed along the spanwise direction, indicating that the dynamics are effectively two-dimensional.}  
The wave  propagates in the $x$-direction with periodic boundary conditions in both the streamwise ($x$) and spanwise ($y$) directions. No-slip and no-penetration conditions are applied at the bottom boundary. 
We emphasize that the chosen spanwise extent is sufficient to capture the essential dynamics of the breaking wave while maintaining computational efficiency.

 The initial free-surface profile $\eta(x,y)$ follows the previous work from \citet{Chen} and \citet{iafrati2009numerical},  which corresponds to a  third-order Stokes wave solution that neglects depth effects and \textcolor{black}{third-order corrections to the fundamental frequency component}, computed up to third order in the wave amplitude  $a$. The surface elevation is prescribed as
\begin{equation}
\eta(x,y) = \frac{a}{\lambda} \left( \cos(kx) + \frac{1}{2} \varepsilon \cos(2kx) + \frac{3}{8} \varepsilon^2 \cos(3kx) \right),
\end{equation}\label{Third order Stokes wave}
where $k = 2\pi/\lambda$ is the  wavenumber, and $\varepsilon = ak$ is the initial wave steepness.
As noted by  \citet{iafrati2009numerical}, the omission of the \textcolor{black}{third-order correction term} has a negligible effect on the resulting dynamics; we therefore adopt this initialization and focus on surfactant-induced effects.
The initial velocity field is obtained under the assumption of a free surface (e.g., negligible air pressure). The velocity profile in the water is derived from the velocity potential associated with the free-surface profile and is expressed as\begin{equation}
u = \Omega a e^{kz} \cos(kx), \qquad v = \Omega a e^{kz} \sin(kx),
\end{equation}
where $\Omega = \sqrt{gk(1 + \varepsilon^2)}$ accounts for the nonlinear correction by removing higher order terms \citep{whitham2011linear,iafrati2009numerical}. We note that other studies have considered more physically detailed conditions, in which a velocity profile in the air is also defined \textcolor{black}{\citep{deike2015capillary}}. Finally, we note that  surface tension forces are not included in the initializations, but their influence will become relevant once the wave propagates in the streamwise direction.

The dimensionless parameters used in this study are consistent with experimentally realizable systems. The  density and viscosity ratios are set to $\rho_g / \rho_l = 1/850$ and $\mu_g / \mu_l = 17.4 \times 10^{-6} / 8.9 \times 10^{-4} = 1.96 \times 10^{-2}$, respectively,  based on standard properties of air and water. The water depth is set to  $h = \lambda/2$, consistent with the configuration used by \textcolor{black}{\citet{deike2015capillary}}, who showed that this ratio has minimal influence on wave evolution in the regimes of interest. Following  \citet{iafrati2009numerical}, we fix the Weber and Reynolds numbers at $ We = 100 $ and $ Re = 10^4 $. The chosen  Weber number corresponds to water waves with a wavelength of approximately $27$ cm, corresponding to Reynolds number of $ Re \approx 4.4 \times 10^5$. Resolving flows at such high Reynolds would require prohibitively fine resolution, which is not feasible within the constraints of our fixed-grid approach. While adaptive mesh refinement has enabled higher Re simulations in other studies \citep{Deike_prf_2018,mostert2022high}, our focus is limited to regular \textcolor{black}{(non-breaking)} and weakly spilling breakers. The chosen parameters therefore strike a balance between computational feasibility and physical realism, while capturing the dominant surfactant-laden dynamics.
In this work, we have considered insoluble surfactants whose critical micelle concentration (CMC), i.e. $\Gamma_\infty \sim \mathcal{O}(10^{-6})$ mol m$^{-2}$  for NBD-PC (1-palmitoyl-2-{12-[(7-nitro-2-1,3-benzoxadiazol-4-yl)amino]dodecanoyl}-sn-glycero-3 -phosphocholine); thus, we have explored the range of $0.1 <\beta_s<0.5$ which corresponds to \CRCA{$2.9 \times 10^{-6} < \mathrm{CMC} < 1.4 \times 10^{-5}~\mathrm{mol\,m^{-2}}$}.
%
 \textcolor{black}{The values $\beta_s = 0.3$ and $\beta_s =0.5$ were selected to ensure that the interface remains strictly below the critical value of the Langmuir equation of state, so that it never enters the saturation (cutoff) regime. This guarantees that the surface tension varies monotonically with surfactant concentration and that Marangoni stresses remain active throughout the simulation. To further avoid operating near the saturation limit of the Langmuir equation of state, where Marangoni stresses are suppressed, the initial surfactant concentration is set to $\Gamma_0 = \Gamma_\infty/4$ in all simulations, ensuring that both surface-tension reduction and Marangoni stresses remain active.
Finally, the initial concentration of the surfactant-laden cases is   initialized uniformly along the interface as in previous work by \citet{Constante-Amores_prf_2020,constante2023impact,kamat_2020}. 
We note that a real Stokes wave at the onset of breaking may possess pre-existing surfactant gradients; our choice of a uniform initial concentration is an idealization. } We have set $Pe_s=10^3$ following 
\citet{batchvarov2020effect} and \citet{Constante-Amores_prf_2020}, who showed that the interfacial dynamics are weakly-dependent on $Pe_s$ beyond this value. 
The simulations presented here are mesh independent. A resolution of 
$\lambda/\Delta {\bf x}=512$ was found sufficient to ensure that further mesh refinement does not alter the numerical results. This resolution is consistent with that used in previous studies at the same dimensionless parameters \citep{iafrati2009numerical}.
We also note that the simulations in the plunging regime are terminated  upon the onset of plunging. Although the numerical framework is capable of resolving  post-breaking behavior, this is not the focus of the current study. Further investigation would require refining the simulations, which would significantly increase computational cost, and \textcolor{black}{consequently}, we do not explore this avenue here.

The numerical validation of the surfactant equations has been previously presented in \citet{Shin_jcp_2018}. 
Extensive mesh studies for surface-tension-driven phenomena using the same computational method can be found elsewhere, e.g. \citep{batchvarov2020threedimensional,batchvarov2020effect,Constante-Amores_prf_2020,constante_jets,sheet_constante,constante_2023}.
Additionally,  liquid volume and surfactant mass conservation are met under errors of less than $10^{-1}\%$, and $10^{-2}\%$, respectively.


\section{Results \label{Results}}

\begin{table}
\centering
\begin{tabular}{ccccc}
$\epsilon$ & $\beta_s$ & Regular & Spilling & Plunging \\
\hline
0.3 & Surfactant-free & X & & \\
0.3 & \textcolor{black}{0.3} & X & & \\
0.3 & \textcolor{black}{0.5} & X & & \\
0.324 & Surfactant-free & & X & \\
0.324 & \textcolor{black}{0.3} & & X & \\
0.324 & \textcolor{black}{0.5} & & X & \\
0.33 & Surfactant-free & & X & \\
0.33 & \textcolor{black}{0.3} & & X & \\
0.33 & \textcolor{black}{0.5} & & X & \\
0.35 & Surfactant-free & & X & \\
0.35 & \textcolor{black}{0.3} & &X &  \\
0.35 & \textcolor{black}{0.5} & &X &  \\
0.37 & Surfactant-free & &  &X \\

\end{tabular}
\caption{ Wave behavior as a function of wave steepness $\epsilon$ and surfactant parameter $\beta_s$. \textcolor{black}{The steepness $\epsilon$ denotes the initial Stokes-wave
steepness prescribed in equation (2.8). The surfactant parameter $\beta_s$ denotes surfactant elasticity.}
} \label{wave_behavior}
\end{table}

Table \ref{wave_behavior} summarizes the observed wave behavior as a function of the steepness parameter $\epsilon$ with $Re=10^4$ and $We=100$.  
 We observe that for surfactant-free cases, the wave remains regular for $\epsilon = 0.3$, transitions to a spilling-type breakup for $\epsilon = 0.324$, and finally exhibits plunging-type breakup for $\epsilon = 0.37$.  
These findings agree with previous work by \citet{iafrati2009numerical} and \citet{Grue}. We have  included simulations with  $\epsilon=0.324$, which aligns with the onset from regular to spilling-type breaking  predicted when accounting for the secular growth term $\epsilon^3/8$ by \citet{Grue}. It should be noted that for $\epsilon = 0.37$, the simulation was terminated just prior to the breakup. This choice was made to focus the analysis on the wave behavior leading up to the event. A detailed study of the post-breakup dynamics in the  plunging regime is beyond the scope of this study.

\begin{figure}
 \centering
 \begin{tabular}{c}
 \includegraphics[width=0.8\textwidth]{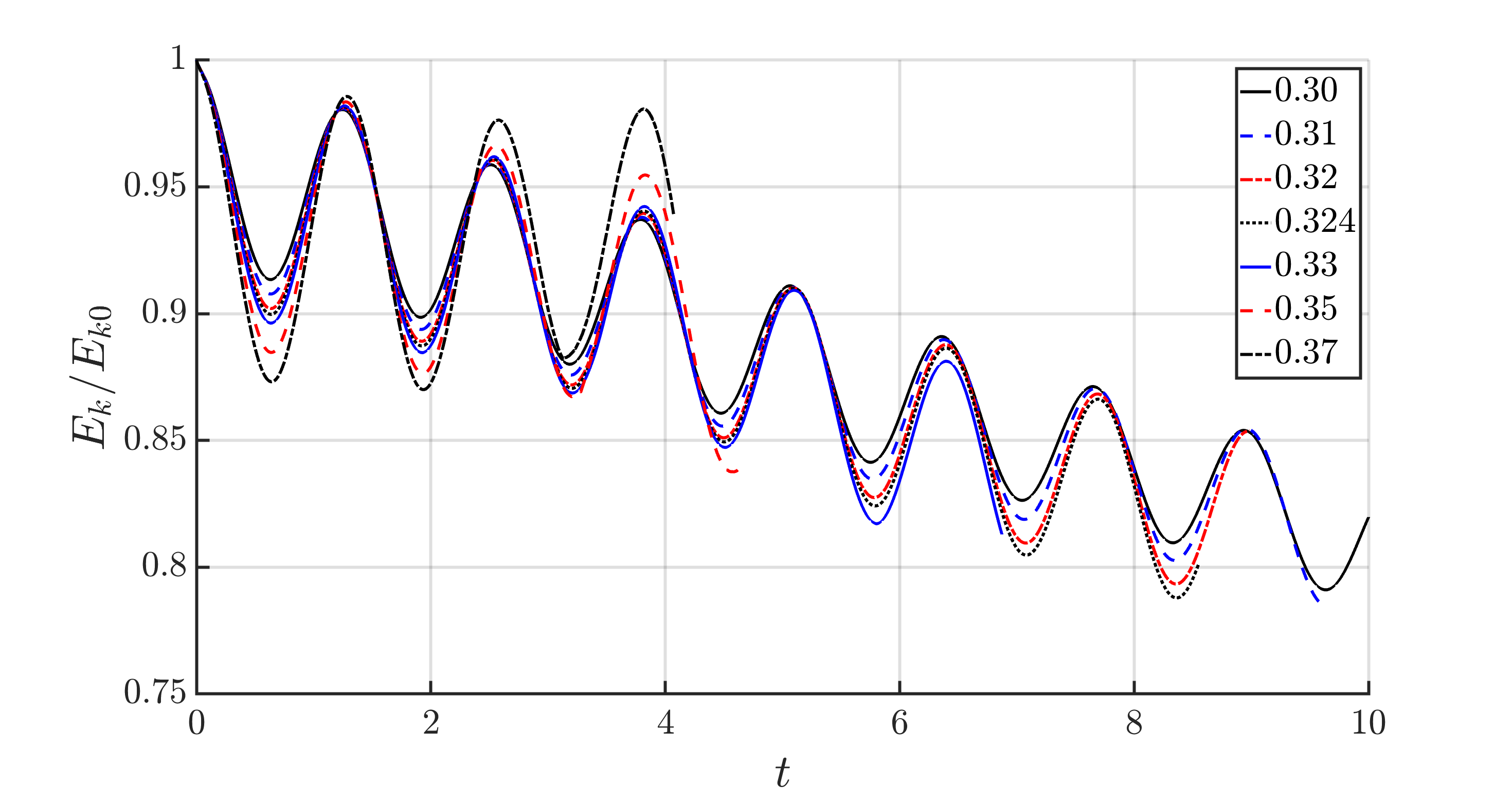}
 \end{tabular}
\caption{\textcolor{black}{Time evolution of  kinetic energy  as a function of the wave steepness $\epsilon$ for  surfactant-free cases.}}\label{Ek_clean}
\end{figure}

To investigate the role of wave breaking in energy dissipation, we track the temporal evolution of the kinetic energy. Figure~\ref{Ek_clean} presents the \textcolor{black}{ kinetic energy  normalised by its initial value $\textcolor{black}{E_{k0}}$}, defined as  $E_k/\textcolor{black}{E_{k0}}=\rho\int_v\textbf{u}^2/2dV$, as a function of the initial wave steepness $\epsilon$. 
The figure reveals distinct trends corresponding to the different breaking regimes,
highlighting the increasing intensity of energy decay with more intense wave breaking.
In the regular wave regime, a nearly constant decay rate is observed. The theoretical decay of mechanical energy due to viscous damping is given by $E_{mech}(t)=E_{mech}(0)e^{-2\gamma t}$, where $E_{mech}$ is the mechanical energy (e.g., equal to twice the kinetic energy for small-amplitude waves) and $\gamma=2k^2\mu_l/\rho_l$ is the damping coefficient 
 \citep{landau1987fluid} (see Appendix D for more detail). 
 For the wave with steepness   $\epsilon=0.3$, the numerical simulation results in  $\gamma=0.0091$, which is in good agreement with the theoretical estimate 
 $\gamma\approx0.0079$. 
At higher steepness $\epsilon=0.37$, a monotonic decrease in kinetic energy is observed, preceded by a phase of increasing amplitude and enhanced dissipation. Although the simulation is terminated just before the onset of plunging-type breakup, the flow already exhibits intensified velocities and steep velocity gradients near the interface, resulting in large kinetic energy and enhanced energy dissipation. \textcolor{black}{
Nonlinear energy exchange in the finite-amplitude wave produces oscillations in $E_k/E_{k0}$, so the linear dissipation theory applies only to the average decay rate rather than the instantaneous dynamics.
}

\begin{figure}
\centering
\begin{tabular}{ccc}
\hline
$\epsilon=0.3$&$\epsilon=0.33$&\textcolor{black}{$\epsilon=0.35$}\\
\hline
$\text{Surfactant-free}$&&\\
\includegraphics[width=0.3\textwidth]{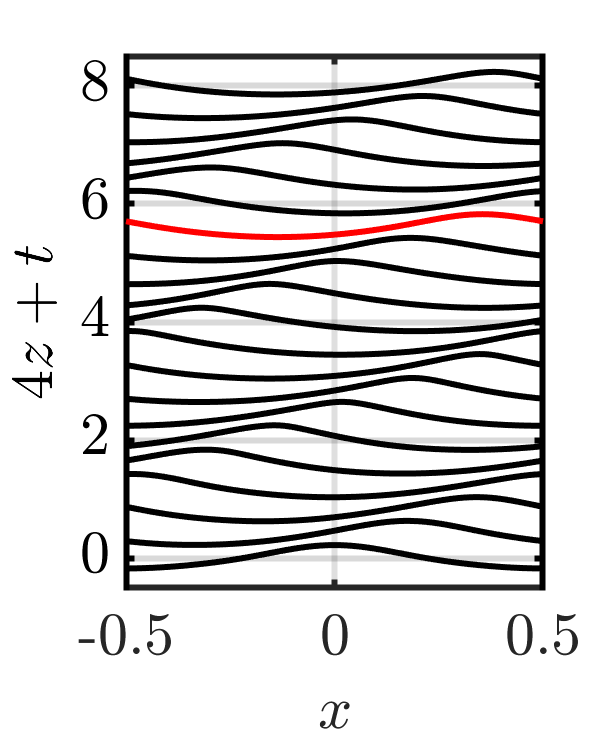}&
\includegraphics[width=0.3\textwidth]{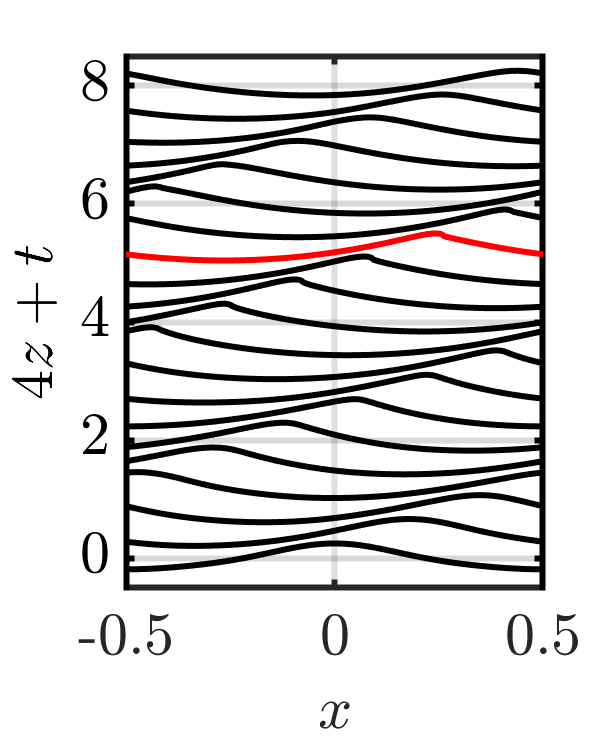}&
\includegraphics[width=0.3\textwidth]{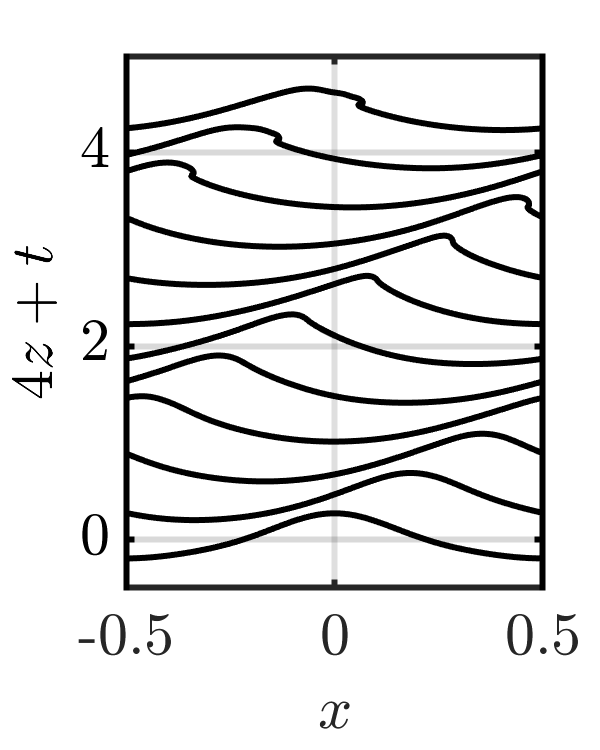}\\
(a) & (b) &(c) \\ 
$\beta_s=0.3$&&\\
\includegraphics[width=0.3\textwidth]{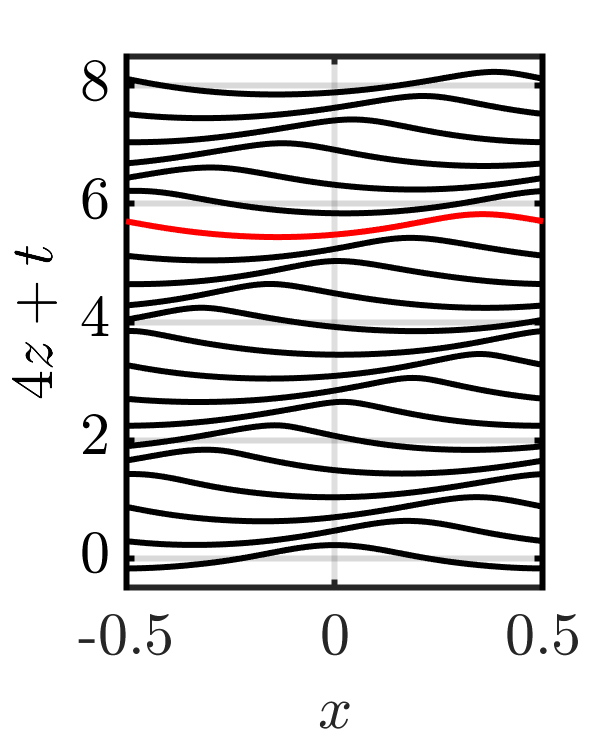}&
\includegraphics[width=0.3\textwidth]{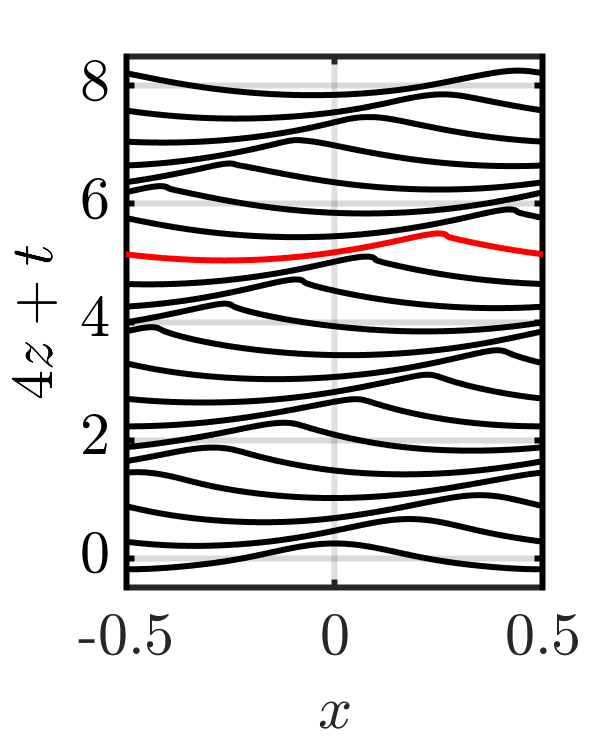}&
\includegraphics[width=0.3\textwidth]{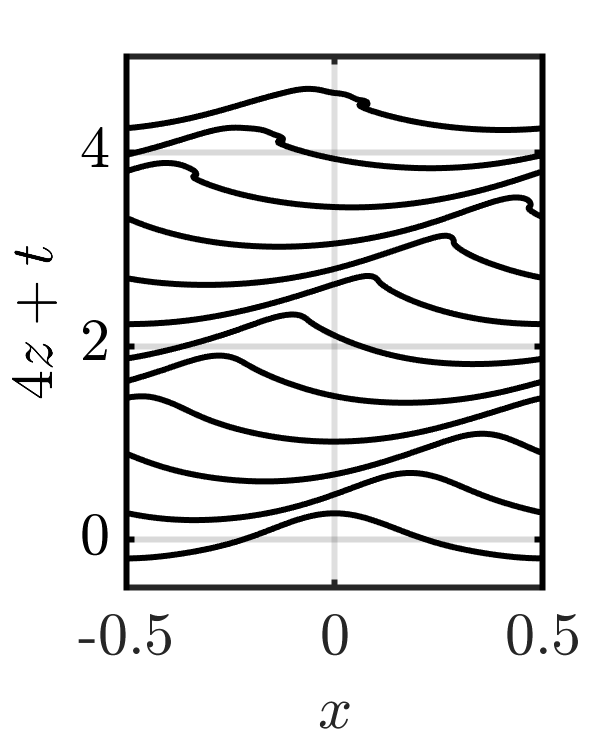}\\
(d) & (e) &(f) \\ 
$\beta_s=0.5$&&\\
\includegraphics[width=0.3\textwidth]{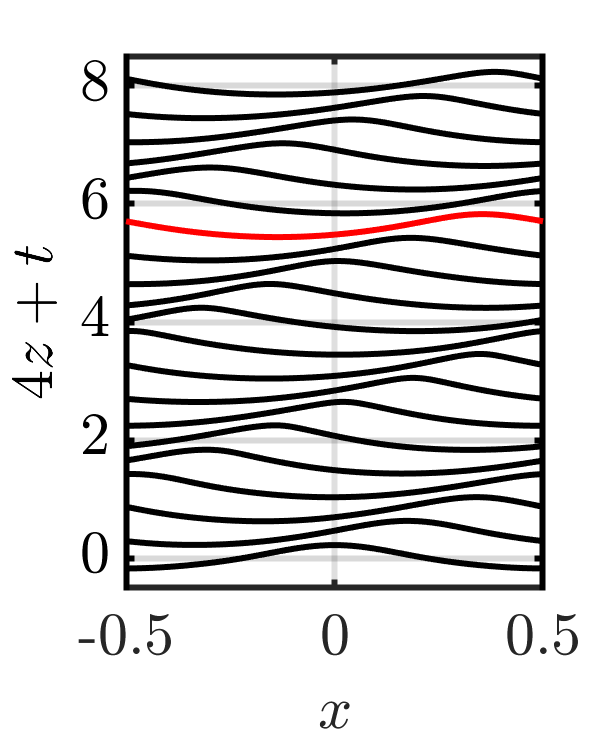}&
\includegraphics[width=0.3\textwidth]{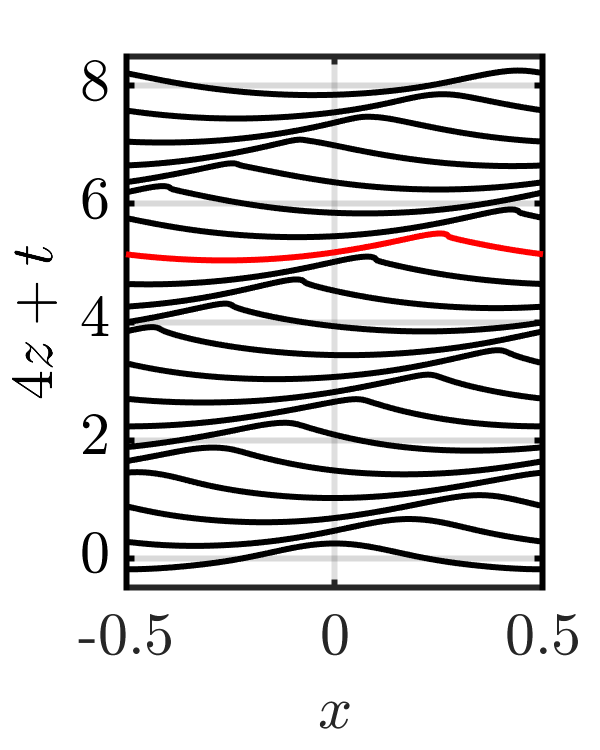}&
\includegraphics[width=0.3\textwidth]{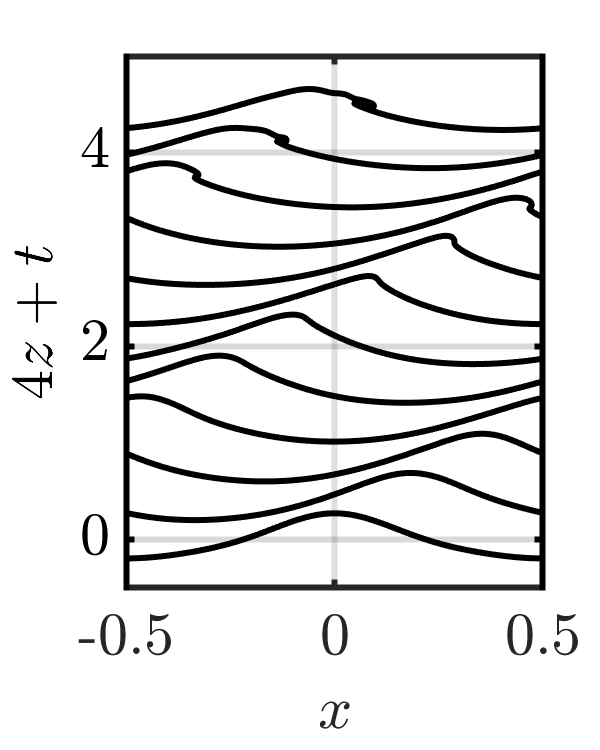}\\
(g) & (h) &(i) \\ 
\end{tabular}
\caption{Time-space plots of the interface in  the $(x-z)$ plane ($y=\lambda/8$)  at equal time intervals for varying wave steepness $\epsilon$, and surfactant elasticity $\beta_s$. The vertical coordinates (y-axis) are stretched by a factor of 4 and shifted upward by the corresponding time to better illustrate the interface evolution.
Colored lines mark selected times analyzed in detail below.} \label{interface_evolution}
\end{figure}

The changes in wave morphology that accompany these energy dissipation trends are illustrated in figure~\ref{interface_evolution}a-c, which  show the spatiotemporal evolution of the interface in the $(x\text{--}z)$ plane at $y = \lambda/8$ for initial steepness values \textcolor{black}{$\epsilon = [0.3,\ 0.33,\ 0.35]$}. 
At $\epsilon = 0.3$, the wave evolution remains regular over time. As the steepness increases to \textcolor{black}{$\epsilon = 0.33$} , weak signs of spilling emerge in the surfactant-free case, which become more pronounced at 
\textcolor{black}{$\epsilon = 0.35$}, where the formation of a forward-facing bulge near the wave crest marks the onset of spilling. The leading edge of this bulge, known as the toe, curves inward due to surface tension effects, consistent with the mechanism described by \citet{duncan2001spilling}. Capillary waves are also form upstream of the toe, as reported by \citet{duncan1999gentle}. 
\textcolor{black}{For gravity-capillary waves, \citet{deike2015capillary} reported that the breaking threshold depends primarily on the Bond (or Weber) number and the steepness. In the large Weber number regime relevant to our simulations ($We =100$, $Bo \sim 253$), their scaling predicts a critical steepness in the range $\epsilon_{\mathrm{crit}} \approx 0.32$. The transition observed here in the clean-wave case at $\epsilon_{\mathrm{crit}} \approx 0.324$ therefore lies within the gravity-capillary breaking regime reported by \citet{deike2015capillary}. We note, however, that their simulations were performed at a significantly
higher Reynolds number ($Re \approx 4\times10^{4}$) compared with the present $Re = 10^{4}$, so that the agreement should not be interpreted as a strict quantitative match but rather as a consistency of physical regimes.}

Figure~\ref{interface_evolution}(d--i) shows the effect of insoluble surfactants on the wave dynamics.
At $\epsilon = 0.3$ (see panels \ref{interface_evolution}d-g), the addition of surfactant produces minimal changes compared to the surfactant-free case. This is expected, as the steepness remains below the critical value for Stokes waves  ($\epsilon_c \approx 0.32$); although  surfactants reduce the surface tension, it is insufficient to generate a noticeable wave deformation. \textcolor{black}{When the wave enters the spilling regime, there are no visually obvious differences between surfactant cases and the surfactant-free case since the spilling signatures remain weak at this stage. The enhancement of spilling signs can be observed for higher steepness ($\epsilon = 0.35$), with the effect being more pronounced at higher values  of $\beta_s$ (see panels \ref{interface_evolution}f and \ref{interface_evolution}i).}


\begin{figure}
\centering
\begin{tabular}{ccc}
\hline
$\epsilon=0.3$&$\epsilon=0.33$& \textcolor{black}{$\epsilon=0.35$}\\
\hline
$\beta_s=0.3$&&\\
\includegraphics[width=0.33\textwidth]{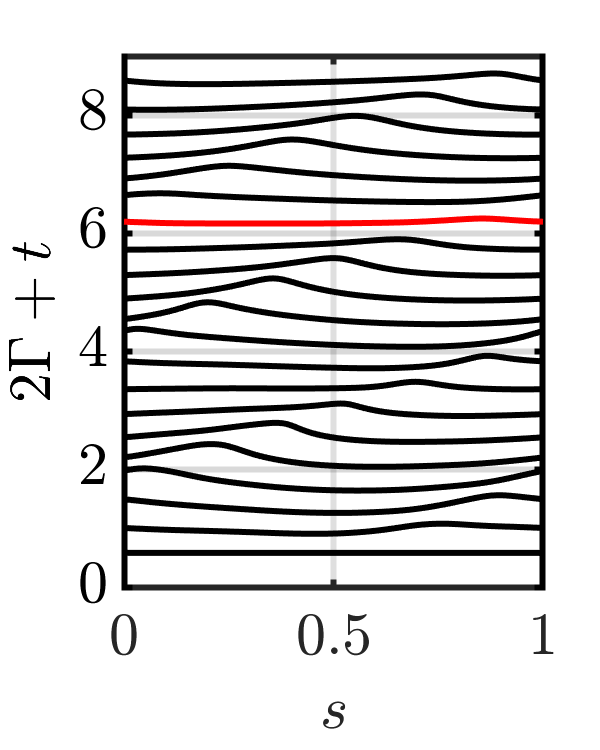}&
\includegraphics[width=0.33\textwidth]{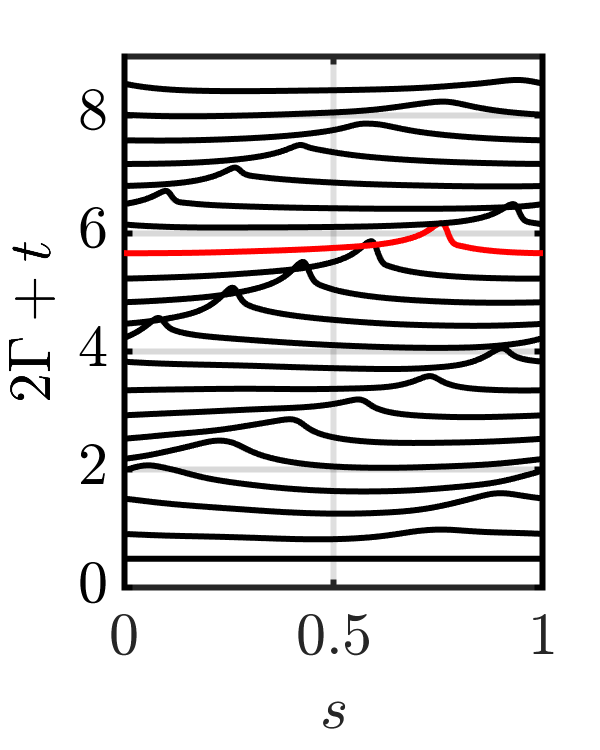}&
\includegraphics[width=0.33\textwidth]{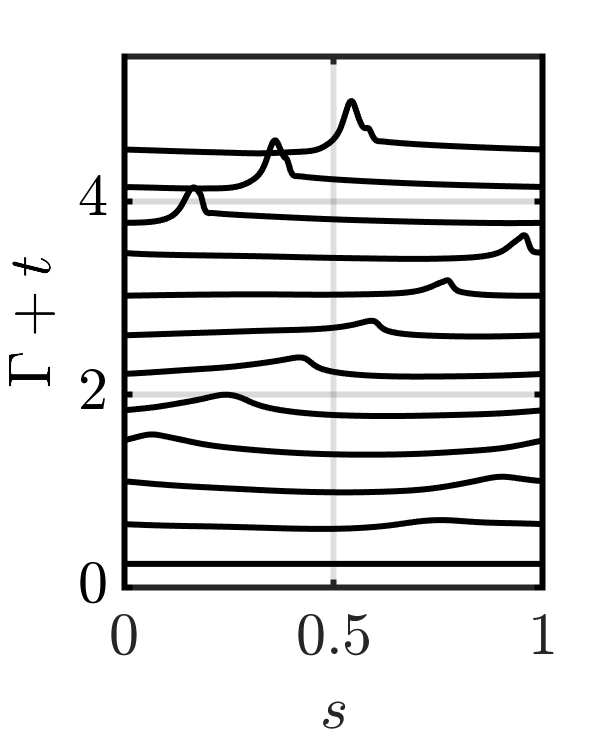}\\
(a) & (b) & (c)\\
$\beta_s=0.5$&&\\
\includegraphics[width=0.33\textwidth]{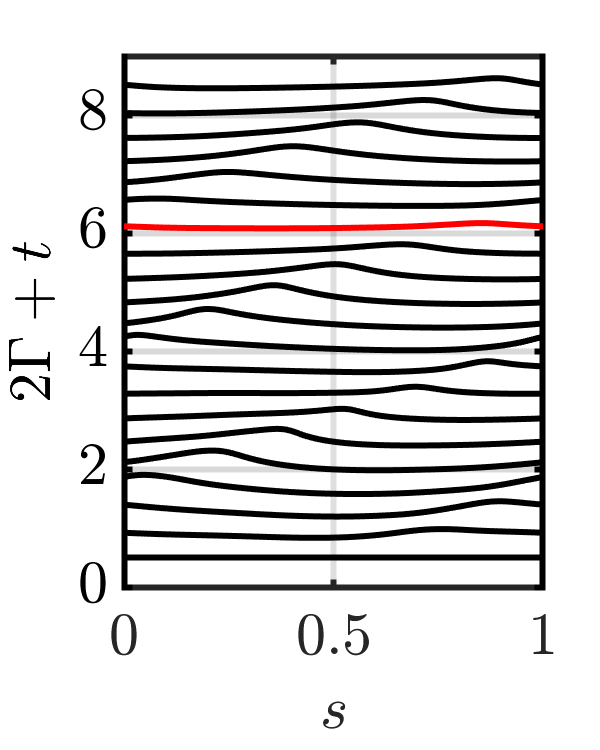}&
\includegraphics[width=0.33\textwidth]{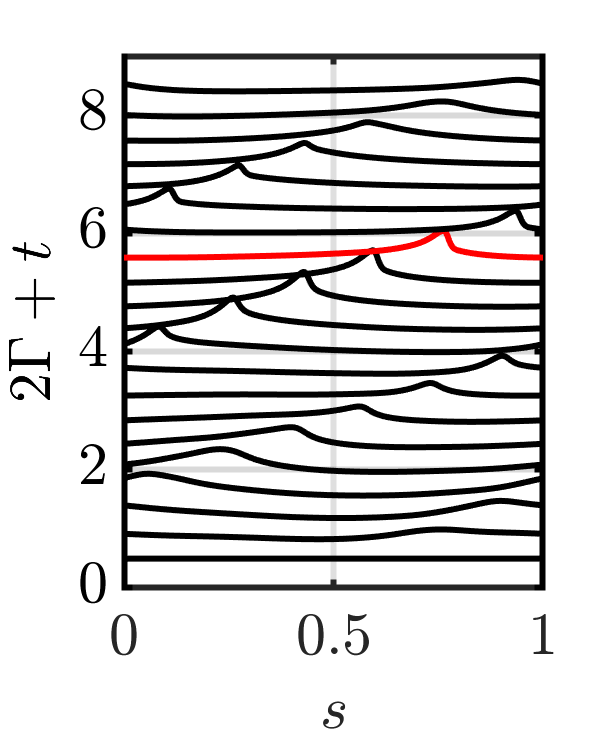}&
\includegraphics[width=0.33\textwidth]{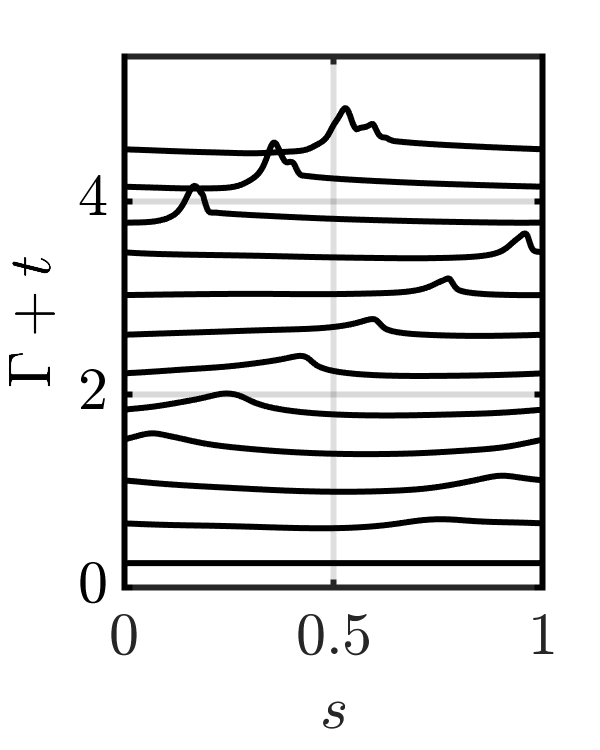}\\
(d) & (e) & (f)\\
\end{tabular}
\caption{Time-space plots of  normalized surfactant concentration $\Gamma$ at  time intervals as a function of the steepness $\epsilon$, and $\beta_s$. The data are plotted against arclength $s \in [0,1]$, with vertical coordinates representing the normalized surfactant concentration $\Gamma$ shifted by the corresponding time to clearly show the evolution of surfactant distribution. 
For the surfactant-laden cases, $Pe_s=10^3$ and \CRCA{$\Gamma=\Gamma_\infty/4$}. Colored lines mark selected times analyzed in detail below.} 
    \label{surfactant_profiles}
\end{figure}

To elucidate the coupling between the interface dynamics and surfactant concentration, figure~\ref{surfactant_profiles} presents the spatiotemporal evolution of  $\Gamma$ at the same time  as in figure~\ref{interface_evolution}. 
A clear correlation emerges between the interfacial shape and the distribution of $\Gamma$. 
 As the wave evolves, surfactant initially accumulates at troughs due to surface convergence, then shifts toward the   crests (where the interface exhibits divergent flow). Once the crest is formed, the local flow divergence redistributes $\Gamma$,  causing it to diffuse away gradually. Comparing cases with $\beta_s = 0.3$ and \CRCA{$\beta_s = 0.5$}, the crest in the higher $\beta_s$ leans forward more prominently, indicative of surfactant-induced modifications to the interfacial dynamics and wave deformation. The largest accumulation of surfactant concentration  occurs in the \textcolor{black}{$\epsilon = 0.35$} case,
 \textcolor{black}{ where enhanced crest steepening and incipient spilling intensify interfacial compression, leading to sharper concentration gradients and increased accumulation of $\Gamma$.  Noticeably, a double-peaked distribution of  $\Gamma$  emerges (see figure~\ref{surfactant_profiles}c and  \ref{surfactant_profiles}f), 
 originates from interfacial overturning (see 
  figure~\ref{interface_evolution}f and \ref{interface_evolution}i).} 

\begin{figure}
    \centering
\begin{tabular}{cc}
\includegraphics[width=0.48\textwidth]{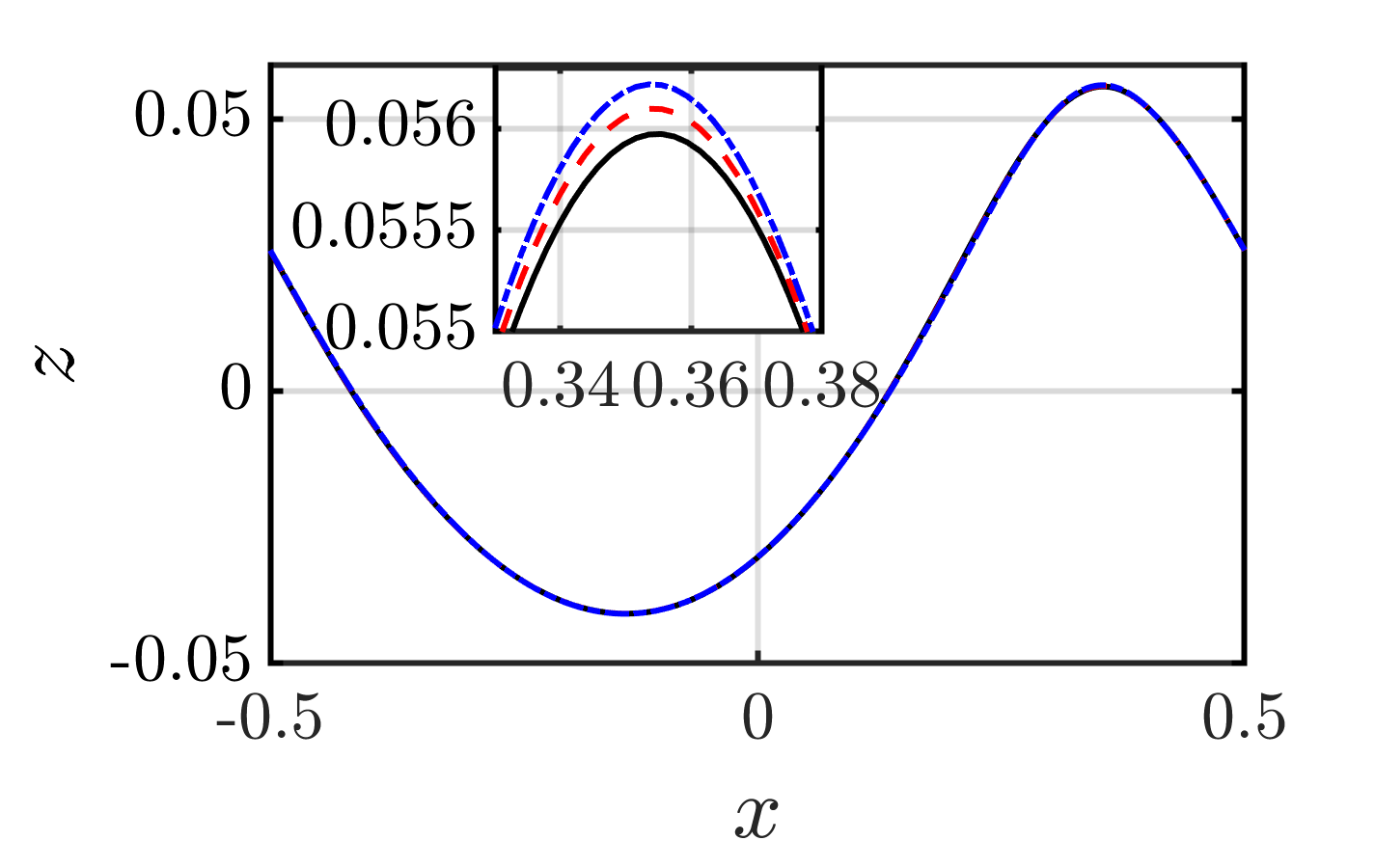}& 
\includegraphics[width=0.48\textwidth]{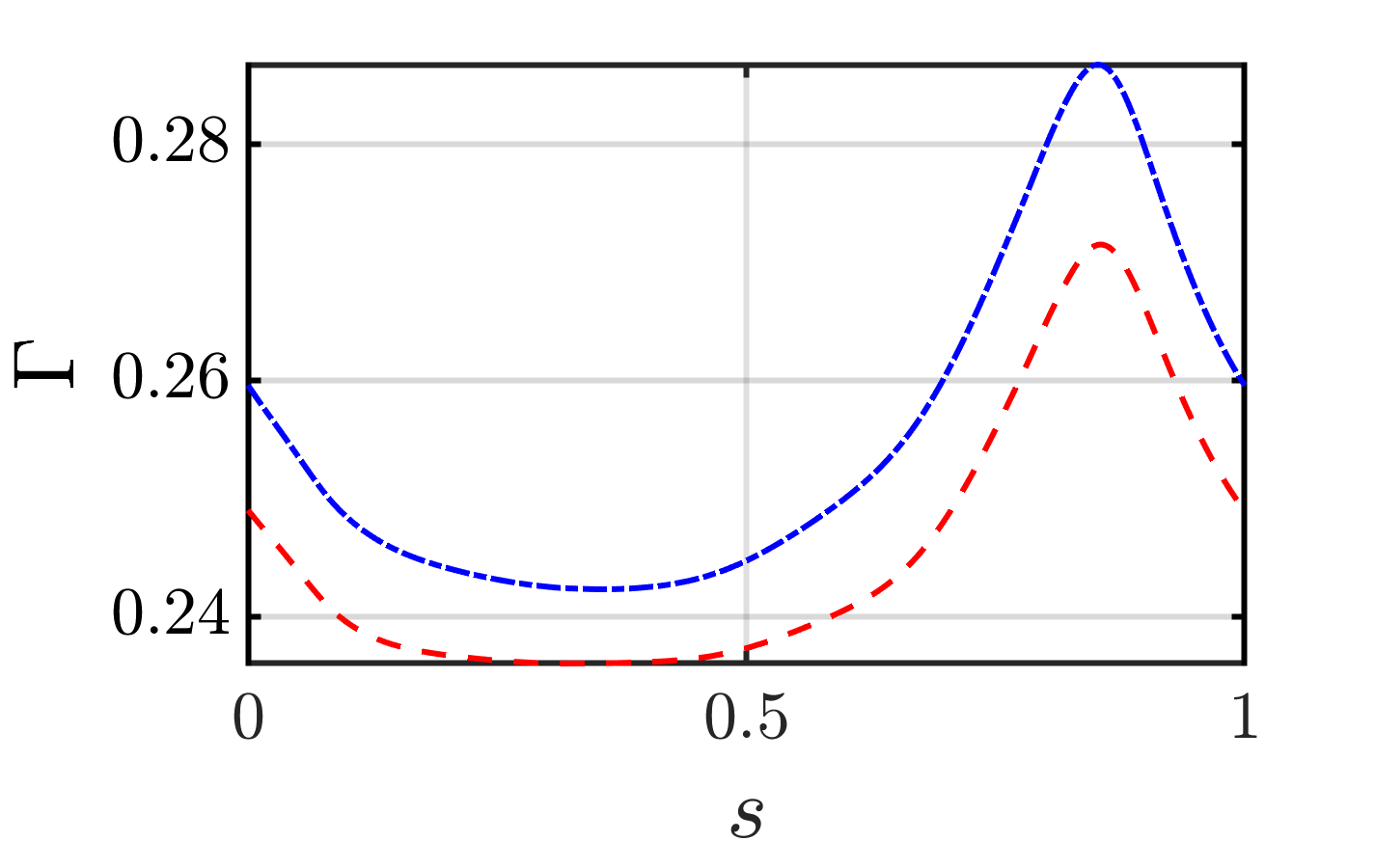}
\\
(a) & (b)\\
 \includegraphics[width=0.48\textwidth]{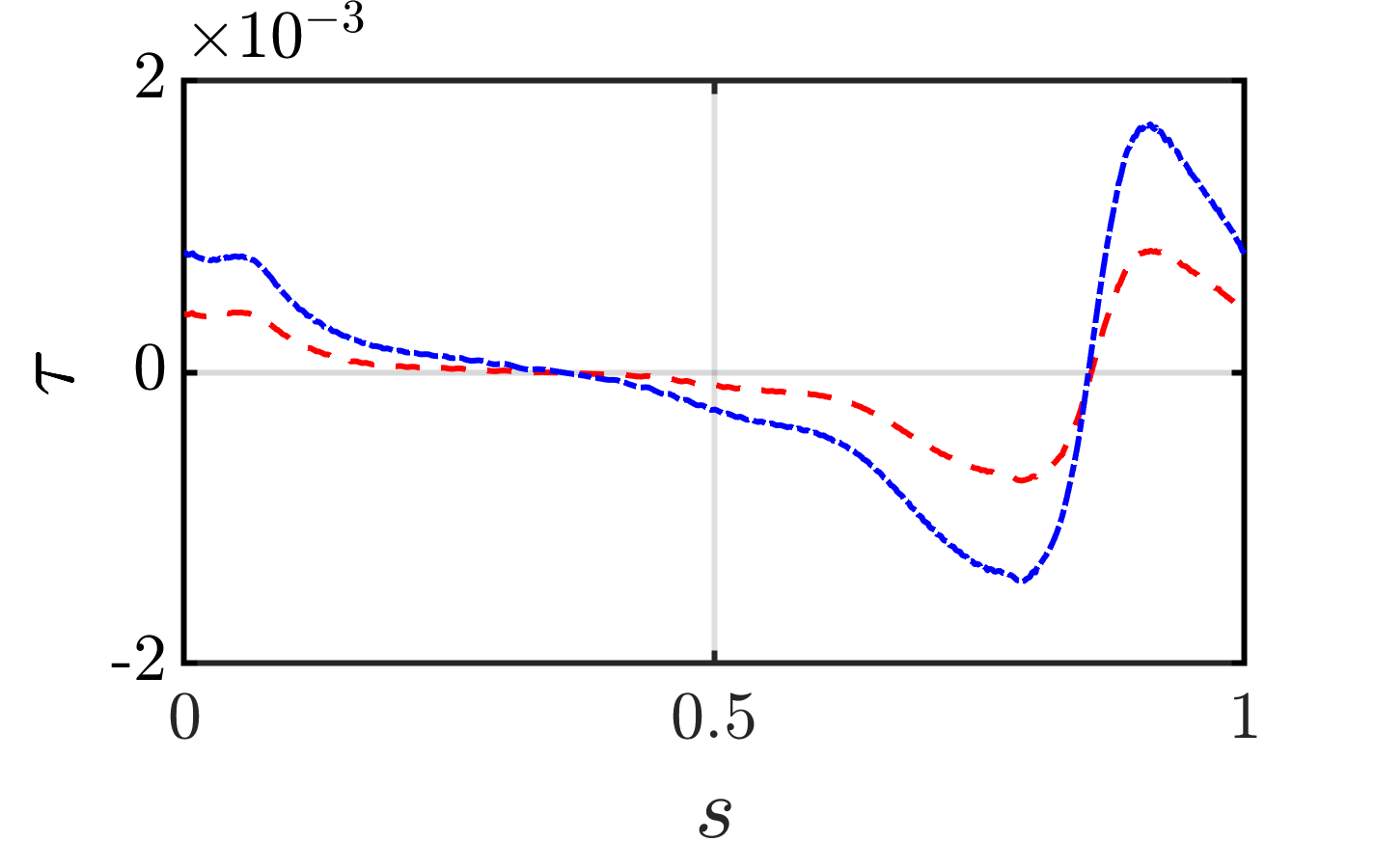}&
\includegraphics[width=0.48\textwidth]{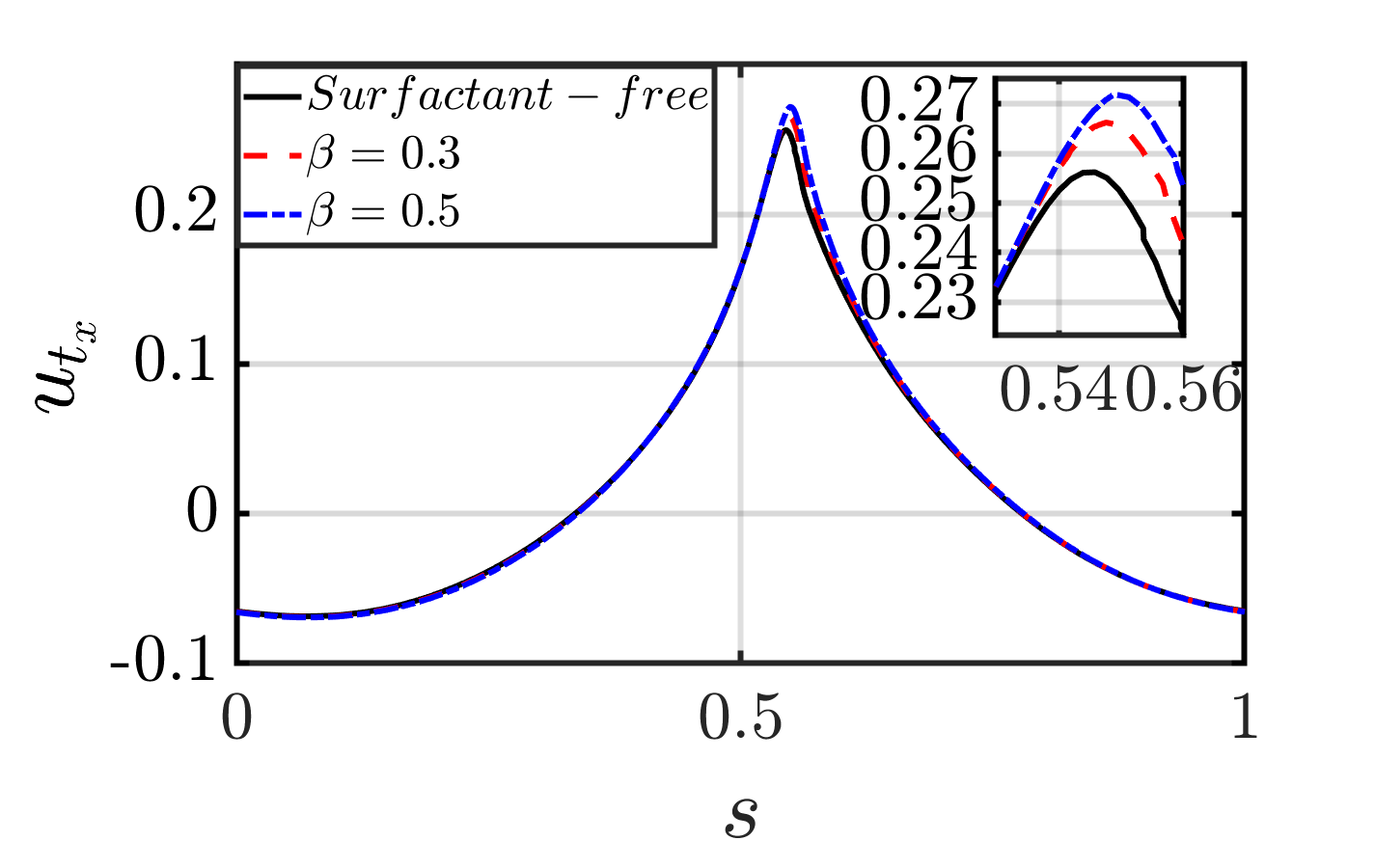}\\
(c) & (d)\\
\end{tabular}
\caption{Effect of the elasticity parameter $\beta_s$ for regular wave regime ($\epsilon=0.3$) at  $t=5.6$. Two-dimensional projections of the interface, $\Gamma$, $Ma$ and $u_{t_x}$ in the $(x–z)$ plane ($y=\lambda/8$) are shown in (a–d), respectively.  Note that the abscissa in (a) corresponds to the $x$ coordinate, and in (b–d) to the arc length, $s$.    }
    \label{ep_0d3_gamma_0d25}
\end{figure}

We now examine the coupling between the interface location, surfactant concentration $\Gamma$, Marangoni stress, and tangential velocity in the streamwise direction $u_{tx}$. To elucidate the underlying mechanisms, we select representative time instances at which wave features are most pronounced. These times,  corresponding to different steepness regimes,  are indicated in different colors in figure \ref{interface_evolution} and \ref{surfactant_profiles}.
We begin with the regular wave regime. For $\epsilon = 0.3$, figure~\ref{ep_0d3_gamma_0d25}a shows that  the presence of surfactant has a minimal effect on  the interface location at late times ($t = 5.6$). 
However, figure~\ref{ep_0d3_gamma_0d25}b reveals that the surfactant concentration peaks at the wave crest, where surface tension is locally minimized. The resulting Marangoni stresses, driven by surface tension gradients, generate outward tangential flow away from the crest, i.e., $\CRCA{\tau} > 0$ on the forward part of the crest and $\CRCA{\tau} < 0$ on the rear  part of it. These Marangoni stresses  act to suppress further steepening of the wave by opposing interfacial deformation, thereby reducing the wave amplitude at late times (not shown).
The influence of Marangoni stresses on the surface flow is evident in figure~\ref{ep_0d3_gamma_0d25}d, which  shows that $u_{tx}$ is significantly enhanced near the crest in the presence of surfactants, compared to the clean case, particularly for the high elasticity case ($\beta_s = \CRCA{0.5}$).
Despite the enhanced tangential velocity and the non-uniform surfactant distribution, the interface displacement remains small at this particular time. At $\epsilon = 0.3$, the inertial forces associated with wave propagation are insufficient to drive significant interface deformation or wave overturn, and  the flow remains in a weakly nonlinear regime, where the balance between gravity and surface tension dominates. Moreover, surface tension remains relatively strong and effectively resists interface steepening. As a result, the Marangoni-induced tangential flow primarily redistributes momentum along the interface without producing noticeable changes in the wave shape.

\begin{figure}
    \centering
\begin{tabular}{cc}
\includegraphics[width=0.5\textwidth]{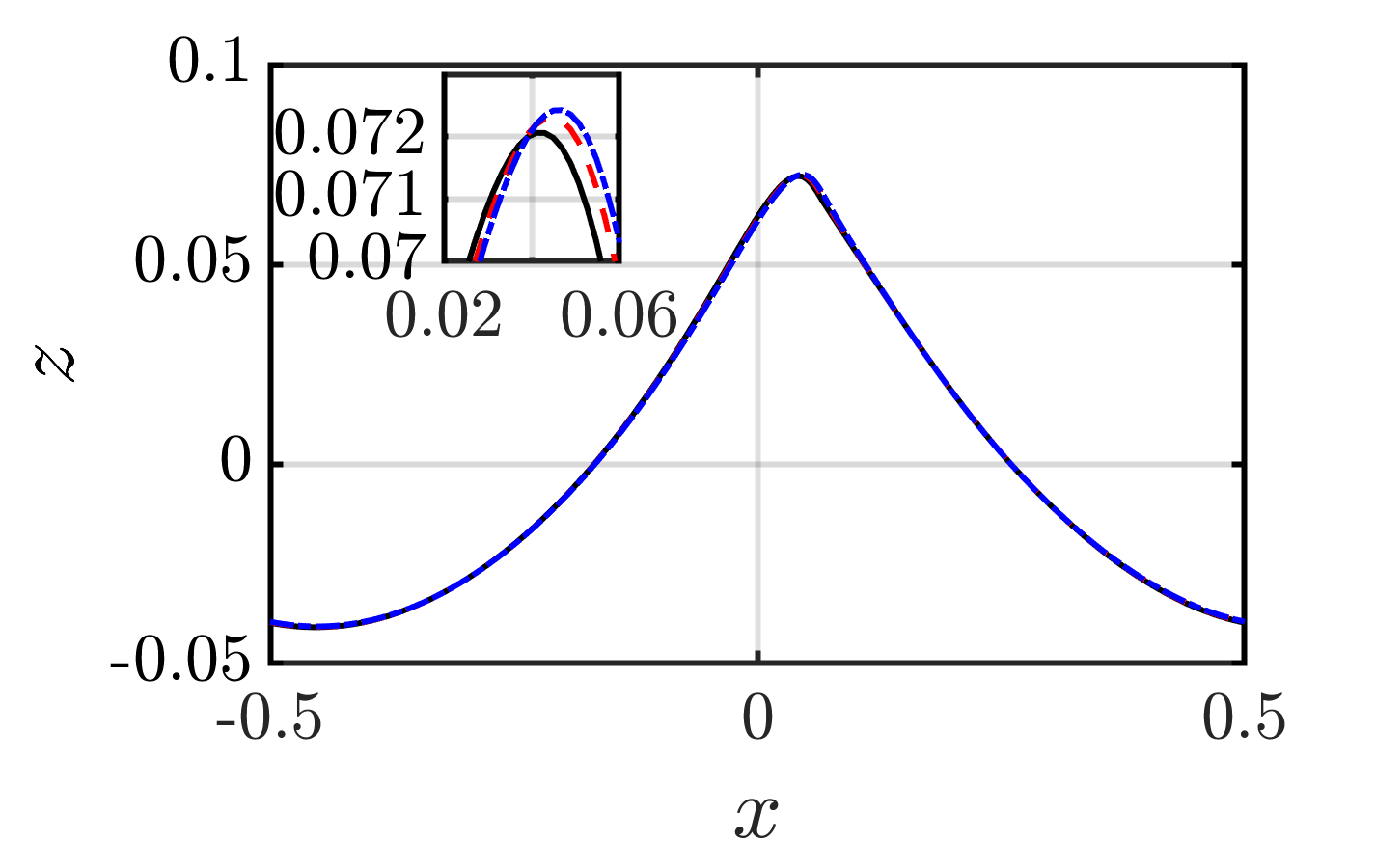} & \includegraphics[width=0.5\textwidth]{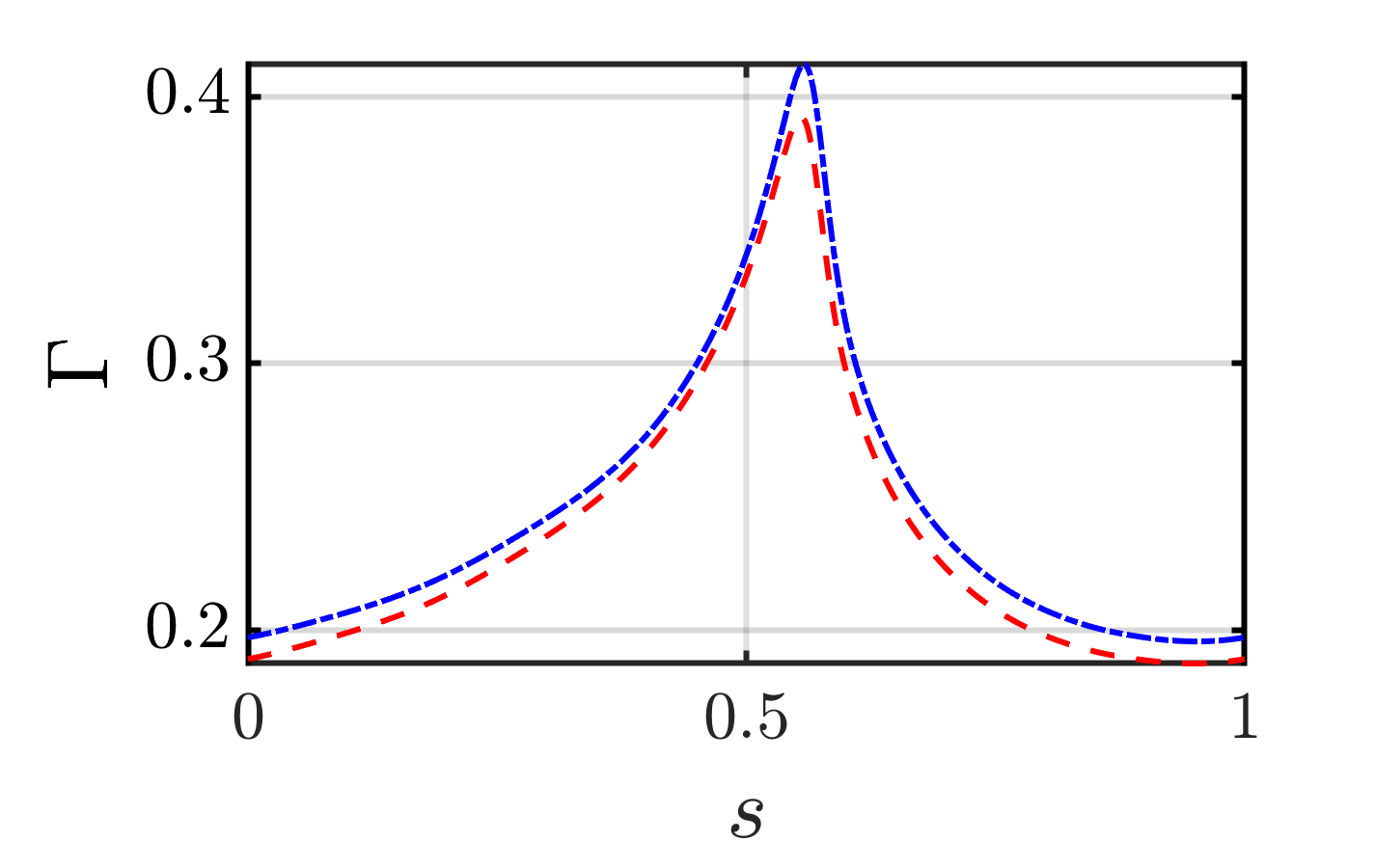} \\
(a) & (b) \\
\includegraphics[width=0.5\textwidth]{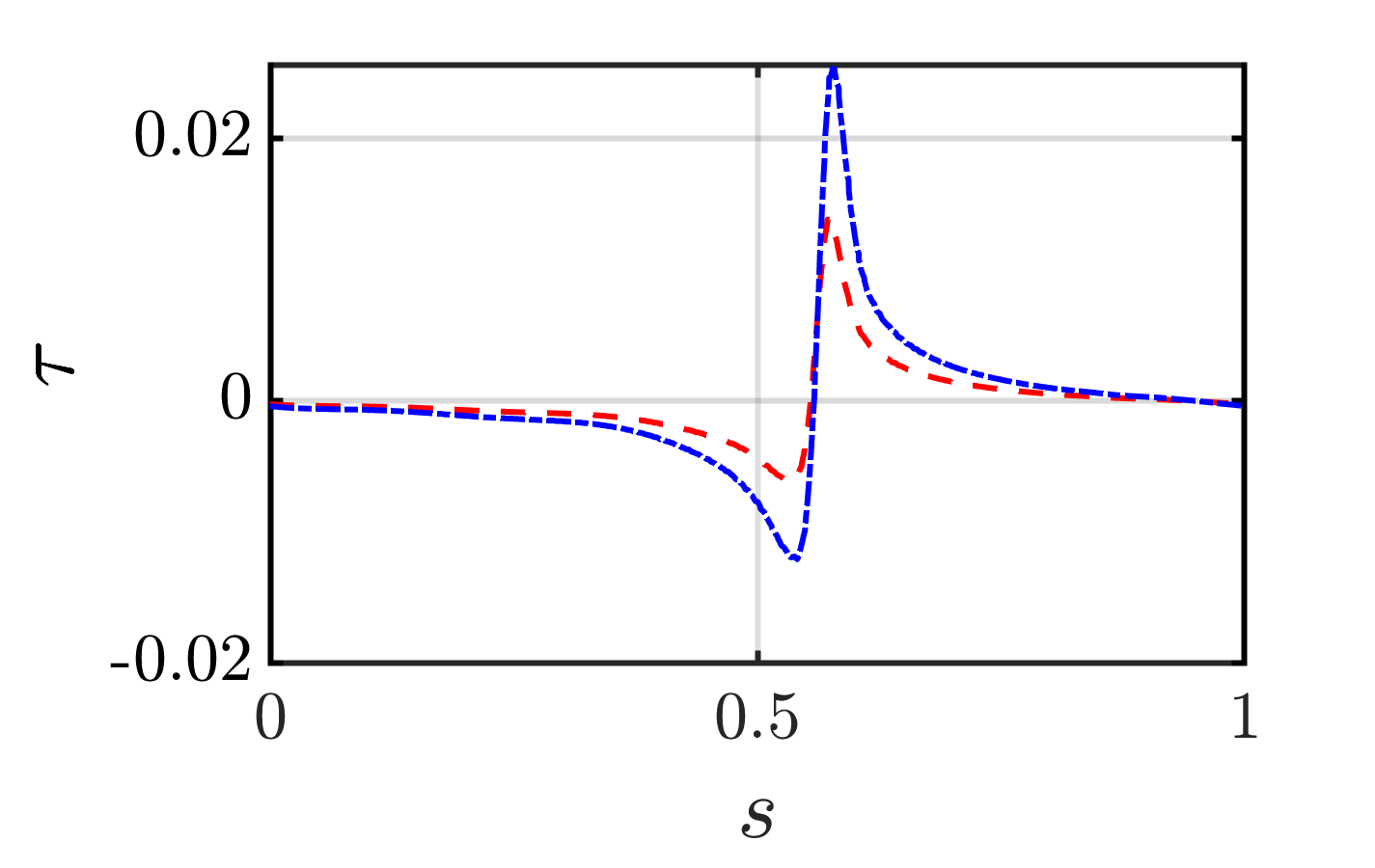} & \includegraphics[width=0.5\textwidth]{PIC_REV/Tangential_velocity_vs_arclength_for_peak_0d324.png} \\
(c) & (d) \\
\hline
\includegraphics[width=0.5\textwidth]{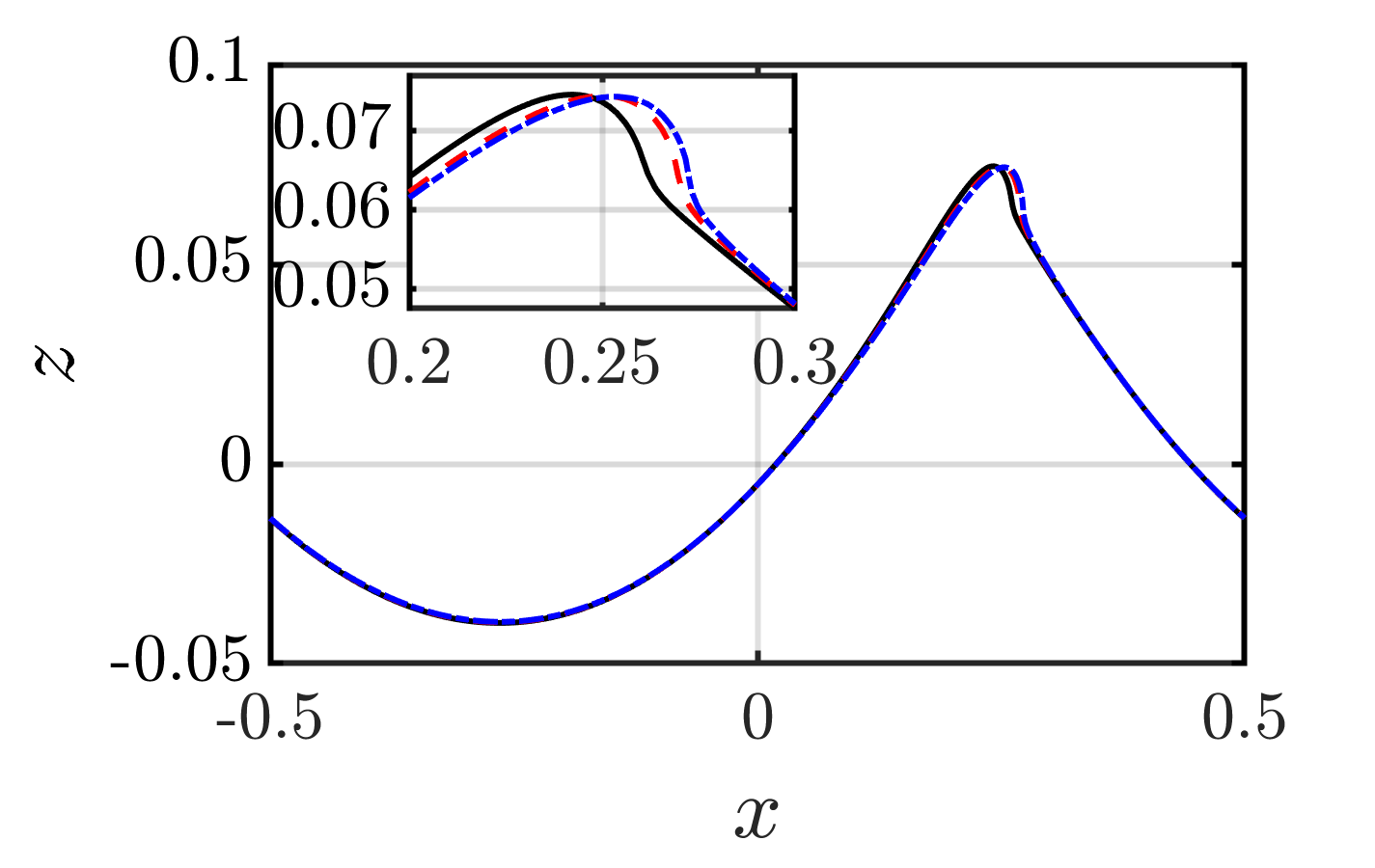} & \includegraphics[width=0.5\textwidth]{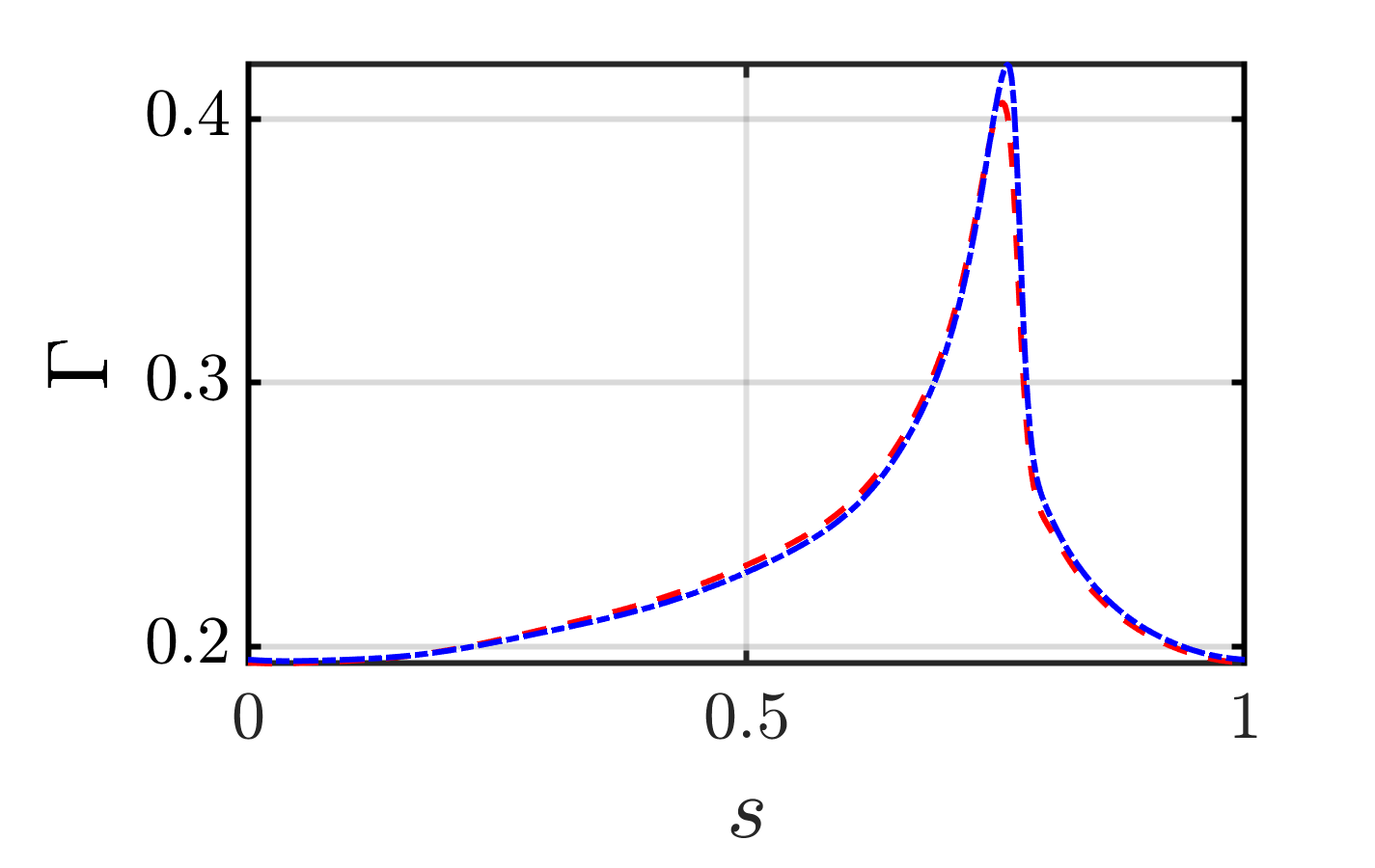} \\
(e) & (f) \\
\includegraphics[width=0.5\textwidth]{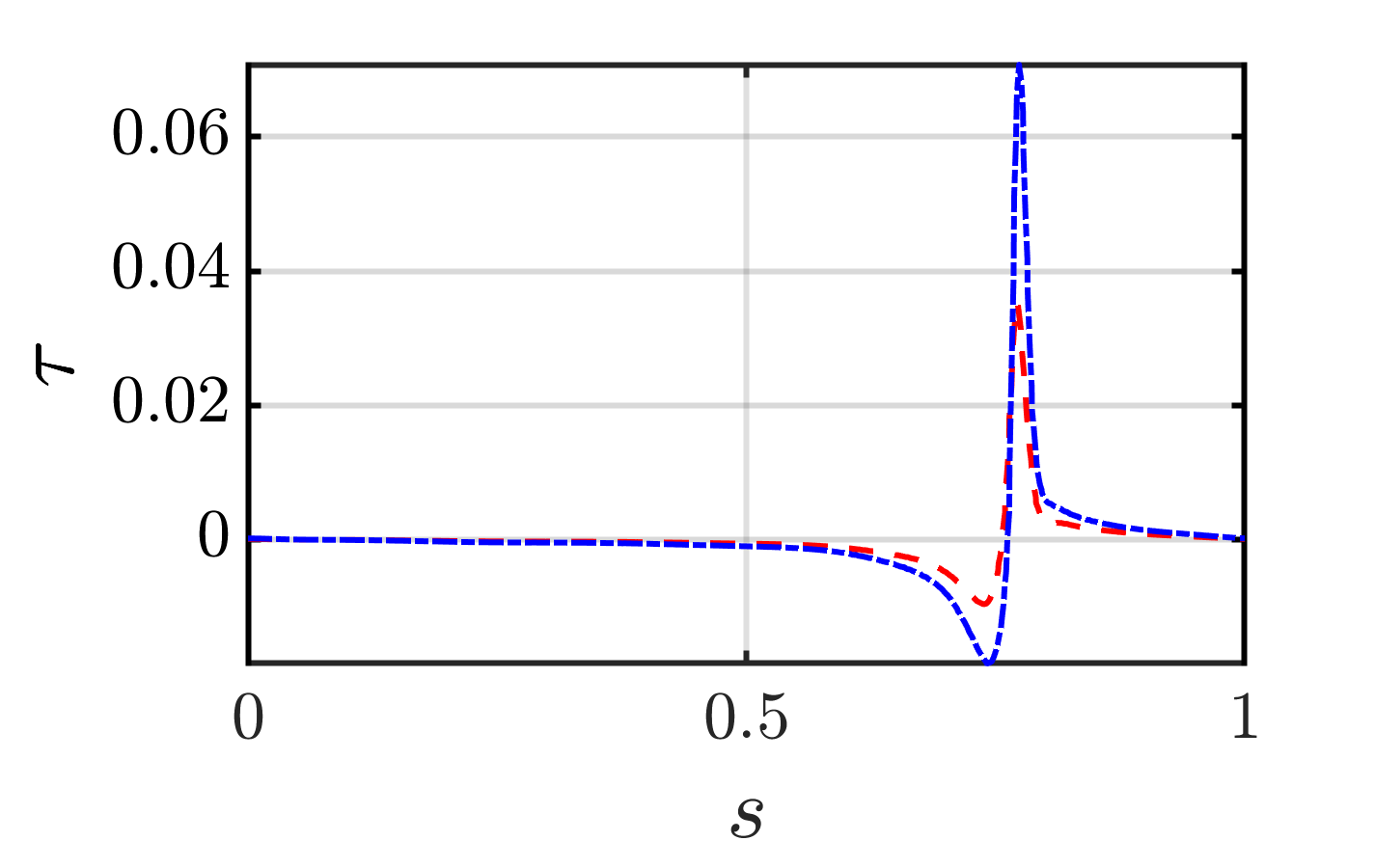} & \includegraphics[width=0.5\textwidth]{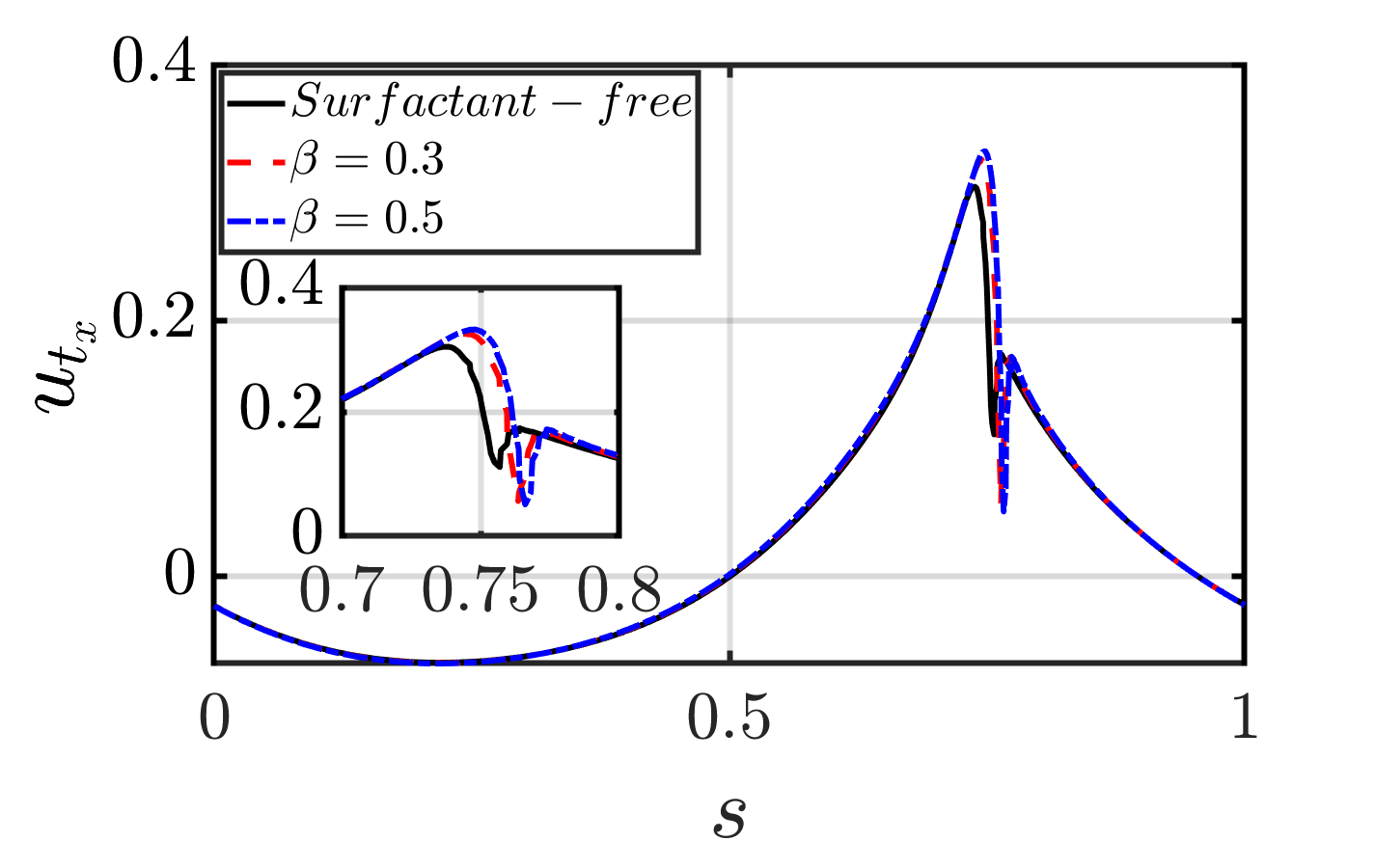} \\
(g) & (h) \\
\end{tabular}
\caption{Effect of the elasticity parameter for the spilling regime for  $\epsilon=0.324$  at $t=4.8$,  and $\epsilon=0.33$  at $t=5.2$. Two-dimensional projections of the interface, $\Gamma$, $Ma$ and $u_{t_x}$ in the $(x–z)$ plane ($y=\lambda/8$) are shown.  Panels (a–d) correspond to $\epsilon = 0.324$, and panels (e–h) to $\epsilon = 0.33$.} 
\label{fields_spilling}
\end{figure}

We next consider the spilling regime; the interfacial location, $\Gamma$, $\CRCA{\tau}$ and $u_{tx}$  are shown in figure~\ref{fields_spilling}.
For $\epsilon = 0.324$, the addition of surfactant significantly modifies the  shape and position of  the wave crest (see figure~\ref{fields_spilling}a).  Surfactants also accumulate near the wave crest, leading to a \CRCA{slightly} sharper, and more forward-leaning profile (figure~\ref{fields_spilling}b). This effect becomes more pronounced with increasing surfactant elasticity, which enhances interfacial gradients and promotes stronger Marangoni stresses. The resulting non-uniform surfactant distribution leads to localized peaks in Marangoni stresses  near the crest (figure~\ref{fields_spilling}c), and larger tangential velocities $u_t$ (figure~\ref{fields_spilling}d).
%
At slightly higher steepness, $\epsilon = 0.33$, the addition of surfactant  induces significant changes in the interfacial dynamics at $t=4.8$. As shown in figure~\ref{fields_spilling}e, a small forward jet forms near the crest.
\citet{liu2006experimental} observed experimentally that, in surfactant-laden spilling breakers, the bulge shape flattens and becomes elongated along the slope of the front face of the wave.
We observe \textcolor{black}{similar} qualitative behavior in our simulations, 
\textcolor{black}{ and acknowledge that  quantitative agreement is not expected due to differences in Reynolds number and surfactant solubility, the geometric trends are consistent with the experimental observations (more details are given in Appendix E).}
We also note that  \citet{erinin2023effects} observed experimentally that the addition of  surfactants leads to a jet which curls inward for plunging breakers. 
%
The formation of this forward jet significantly reshapes both the surfactant concentration and tangential velocity fields. For $\beta_s = 0.5$, the surfactant distribution profile (figure~\ref{fields_spilling}f) peaks in the wave location, with Marangoni stress are larger in the front of the crest.
The resulting surfactant gradient enhances the Marangoni stress (figure~\ref{fields_spilling}g), promoting jet elongation, and increases with $\beta_s$. A negative Maranogoni peak is observed which promotes this  elongation.
%
Figure~\ref{fields_spilling}h shows the tangential velocity field 
$u_{tx}$ along the interface.
\CRCA{In the surfactant-free case, the tangential surface velocity first accelerates along the front face of the wave but undergoes a localized deceleration near the jet base, where  momentum is redirected into the forming jet. 
Surfactants redistribute surface momentum by promoting upstream acceleration along the wave front while inducing a sharply localized deceleration near the jet base through Marangoni stresses acting over a short arclength. By partially rigidifying  the interface near the jet base, surfactants promote upstream surface stretching, which favors elongation of the bulge along the wave slope and suppresses its vertical growth, consistent with experimental observations.}



\begin{figure}
    \centering
\begin{tabular}{ccc}
         \includegraphics[width=0.33\textwidth]{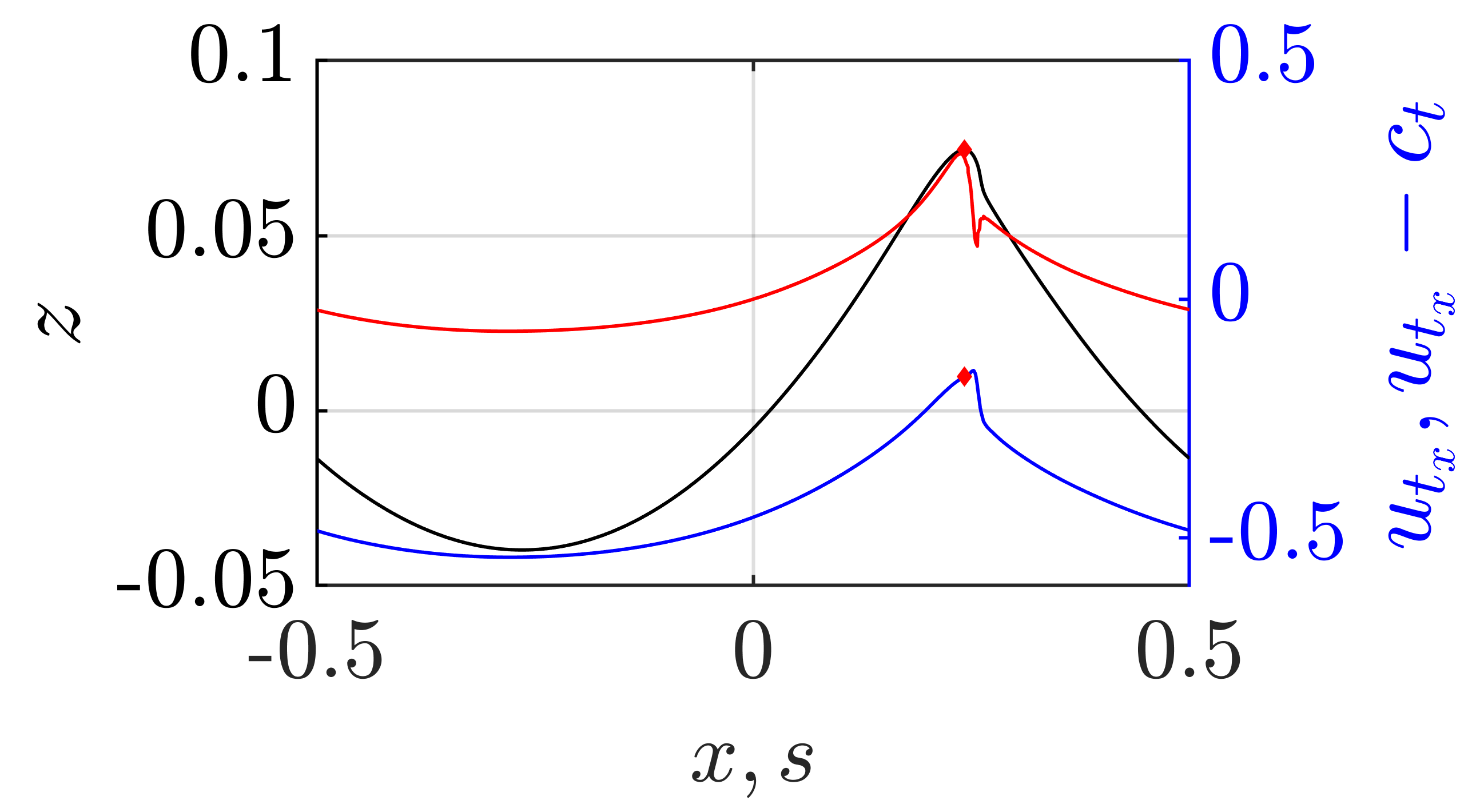} &  
        \includegraphics[width=0.33\textwidth]{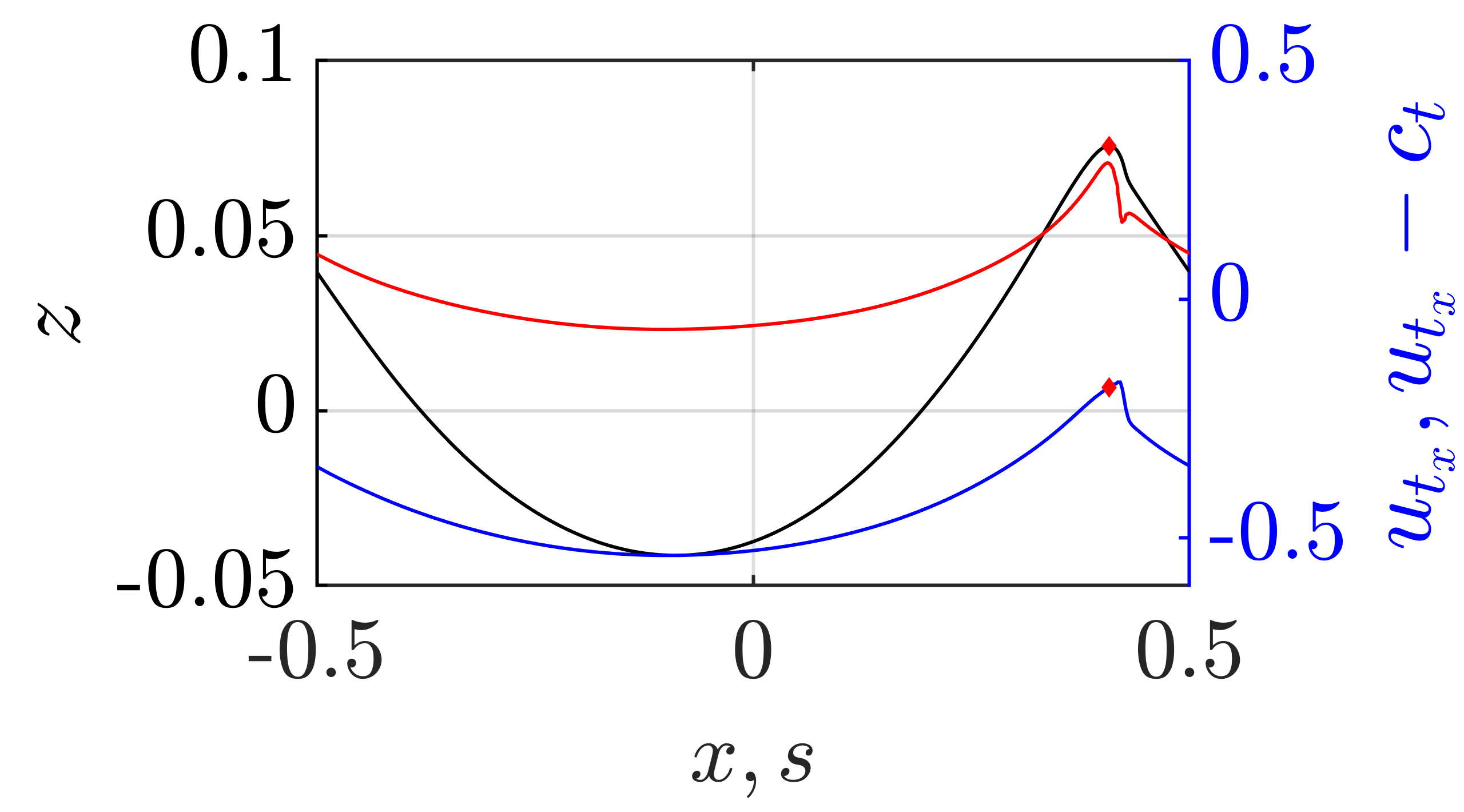} &  
         \includegraphics[width=0.33\textwidth]{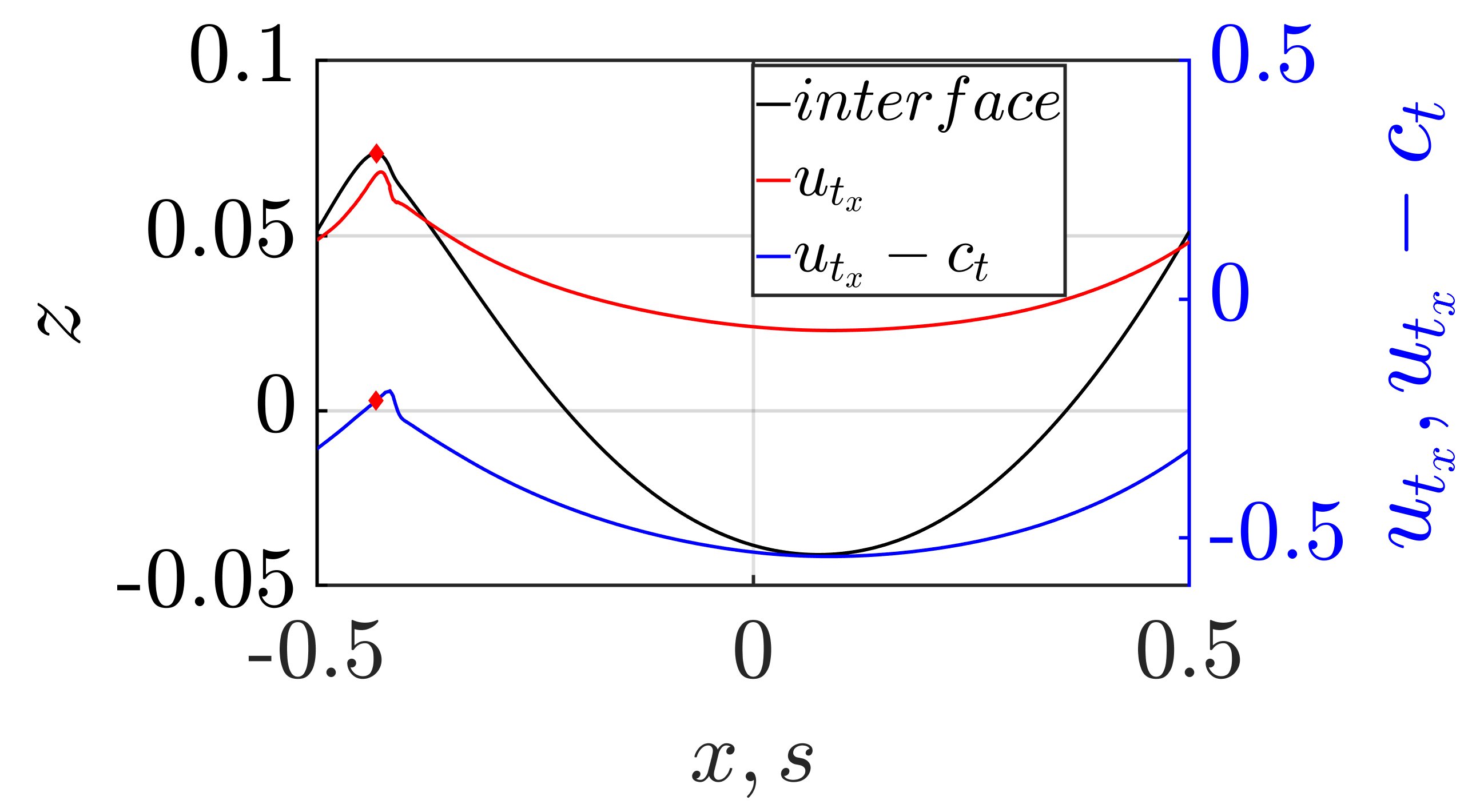} \\
(a) & (b) & (c)\\
        \includegraphics[width=0.33\textwidth]{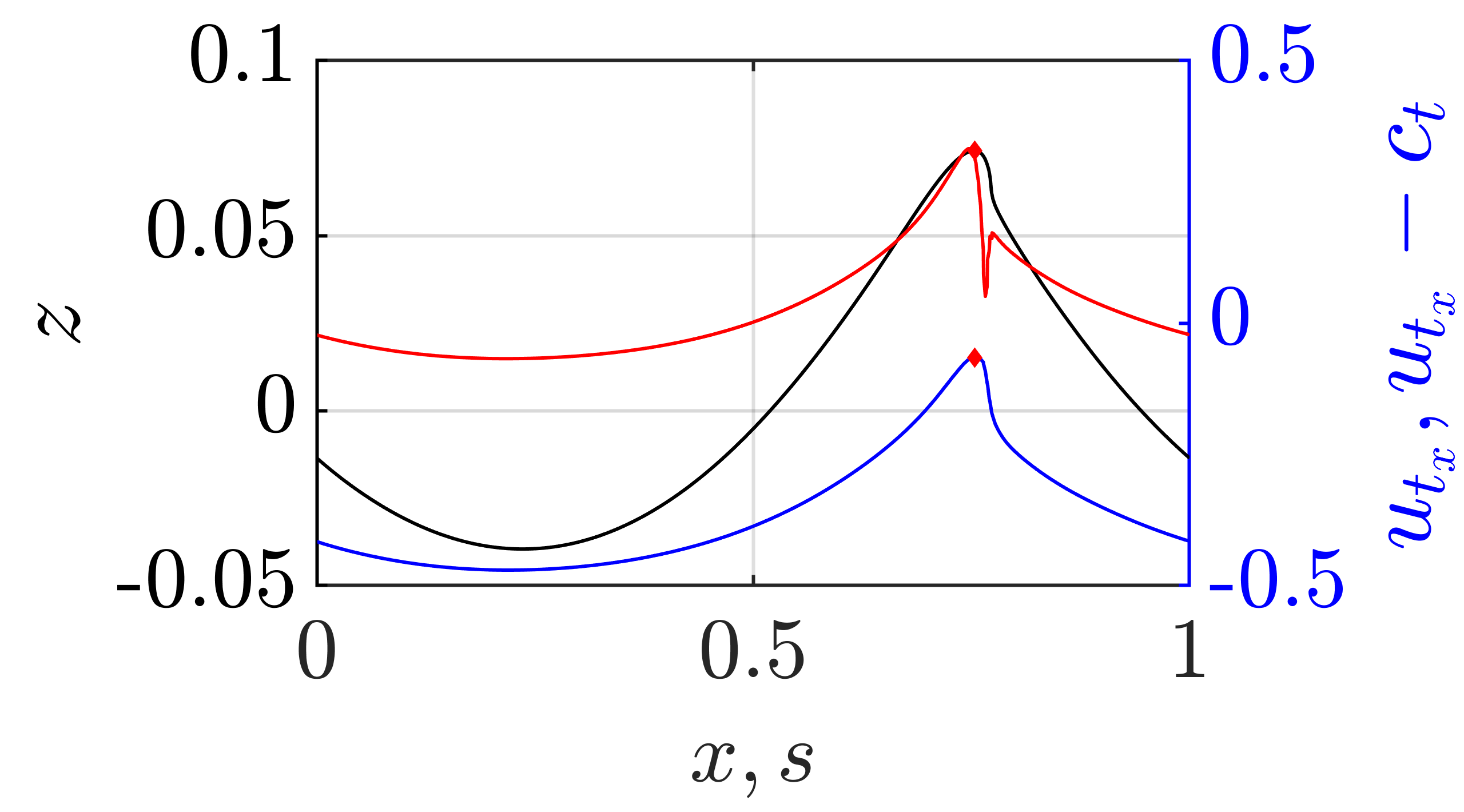}& 
        \includegraphics[width=0.33\textwidth]{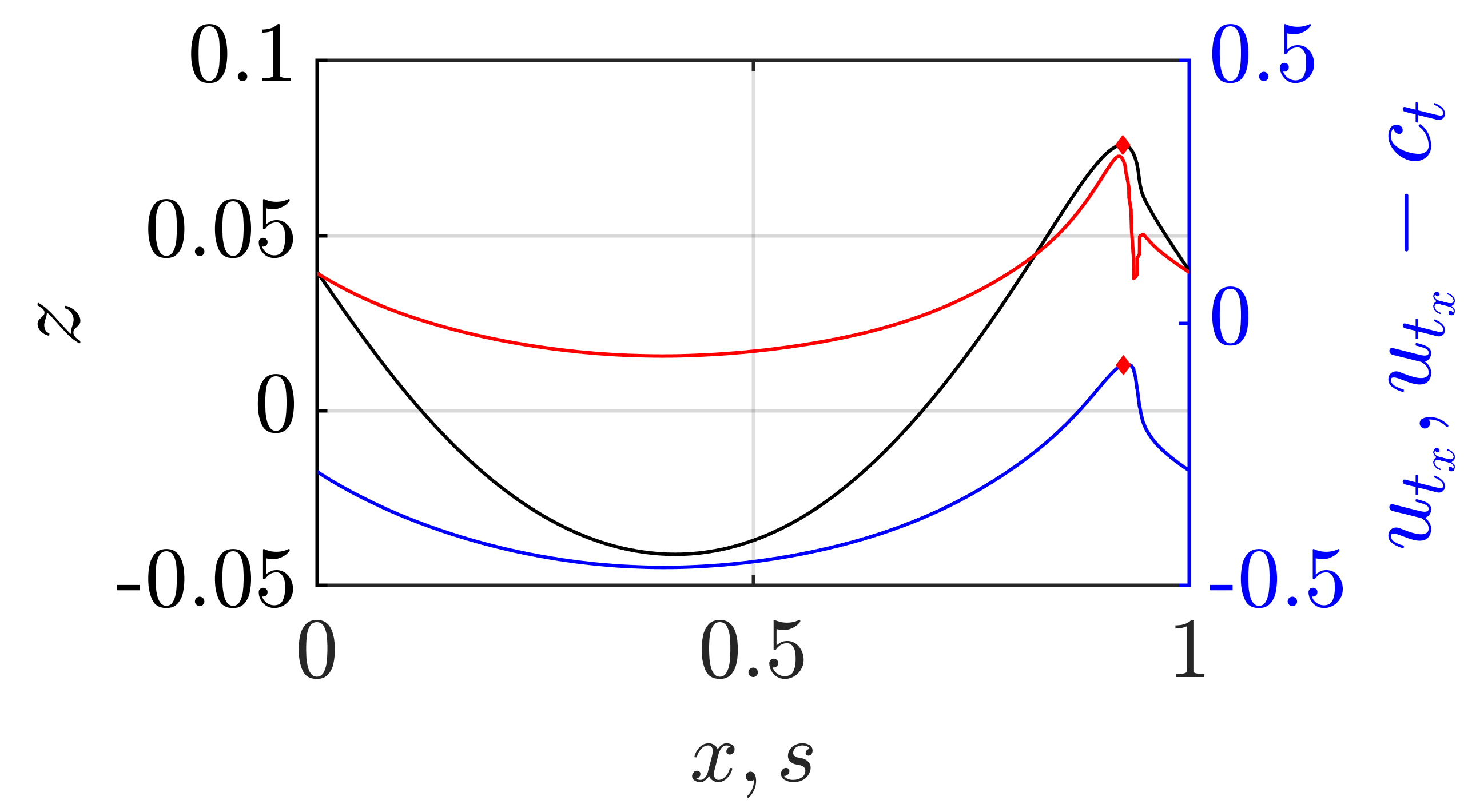}& 
        \includegraphics[width=0.33\textwidth]{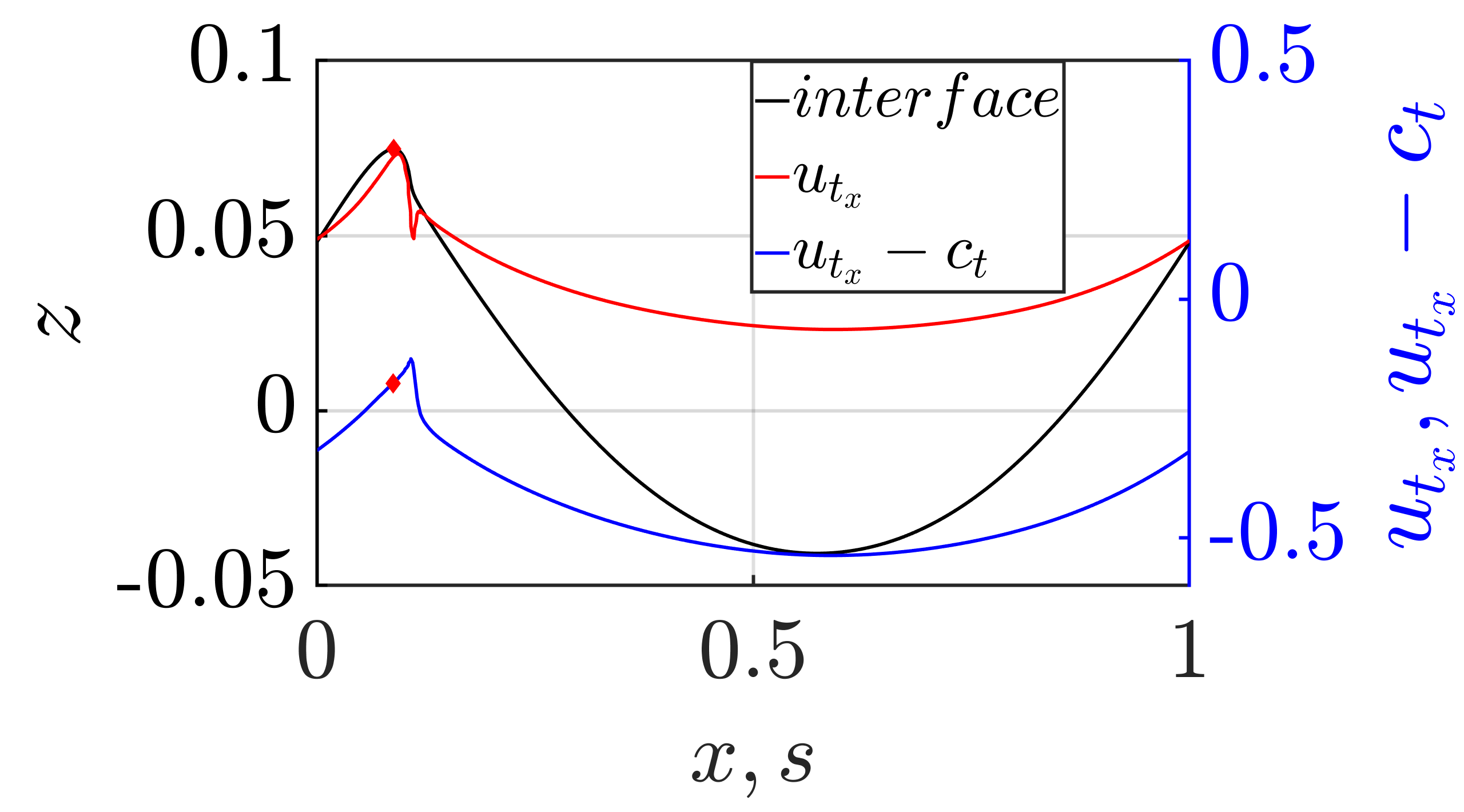}\\

 (d) & (e) & (f)\\
        \includegraphics[width=0.33\textwidth]{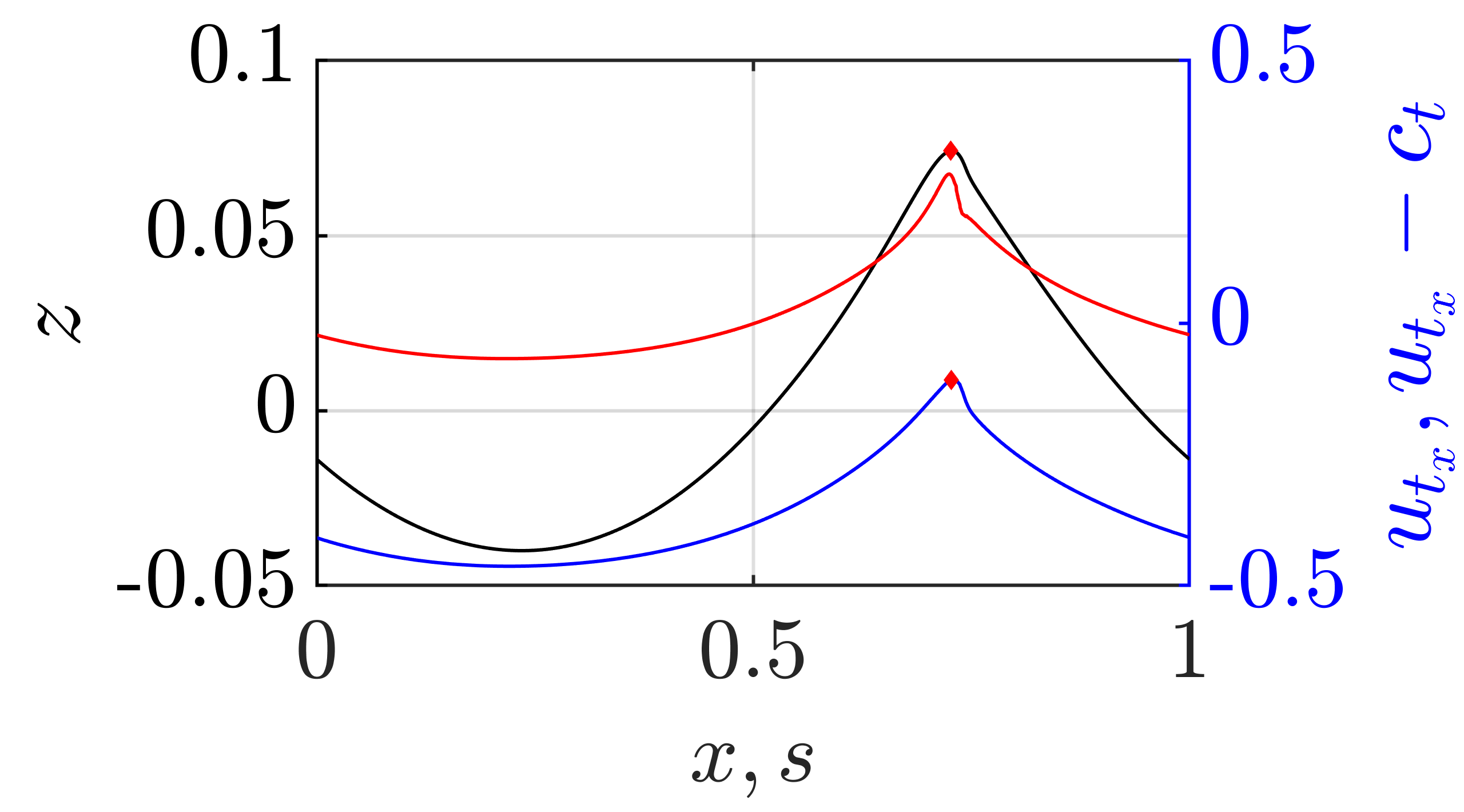}& 
        \includegraphics[width=0.33\textwidth]{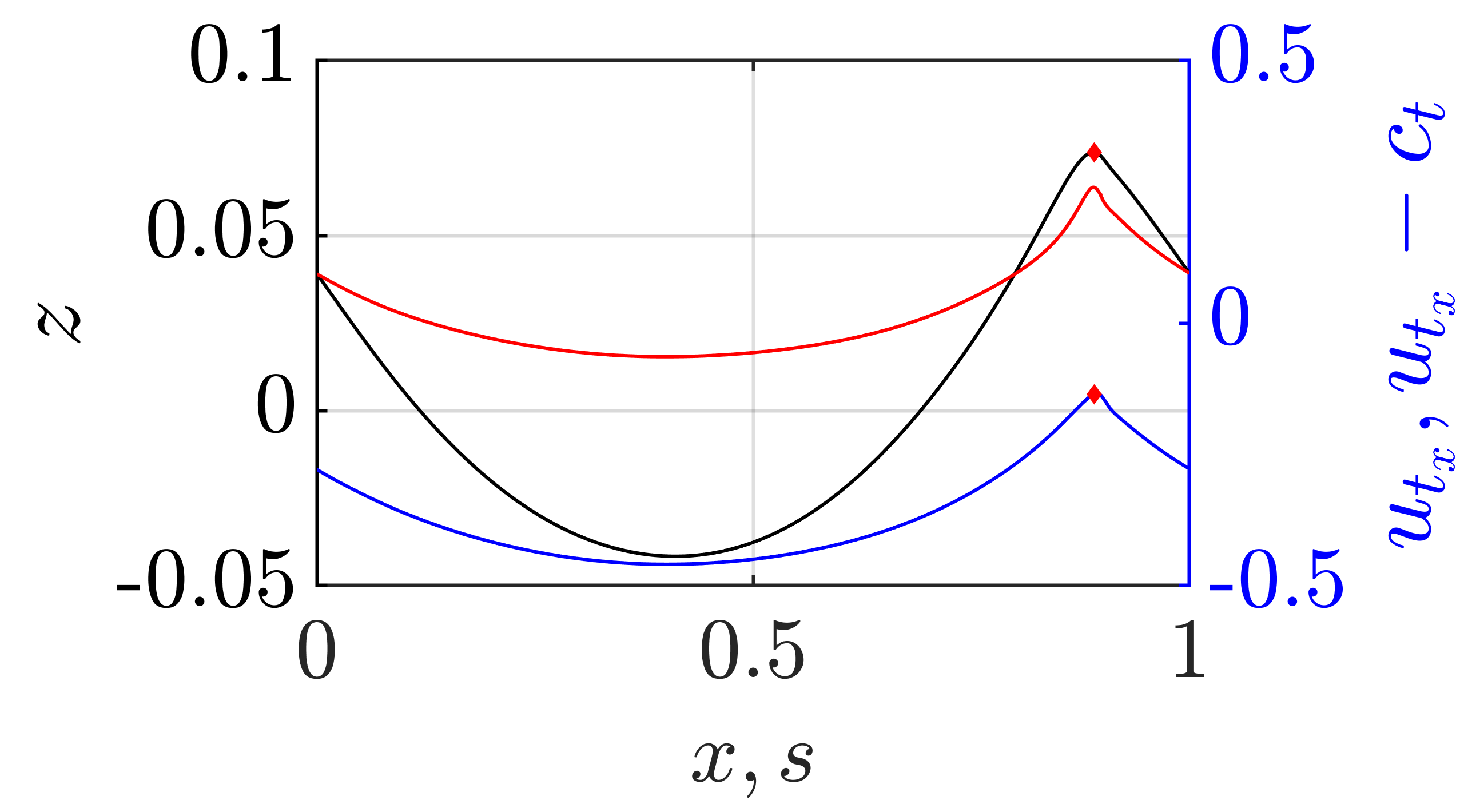}& 
        \includegraphics[width=0.33\textwidth]{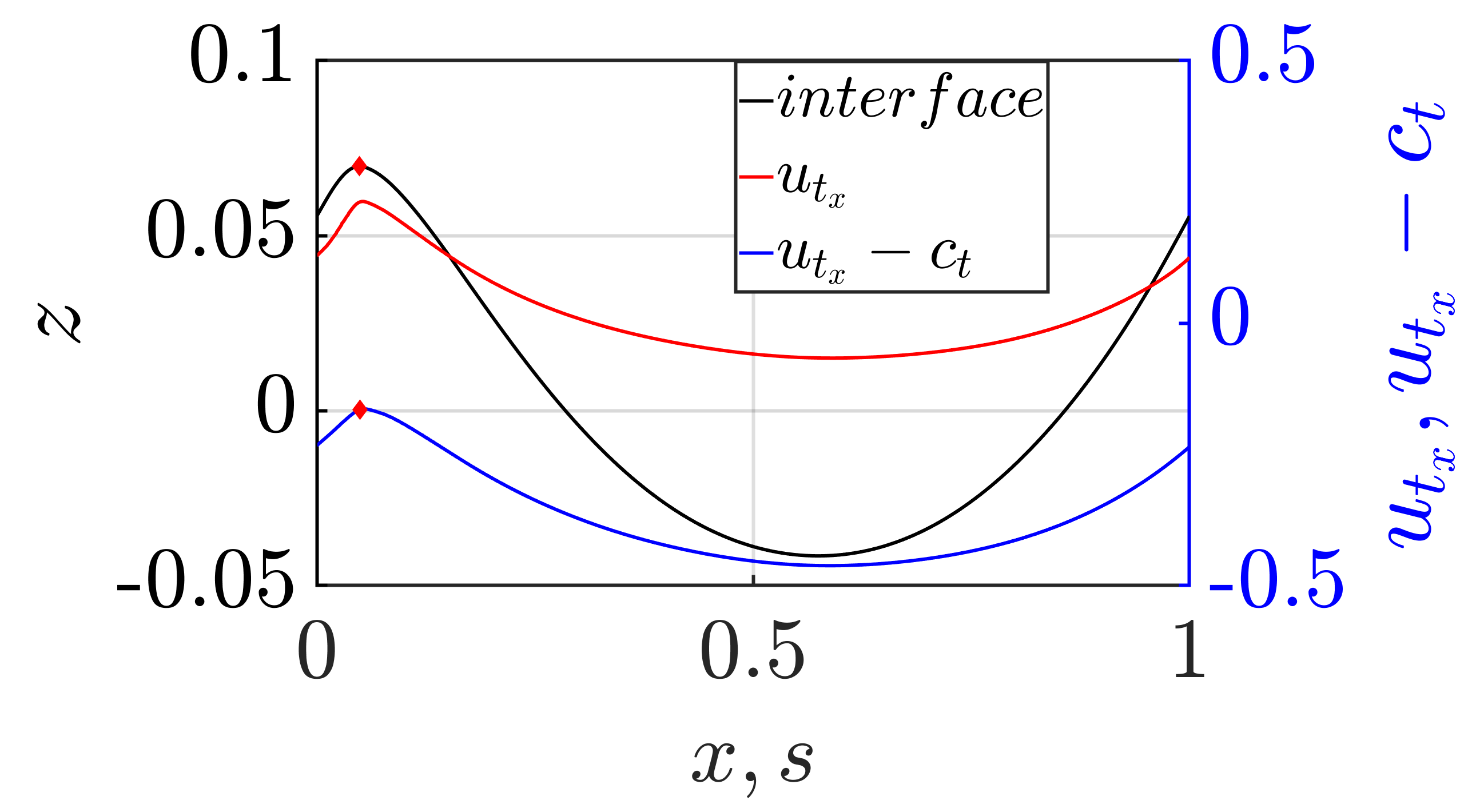}\\

 \textcolor{black}{(g)} & \textcolor{black}{(h)} & \textcolor{black}{(i)}\\
\end{tabular}
    \caption{Two-dimensional projections of the interface  location and tangential velocity in the reference frame of the wave crest for surfactant-free (top panels) with $\epsilon=0.33$; surfactant-laden case  with $\beta_s=0.5$ (middle panels) \textcolor{black}{ and surfactant-laden without Marangoni stresses (i.e.,  $\nabla_s \tilde{\sigma}=0$ in equation \ref{NS_Eq}) with $\beta_s=0.5$ (bottom panels)} at times $t=[5.2, 5.6, 6.0]$, corresponding to columns one-to-three, respectively. Here, $c_t$ denotes the tangential component of the average interfacial velocity.}\label{tangential_velocity_c}
\end{figure}

Finally, in figure~\ref{tangential_velocity_c}, we examine the effect of surfactant in greater detail by analyzing the tangential velocity in the reference frame of the wave crest for both surfactant-free and surfactant-laden cases with $\epsilon=0.33$ and $\beta_s=0.5$ at $t = [5.2, 5.6, 6.0]$.
\textcolor{black}{During the onset of spilling, a  local minimum develops in the tangential velocity profile near the base of the crest. A feature which is enhanced in the presence of surfactants   for spilling breakers. In the surfactant-free case, this secondary peak decays rapidly, and has disappeared by 
$t=6.0$ (see figure \ref{tangential_velocity_c}c). In contrast, for the surfactant-laden case the secondary peak remains clearly visible at the same instant (see figure \ref{tangential_velocity_c}f), indicating a prolonged effect of surfactant-driven Marangoni stresses  near the crest. Although the crest speed  is constant, the persistence and increased magnitude of in the surfactant case reflect enhanced tangential shear along the interface. The extended lifetime and strengthened secondary peak therefore provide direct evidence that surfactants promote and intensify spilling by sustaining near-crest shear through Marangoni stresses.}

\textcolor{black}{
To isolate the role of Marangoni stresses in the interfacial dynamics, we perform additional simulations in which Marangoni stresses are suppressed by imposing $\nabla_s \sigma = 0$ in equation~\ref{NS_Eq}. This constraint allows surfactant to remain  non-uniformly distributed along the interface while preventing the generation of Marangoni stresses (a similar approach has been done by \citet{kamat_2020} and \citet{sheet_constante,constanteamores2020bb}).
Figure~\ref{tangential_velocity_c} g-i shows the cases for $\nabla_s \sigma = 0$ with $\epsilon=0.33$ and $\beta_s=0.5$ at times $t=[5.2, 5.6, 6.0]$,  allowing the distinct roles of Marangoni stresses and  surface tension variations can be inferred in the dynamics. Comparing panels \ref{tangential_velocity_c}d-f (Marangoni stresses allowed) with \ref{tangential_velocity_c}g-i (no Marangoni stresses) shows that the presence of Marangoni stresses enhances the second peak in the tangential velocity distribution, which is associated with the onset of spilling. 
In contrast, comparison with the surfactant-free cases (figure \ref{tangential_velocity_c}a-c) shows that 
a local reduction in surface tension, in the absence of Marangoni stresses, suppresses the secondary tangential velocity peak relative to the surfactant-free case.  This demonstrates that a decrease in the local surface tension alone does not promote spilling. Instead, the enhancement of spilling observed in surfactant-laden simulations arises from Marangoni stresses, which generate tangential stress gradients, $\nabla_s \sigma(\Gamma)$, that locally accelerate the interface and amplify the secondary velocity peak.}



\begin{figure}
 \centering
\vspace{0.5cm}
 \begin{tabular}{ccc}
 Surfactant-free ~~~~~~~~~~~~~~~~~~~~    & \CRCA{ $\beta_s = 0.3$} ~~~~~~~~~~~~~~~~~~~~& \CRCA{$\beta_s = 0.5$}~~~~~~~~~~~~~~~~~~~~\\
 \end{tabular}
 \begin{tabular}{c}
 \includegraphics[width=1\textwidth]{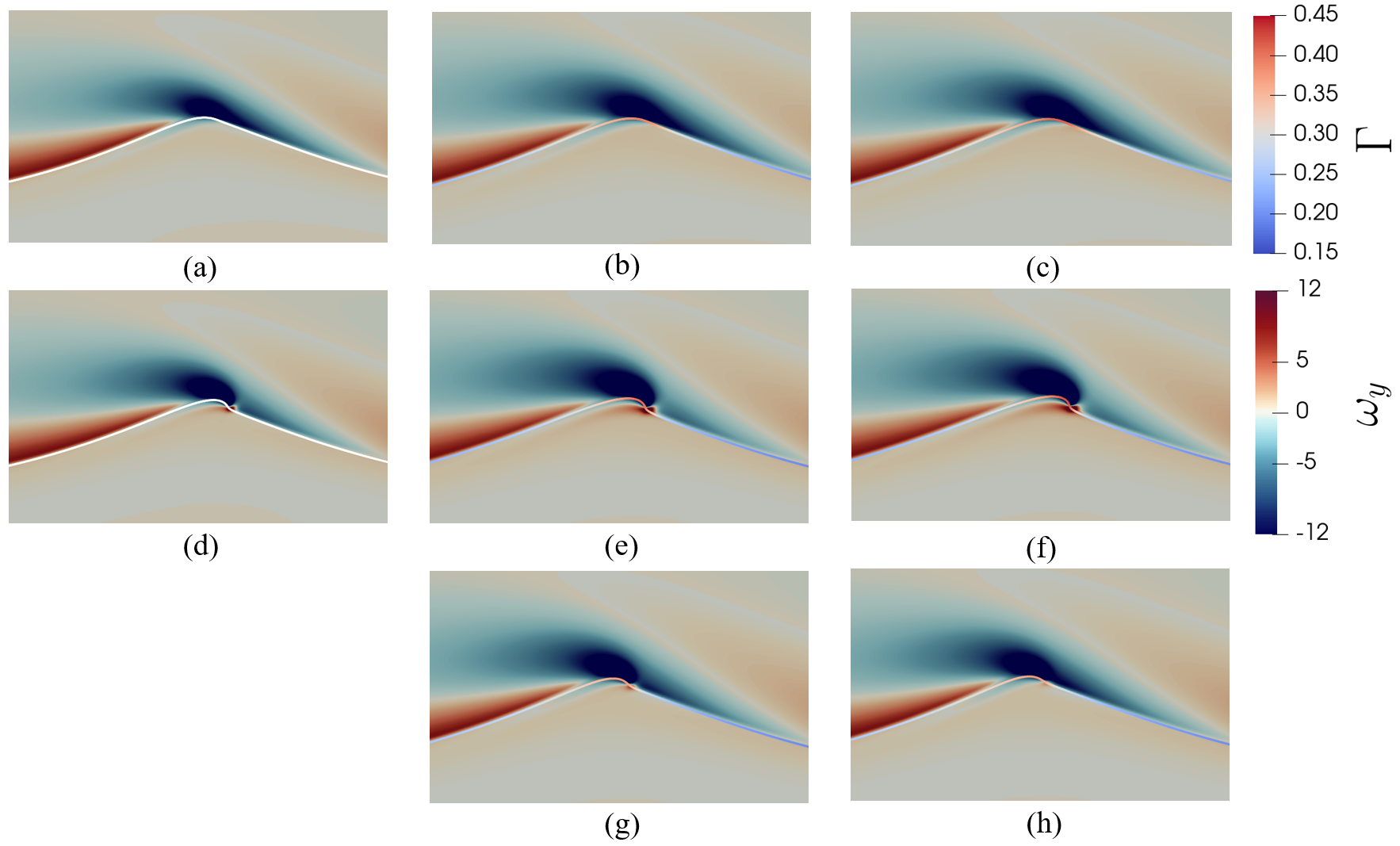}
 \end{tabular}
\caption{Two-dimensional representation of the interfacial location (colored by $\Gamma$) overlaid with the spanwise vorticity component $\omega_y$ in the  $(x-z)$ plane. Panels (a–c) correspond to $\epsilon = 0.324$ at $t = 4.8$, while panels (d–f) correspond to $\epsilon = 0.33$ at $t = 5.2$; \textcolor{black}{and panels (g-h) correspond to $\epsilon=0.33$ at $t = 5.2$ with $\nabla_s \tilde{\sigma}=0$ in equation \ref{NS_Eq}.} Columns from left to right represent the surfactant-free case, \CRCA{$\beta_s = 0.3$}, and $\beta_s = 0.5$, respectively.  }\label{vorticity_plot}
\end{figure}

Figure \ref{vorticity_plot} shows the spanwise vorticity $\omega_y = (\nabla \times \boldsymbol{u})_y$ in a $(x-z)$ plane overlaid with the interface shape for  steepness, $\epsilon = 0.324$ (top panels) and $\epsilon =0.33$ (bottom panels). For each case,  results are shown for the surfactant-free and the surfactant-laden cases with $\beta=0.3$ and $\beta_s=0.5$ (columns one to three, respectively). For $\epsilon = 0.324$, the surfactant-free case exhibits a smooth flow separation near the wave crest, forming a broad region of negative $\omega_y$ loosely coupled to the interface (see figure \ref{vorticity_plot}a).  At moderate elasticity number  \CRCA{$\beta_s=0.3$}, a distinct interfacial shear layer forms beneath the interface due to Marangoni-driven flow, characterized by adjacent  bands of positive and negative $\omega_y$ straddling the interface, a signature  of Marangoni-driven vorticity  generation (see figure \ref{vorticity_plot}b). At higher elasticity \CRCA{$\beta_s=0.5$}, this vorticity  layer becomes more localized, and a secondary vorticity layer of opposite sign develops beneath the interface,  a feature absent in the surfactant-free case. 
For the slightly larger steepness ($\epsilon = 0.33$), separation is more abrupt, and early vortex roll-up occurs near the crest (see figure \ref{vorticity_plot}d). The presence of surfactant significantly amplifies these effects. The interfacial shear layer strengthens, and the vorticity  becomes more intense and confined in regions of high $\Gamma$ (see figure \ref{vorticity_plot}e,f). At  \CRCA{$\beta_s=0.5$, we observe a larger positive vorticity near at the base of the wave, coinciding with the location of strongest surfactant accumulation. This correlation is consistent with enhanced interfacial shear and sharper velocity gradients induced by Marangoni stresses acting along the interface.} 
\textcolor{black}{In addition, figure~\ref{vorticity_plot}g-h shows  the corresponding flow fields when Marangoni stresses are suppressed,
$\nabla_s \sigma = 0$.  In this case, the positive vorticity generated near the crest region is significantly reduced compared to the surfactant-free case. This reduction becomes more pronounced as the elasticity parameter increases from $\beta_s = 0.3$ to $\beta_s = 0.5$, reflecting the stronger decrease in local surface tension at the crest for a given surfactant concentration.
These results indicate that a reduction in surface tension alone does not enhance the shear layer at the crest; instead, it weakens vorticity generation. The amplification of vorticity observed in the full surfactant cases therefore requires the presence of Marangoni stresses, which introduce tangential stress gradients along the interface.}


Finally, we examine the role of Marangoni stresses in the generation of vorticity at the gas–liquid interface. Vorticity production at free surfaces has been the subject of   several studies \citep{longuet_1992,Cresswell,Peck_1998,lundgren_1999,brons_2014,constante_coales},  and is particularly relevant in air–water systems where the free-surface approximation is appropriate. In this work, we adopt a two-dimensional version of the generalized formulation derived by \citet{constante_2023}, who described the evolution of circulation at a deformable interface within a three-dimensional control volume in the presence of surfactants. For brevity, we present only the final form of the governing equation used in the analysis in the main text; a full derivation of the two-dimensional form is provided in Appendix. The  circulation is given  by
\begin{align}
\frac{D}{Dt}\!\left(\int_A \omega\,dA + \int_a^b [[\mathbf{u}\!\cdot\!\mathbf{t}]]\,ds\right)
&= \oint_C \nu\,\nabla\omega\!\cdot\!\mathbf{n}_c\,ds
     - \int_a^b \kappa\, [[(\mathbf{u}\!\cdot\!\mathbf{n})(\mathbf{u}\!\cdot\!\mathbf{t})]]\,ds
\nonumber\\[4pt]
&\quad + \frac{1}{2}\int_a^b \frac{\partial}{\partial s} [[(\mathbf{u}\!\cdot\!\mathbf{n})^{2}]]\,ds
     - \int_a^b \frac{\partial}{\partial s}
       \left[\!\left[\frac{p}{\rho}\right]\!\right] ds.
       \label{eq:circulation_jump_split}
\end{align}

Equation~\ref{eq:circulation_jump_split} describes the rate of change of circulation within  a two dimensional  control volume intersecting an arbitrary fluid interface of area $A$ separating two fluids with different material properties. The left-hand side consists of two terms: the integral of vorticity over  $A$, and the circulation jump along the curve from $a$ to $b$, represented by the jump in tangential velocity $[[\mathbf{u} \cdot \hat{\mathbf{t}}]]$.
However, under the assumption of velocity continuity across the interface in both normal and tangential directions, and given that the unit normal ${\mathbf{n}}$ and tangential ${\mathbf{t}}$ vectors are identical on either side of the interface, the jump in tangential velocity $[[\mathbf{u} \cdot {\mathbf{t}}]]$ vanishes. Similarly, the second term on the right-hand side involving the jump of the product $(\mathbf{u} \cdot {\mathbf{n}})(\mathbf{u} \cdot {\mathbf{t}})$ and the third term involving the jump of $(\mathbf{u} \cdot {\mathbf{n}})^2$ also vanish.
Therefore, the second term on the left-hand side and the second and third terms on the right-hand side of equation~\ref{eq:circulation_jump_split} are identically zero.

\begin{figure}
    \centering
\vspace{0.5cm}
\begin{tabular}{ccc}
Surfactant-free    &  $\CRCA{\beta_s = 0.3}$ & $\CRCA{\beta_s = 0.5}$\\
        \includegraphics[width=0.33\textwidth]{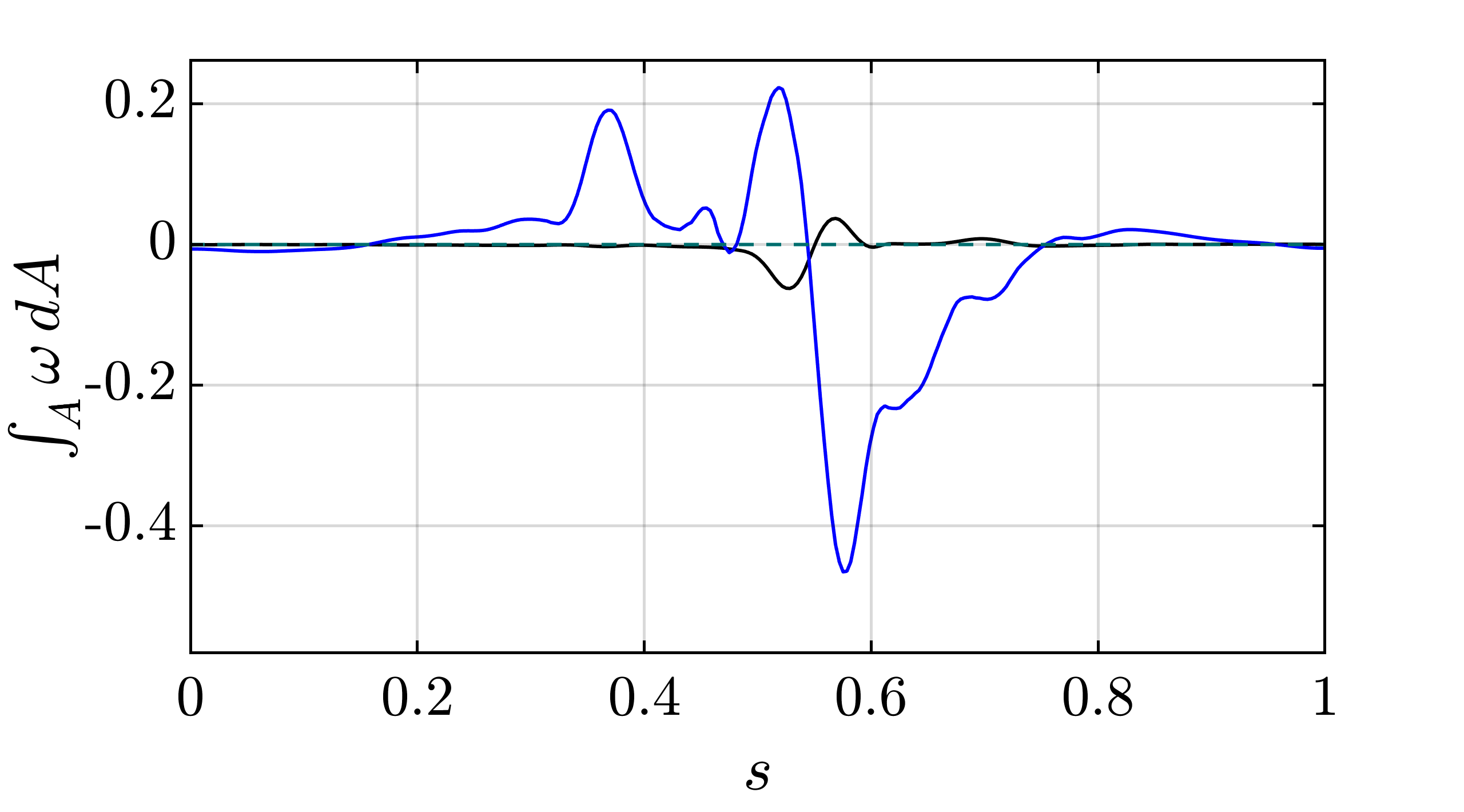}& 
        \includegraphics[width=0.33\textwidth]{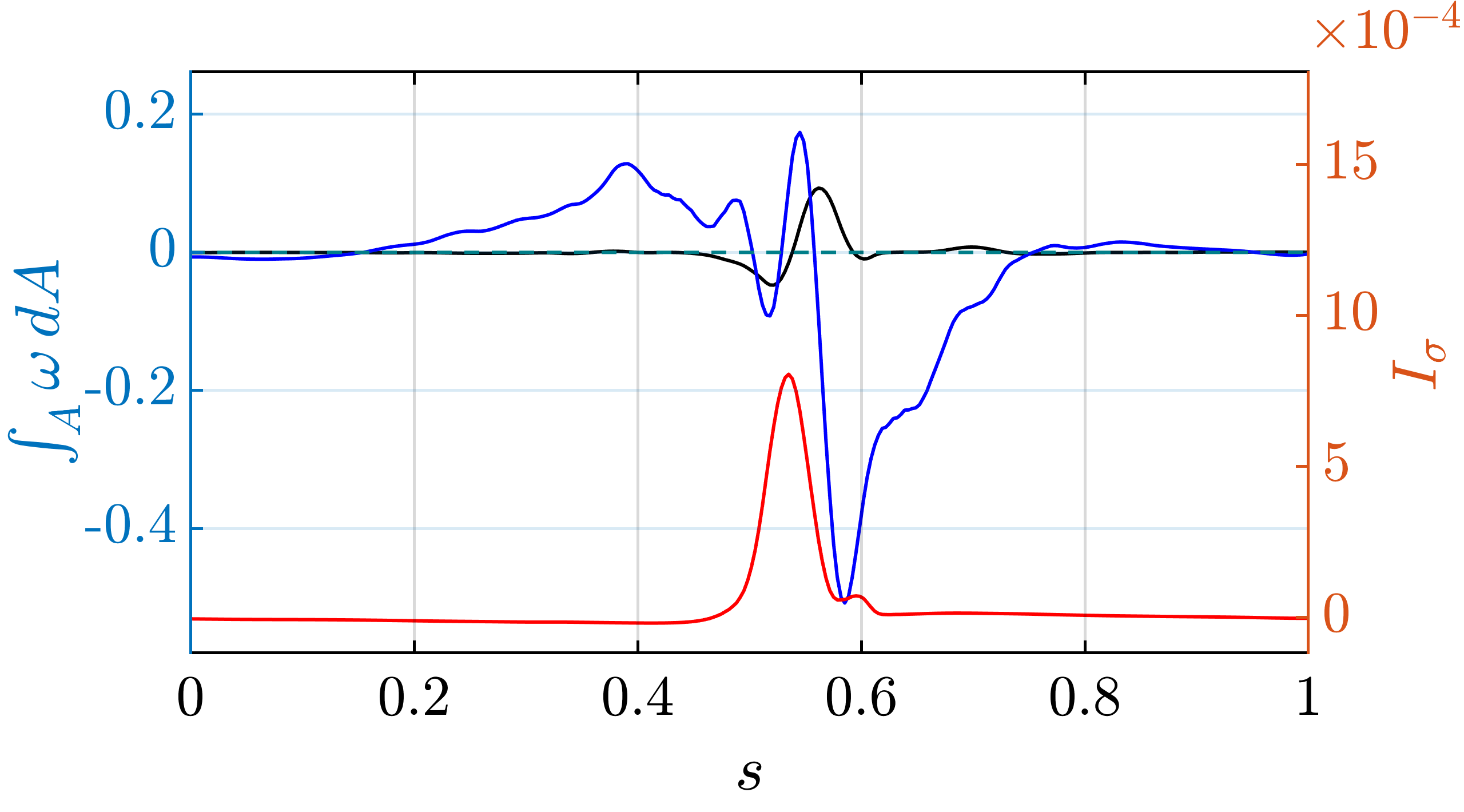}& 
        \includegraphics[width=0.33\textwidth]{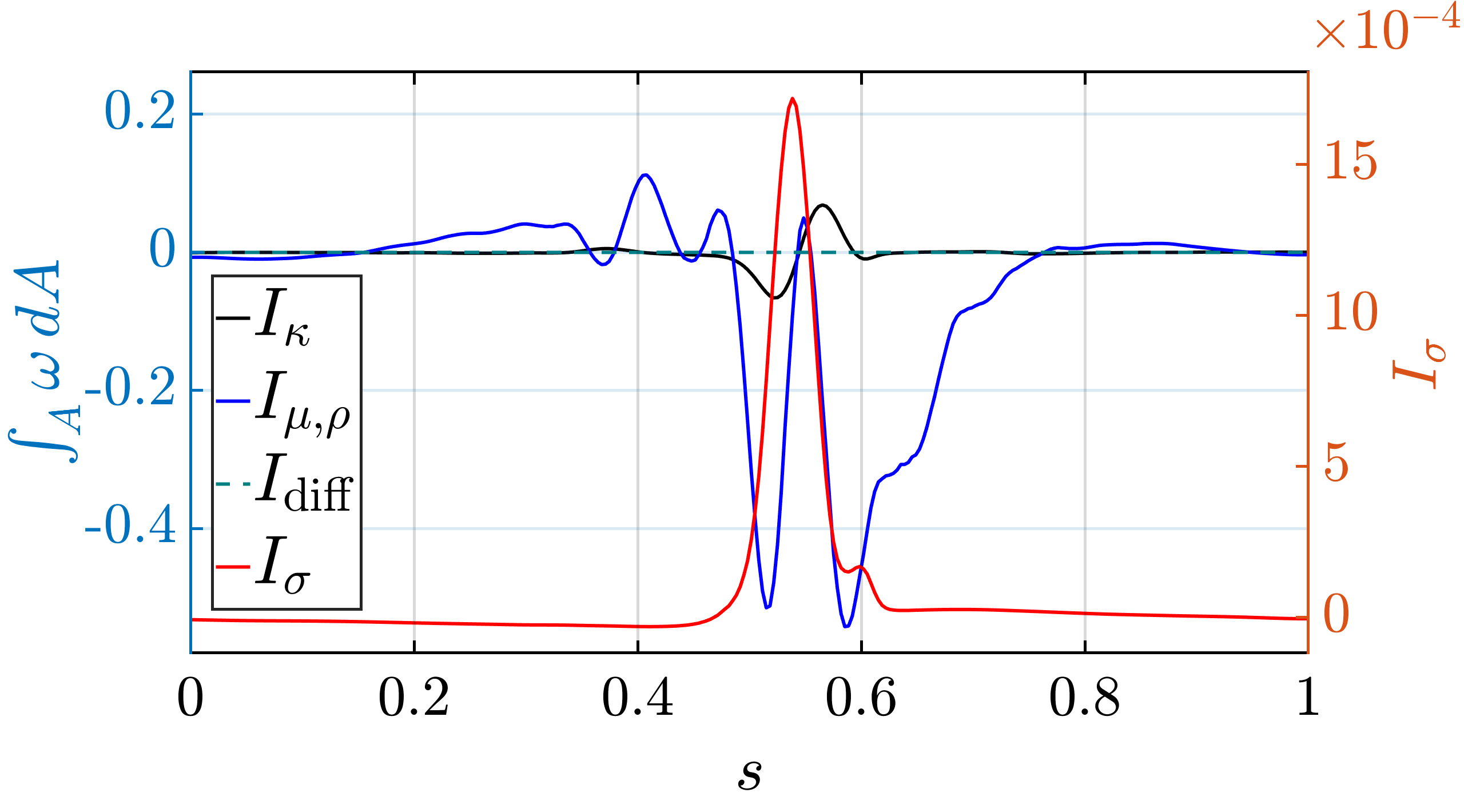}\\
        
(a) & (b) & (c)\\
        \includegraphics[width=0.33\textwidth]{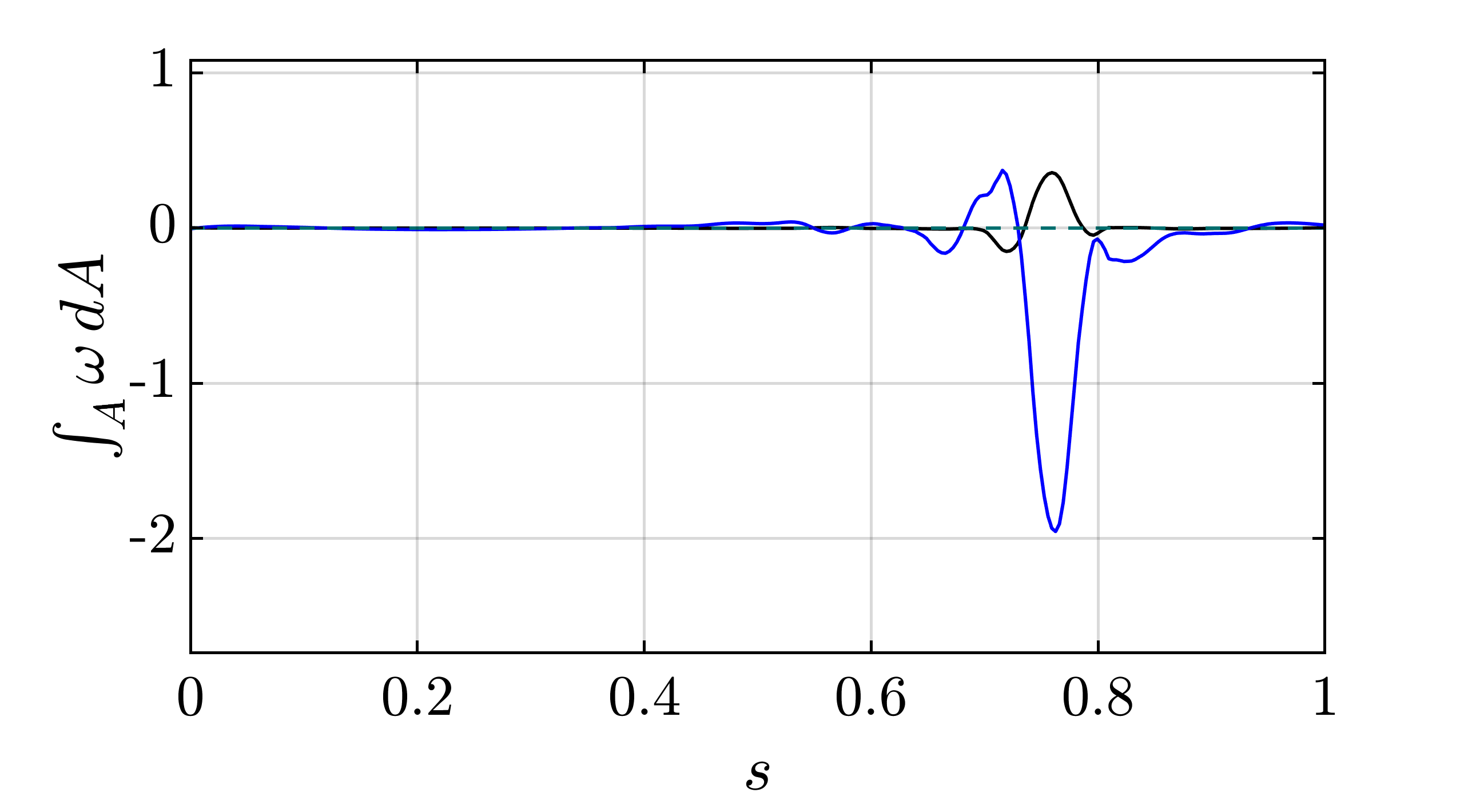}& 
        \includegraphics[width=0.33\textwidth]{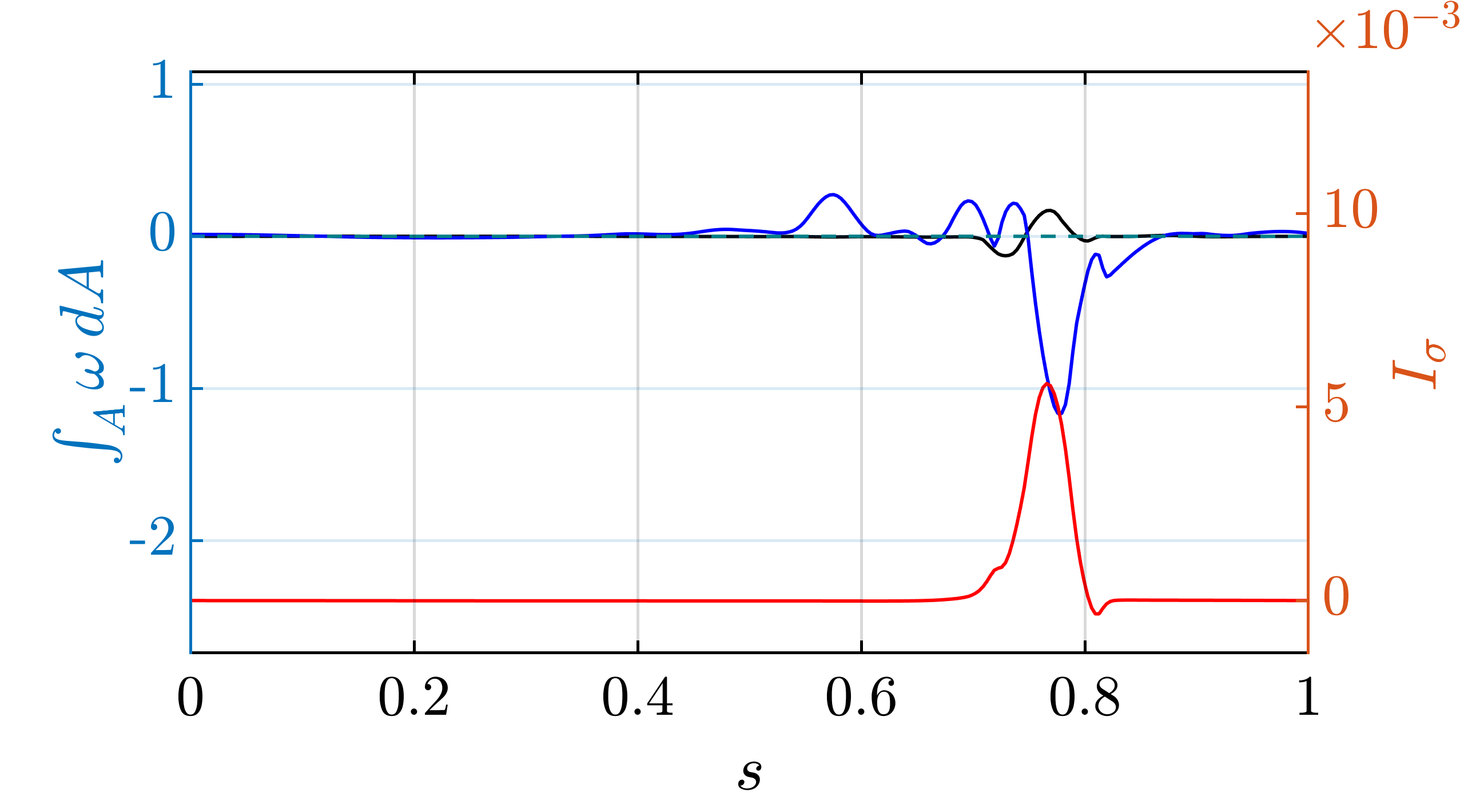}& 
        \includegraphics[width=0.33\textwidth]{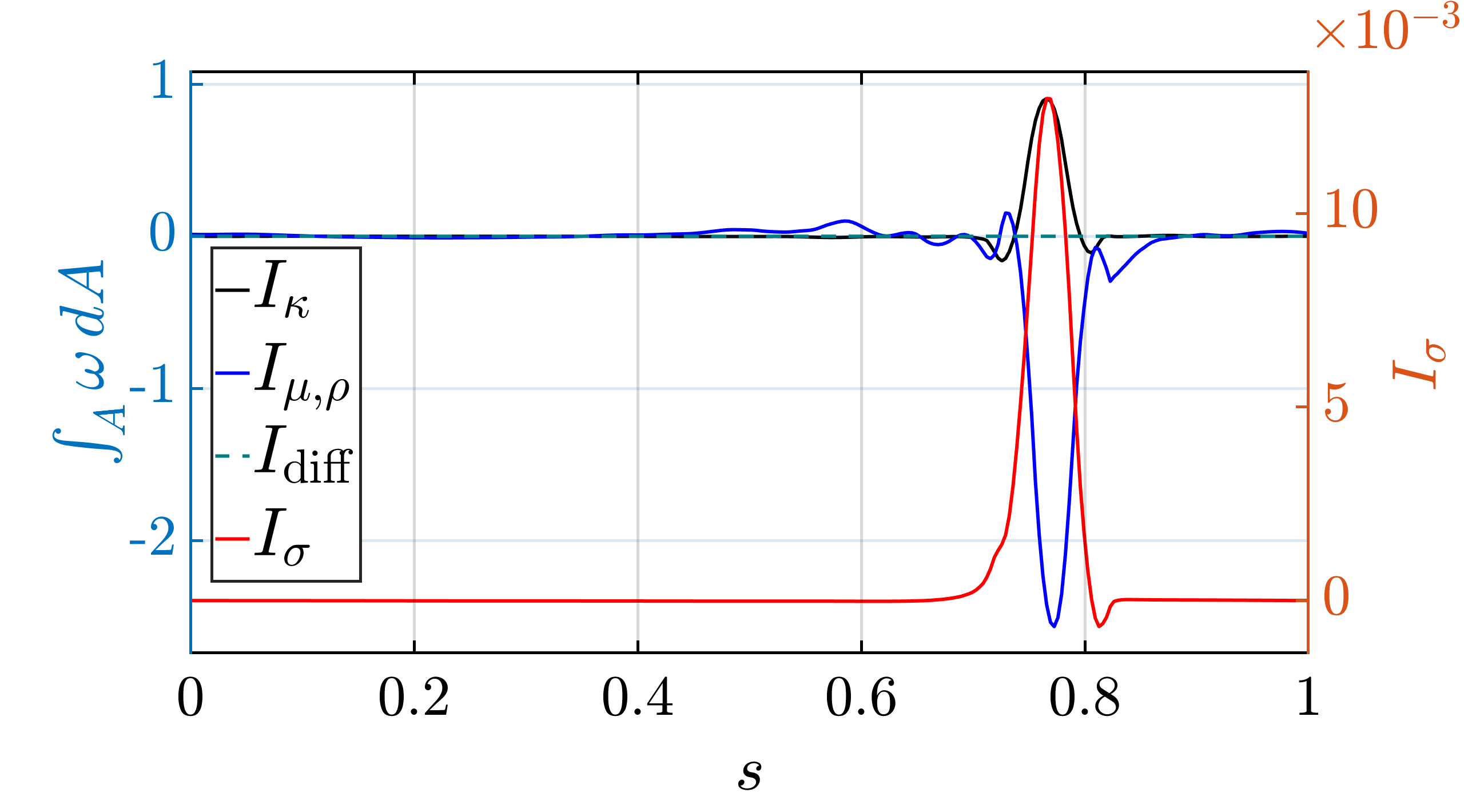}\\

 (d) & (e) & (f)\\
\end{tabular}
    \caption{ Two-dimensional representation of  circulation generation mechanisms along the interface in the $(x-z)$ plane. Panels (a–c) correspond to $\epsilon = 0.324$ at $t = 4.8$, and panels (d–f) to $\epsilon = 0.33$ at $t = 5.2$. Columns from left to right show the surfactant-free case, $\CRCA{\beta_s = 0.3}$, and $\CRCA{\beta_s = 0.5}$, respectively. }\label{vorticity_mechanism}
\end{figure}

The last RHS term of equation \ref{eq:circulation_jump_split} involves a jump in the pressure across the interface. Taking into account the 
jump in the normal stress condition at the interface, we express $\partial[[p/\rho]]/\partial s $  as

\begin{equation}
    \frac{\partial}{\partial s}[[\frac{p}{\rho}]]=-\frac{1}{\rho_1}\frac{\partial}{\partial s}(\kappa\sigma)-\frac{2}{\rho_1}[[\mu\left(\frac{\partial^2}{\partial s^2}(\mathbf{u}\cdot\mathbf{t})+\frac{\partial}{\partial s}(\kappa(\mathbf{u}\cdot\mathbf{n}))\right)]]+[[\frac{1}{\rho}]]\frac{\partial p_2}{\partial s}.
    \label{eq:dp_ds}
\end{equation}
Therefore, the final expression for the circulation   is that
\begin{align}
    \frac{D}{Dt}(\int_A \omega\, dA ) 
    &=
    \oint_C \nu \nabla \omega \cdot \mathbf{n}\, ds + \int_a^b \frac{\sigma}{\rho_2}\frac{\partial \kappa}{\partial s} ds  + \int_a^b \frac{\kappa}{\rho_2}\frac{\partial \sigma}{\partial s} ds  \nonumber\\
    &\quad
    +\int_a^b \left( \frac{2}{\rho_1}[[\mu]]\left(\frac{\partial^2}{\partial s^2}(\mathbf{u}\cdot\mathbf{\hat{t}})+\frac{\partial}{\partial s}(\kappa(\mathbf{u}\cdot\mathbf{\hat{n}}))\right)-[[\frac{1}{\rho}]]\frac{\partial p_2}{\partial s}\right) ds
    \label{eq:final},
\end{align}
where the terms of the RHS  represent mechanisms that drive the net evolution of circulation. These include the viscous diffusion,  curvature gradients,  surface tension gradients along the interface (including Marangoni  effects), jumps in viscosity and pressure across the interface. 
\textcolor{black}{We note that Marangoni stresses and variations in the of surface tension are distinct in previous equation,  Marangoni stresses arise from the surface tension gradient term, $\partial_s \sigma$, which generates tangential vorticity, whereas the magnitude of surface tension enters through the curvature term, $\sigma\partial_s \kappa$, and modifies the flow via capillary pressure. The results presented below indicate that the observed enhancement is associated with the former mechanism.}

Figure~\ref{vorticity_mechanism} shows the RHS terms of equation \ref{eq:final}, evaluated along the arclength $s$ in an $(x-z)$ plane located at the centerline of the domain, for the  two spilling-like regimes considered in this work,  $\epsilon = 0.324$ (top panels) and $\epsilon =0.33$ (bottom panels). 
 For each case,  results are shown for the surfactant-free and the surfactant-laden cases with \CRCA{$\beta=0.3$} and \CRCA{$\beta_s=0.5$} (columns one to three, respectively).
 This allows identification of the dominant physical mechanisms contributing to circulation generation. The viscous diffusion term is negligible in all conditions, confirming that interfacial mechanisms dominate circulation production.  Across all cases, the primary contributor is the term associated with jumps in viscosity and density, which is consistent with the large density ratio in the system. \CRCA{Notably, the magnitude of this term increases in the presence of surfactants,  indicating more intense interfacial activity. Contributions from surface tension gradients (Marangoni stresses) become significant only  at the largest elasticity number for $\epsilon =0.33$, and  they are spatially localized near regions of strong gradients in $\Gamma$, particularly along the wave crest.}


\section{Conclusions \label{Con}}

We present a detailed analysis of the effect of insoluble surfactants on wave breaking dynamics in both regular and spilling regimes, using high-resolution three-dimensional simulations. At $We=100$ and $Re=10^4$, we explore  how wave steepness modulates the flow in the presence of surfactants. The hybrid front-tracking/level-set method employed enables accurate coupling of inertia, capillarity, interfacial diffusion, and Marangoni stresses arising from surfactant-induced surface tension gradients. We find that surfactants have minimal impact on regular breakers, which retain similar shapes to the surfactant-free case, though a gradual interfacial rigidification is observed at longer times. In contrast, for spilling breakers, increasing elasticity induces a marked shift in dynamics\textcolor{black}{, which agrees with previously reported experimental work in the same spilling regime.}
\textcolor{black}{
Surfactants enhance spilling breakers through Marangoni stresses, while surface-tension reduction alone does not reproduce this effect. This is demonstrated by simulations in which Marangoni stresses are suppressed while allowing local variations of surface tension  along the interface.} We also show that surfactants  generate interfacial vorticity through Marangoni stresses driven by surfactant concentration gradients, as quantified by the circulation along the interface.

To our knowledge, this work provides the first direct characterization of the spatiotemporal evolution of surfactant concentration profiles during the spilling transition.
\textcolor{black}{Our results demonstrate that surfactant redistribution and the resulting Marangoni stresses play a key role in shaping the interfacial dynamics in the parameter regime considered here. How these mechanisms extend to higher Reynolds and Weber numbers, where inertia becomes increasingly important, remains an open question and warrants further investigation.}

%
Future studies should extend the present analysis beyond the regular and spilling regimes into the fully plunging regime, where bubble entrainment and droplet ejection result from the dynamics. Capturing these processes requires that all relevant length and time scales be resolved, making adaptive mesh refinement indispensable. Higher Reynolds numbers and greater wave steepness must also be explored to reproduce realistic ocean conditions. In addition, incorporating soluble surfactants, with full adsorption/desorption kinetics and bulk transport, would introduce chemical timescales that could either damp or amplify wave breaking events.\\

Declaration of Interests. The authors report no conflict of interest. \\

\subsection*{Acknowledgments }
This research used the Delta advanced computing and data resource which is supported by the National Science Foundation (award OAC 2005572) and the State of Illinois. CRCA acknowledges with gratitude O. K. Matar for insightful discussions on vorticity production, and L. Kahouadji  for  initial discussions. D.J. and J.C. acknowledge support through HPC/AI computing time at the Institut du Developpement et des Ressources en Informatique Scientifique (IDRIS) of the Centre National de la Recherche Scientifique (CNRS), coordinated by GENCI (Grand Equipement National de Calcul Intensif) grant 2026 A0202B06721. The numerical simulations were performed with code BLUE \citep{Shin_jmst_2017}





\section*{Appendix A: Vorticity generation in two-dimensional flows}

\subsection*{Circulation}
Before we examine the sources of vorticity flux near an interface, we must first consider the case of a single fluid and the vorticity equation which is given by
\begin{equation}
    \frac{\partial\boldsymbol{\omega}}{\partial t}+\mathbf{u}\cdot\nabla \boldsymbol{\omega}+\boldsymbol{\omega}\cdot\nabla\mathbf{u}=\nu\nabla^2\boldsymbol{\omega}.
    \label{eq:vorticity_3D}
\end{equation}
Now, for a 2D flow, $\boldsymbol{\omega}\cdot\nabla\mathbf{u}=0$ since $\nabla\mathbf{u}$ exists only in the $(x,y)$ plane whereas $\boldsymbol{\omega}=(0,0,\omega)$. Thus, Eq. (\ref{eq:vorticity_3D}) can be expressed as a 2D equation for the scalar quantity $\omega$:
\begin{equation}
    \frac{D\omega}{Dt}=\nu\nabla^2\omega.
    \label{eq:vorticity_2D}
\end{equation}
From Stokes's theorem, we can write $\oint_C \mathbf{v}\cdot d\mathbf{s}=\int_A (\nabla\times\mathbf{v})\cdot d\mathbf{a}$ for any vector $\mathbf{v}$ where $d\mathbf{s}$ is an infinitesimal displacement along the arclength on a contour $C$ that encloses a simply connected\footnote{Note that this becomes important below when we consider the case of two fluids separated by an interface.} area $A$ in the fluid and $d\mathbf{a}=\mathbf{e_k} dA$ represents an infinitesimal region in this area; $\mathbf{e_k}$ points into the page perpendicularly to the $(x,y)$ plane. Thus, if we set $\mathbf{v}=\mathbf{u}$ then
\begin{equation}
    \oint_C\mathbf{u}\cdot d\mathbf{s}=\int_A (\nabla\times\mathbf{u})\cdot \mathbf{e_k}dA=\int_A \boldsymbol{\omega}\cdot\mathbf{e_k}dA=\int_A\omega dA\equiv \Gamma_A,
\end{equation}
which relates the {\it total circulation} $\Gamma_A$ to $\omega$. By taking a total derivative of $\Gamma_A$ and making use of Eq. (\ref{eq:vorticity_2D}), we can write
\begin{equation}
    \frac{D\Gamma_A}{Dt}=\frac{D}{Dt}\int_A\omega dA=\int_A\frac{D\omega}{Dt} dA=\int_A \nu\nabla^2\omega dA.
\end{equation}
Using the Divergence Theorem, the final term on the LHS of this equation can re-expressed as follows
\begin{equation}
    \int_A\nu\nabla^2\omega dA=\int_A \nabla\cdot(\nu\nabla\omega)dA=\oint_C\mathbf{n}_c\cdot (\nu\nabla\omega)ds,
\end{equation}
where $\mathbf{n}_c$ is the outward-pointing unit normal to the contour $C$.
As a result, the total rate of change of circulation in area $A$, $\Gamma_A$, becomes
\begin{equation}
    \frac{D\Gamma_A}{Dt}=\frac{D}{Dt}\int_A\omega dA=\oint_C \nu \nabla\omega\cdot\mathbf{n}_c ds.
\label{eq:circulation}
\end{equation}
This is an important equation which shows that the change in circulation results directly from the flux of vorticity from the boundary (characterised by the contour $C$).

\subsection*{Near-interface Vorticity jump}
We  consider the case of two fluids separated by an interface. In order to evaluate the LHS of Eq. (\ref{eq:circulation}), we follow an approach that relies on a construction similar to that represented by Fig. 2 in \cite{brons_2014}. 
This is necessary because the area $A$ is no longer simply connected but is instead divided because of the presence of the interface. With reference to  Fig 2 in \cite{brons_2014}, we write
\begin{equation}
    \frac{D}{Dt}\int_{A_1 U A_2}\omega dA=\int_{C_1}\nu_1\nabla\omega_1\cdot \mathbf{n}_c ds + \int_{C_2}\nu_2\nabla\omega_2\cdot \mathbf{n}_cds
    +\int_{C_1'}\nu_1\nabla\omega_1\cdot \mathbf{n}_c ds + \int_{C_2'}\nu_2\nabla\omega_2\cdot \mathbf{n}_cds.
    \label{eq:contour}
\end{equation}
Here, we distinguish between the normal to the contour $C$, $\mathbf{n}_c$, and that to the interface $I$, $\mathbf{n}$.

We now let $\mathbf{n}_c \rightarrow -\mathbf{n}$ from the fluid 2 side, and $\mathbf{n}_c \rightarrow \mathbf{n}$ from the fluid 1 side, which results in
\begin{equation}
    \frac{D}{Dt}\int_{A_1 U A_2}\omega dA=\int_{C_1}\nu_1\nabla\omega_1\cdot \mathbf{n}_c ds + \int_{C_2}\nu_2\nabla\omega_2\cdot \mathbf{n}_cds
    +\int_{C_1'}\nu_1\nabla\omega_1\cdot \mathbf{n} ds - \int_{C_2'}\nu_2\nabla\omega_2\cdot \mathbf{n}ds.
    \label{eq:contour}
\end{equation}
We note that in
region $A_1$, we march in an anti-clockwise direction along contour $C_1'$ where $\mathbf{n}_c$ is aligned with $\mathbf{n}$ and then along contour $C_1$.
In region $A_2$, we march along contour $C_2$ in an anti-clockwise direction and then along contour $C_2'$ where $\mathbf{n}_c$ points out of fluid 2 into fluid 1 in a direction opposite to that of $\mathbf{n}$; the latter explains the minus sign in the fourth term of Eq. (\ref{eq:contour}).

We now let the contours $C_1'$ and $C_2'$ approach the interface $I$ so that the union of $A_1$ and $A_2$ approaches $A$, i.e. $A_1 U A_2 \rightarrow A$, which leads to
\begin{equation}
    \frac{D}{Dt}\int_A\omega dA=\oint_C\nu\nabla\omega\cdot\mathbf{n}_c ds
       -\left(\int_I\nu_2\nabla\omega_2\cdot\mathbf{n}ds-\int_I\nu_1\nabla\omega_1\cdot\mathbf{n}ds\right).
\end{equation}
By introducing the notation $[[q]]=q_2-q_1$ for the jump across the interface of a quantity $q$, this equation can be re-expressed as follows
\begin{equation}
       \frac{D}{Dt}\int_A\omega dA=\oint_C\nu\nabla\omega\cdot\mathbf{n}_c ds-\int^b_a\left[\left[\nu\nabla\omega\cdot\mathbf{n}\right]\right]ds.
    \label{eq:circulation_jump}
\end{equation}

Equation (\ref{eq:circulation_jump}) illustrates the contribution of the jump in the vorticity flux across the interface to the rate of change of the circulation. 
We can calculate this jump by using  
\begin{equation}
   \frac{D}{Dt}(\mathbf{u}\cdot\mathbf{t})+\kappa(\mathbf{u}\cdot\mathbf{n})(\mathbf{u}\cdot\mathbf{t})-\frac{1}{2}\frac{\partial}{\partial s}(\mathbf{u}\cdot\mathbf{n})^2+\frac{\nabla p}{\rho}\cdot \mathbf{t}=\nu \nabla\omega\cdot\mathbf{n},
\end{equation}
which comes from  Eq. (23) in \cite{lundgren_1999}) and Eq. (2.10) in \cite{brons_2014}.
 Then, we have 
\begin{equation}
    \frac{D}{Dt}[[\mathbf{u}\cdot\mathbf{t}]]+\kappa[[(\mathbf{u}\cdot\mathbf{n})(\mathbf{u}\cdot\mathbf{t})]]
    +\frac{\partial}{\partial s}[[\left(\frac{p}{\rho}\right)]]-\frac{1}{2}\frac{\partial}{\partial s}[[(\mathbf{u}\cdot\mathbf{n})^2]]=[[\nu\nabla\omega\cdot\mathbf{n}]];
    \label{eq:jump_1}
\end{equation}
here, we have used the fact that 
$\mathbf{t}\cdot \nabla=(\partial/\partial s)$.
Substitution of Eq. (\ref{eq:jump_1}) into Eq. (\ref{eq:circulation_jump}) leads to
\begin{align}
\frac{D}{Dt}\!\left(\int_A \omega\,dA + \int_a^b [[\mathbf{u}\!\cdot\!\mathbf{t}]]\,ds\right)
&= \oint_C \nu\,\nabla\omega\!\cdot\!\mathbf{n}_c\,ds
     - \int_a^b \kappa\, [[(\mathbf{u}\!\cdot\!\mathbf{n})(\mathbf{u}\!\cdot\!\mathbf{t})]]\,ds
\nonumber\\[4pt]
&\quad + \frac{1}{2}\int_a^b \frac{\partial}{\partial s} [[(\mathbf{u}\!\cdot\!\mathbf{n})^{2}]]\,ds
     - \int_a^b \frac{\partial}{\partial s}
       \left[\!\left[\frac{p}{\rho}\right]\!\right] ds.
        \label{eq:circulation_jump_2}
\end{align}
Equation (\ref{eq:circulation_jump_2}) involves a jump in the pressure across the interface. To make progress, we must first consider the jump in the normal stress condition at the interface expressed by
\begin{equation}
    -p_1+\mu_1 \mathbf{n}\cdot \mathbf{D}_1\cdot\mathbf{n}+\kappa \sigma=-p_2+\mu_2 \mathbf{n}\cdot \mathbf{D}_2\cdot\mathbf{n},
    \label{eq:NSB}
\end{equation}
in which $\sigma$ represents the interfacial tension and $\mu_1$ and $\mu_2$ the dynamic viscosities of fluids 1 and 2, respectively. Rearranging Eq. (\ref{eq:NSB}) yields
\begin{equation}
    [[p]]=-\kappa\sigma-2[[\mu \left(\frac{\partial}{\partial s}(\mathbf{u}\cdot \mathbf{t})+\kappa (\mathbf{u}\cdot\mathbf{n})\right)]];
    \label{eq:pjump_2D}
\end{equation}
here, we have made use of
\begin{equation}
    \mathbf{n}\cdot \mathbf{D}\cdot \mathbf{n}=
    -2\frac{\partial}{\partial s}(\mathbf{u\cdot \mathbf{t}})-2(\mathbf{u}\cdot\mathbf{n})\kappa.
    \label{eq:nDn}
\end{equation}
%
Now, we re-express $[[p/\rho]]$ as follows
\begin{eqnarray}
    [[\frac{p}{\rho}]]=\frac{p_2}{\rho_2}-\frac{p_1}{\rho_1}&=&\frac{p_2}{\rho_2}-\frac{p_2}{\rho_1}+\frac{p_2}{\rho_1}-\frac{p_1}{\rho_1}\nonumber\\
    &=&p_2\left(\frac{1}{\rho_2}-\frac{1}{\rho_1}\right)\nonumber+\frac{1}{\rho_1}(p_2-p_1)\\
    &=&p_2 [[\frac{1}{\rho}]]+\frac{[[p]]}{\rho_1},
\end{eqnarray}
from which we get that
\begin{equation}
    \frac{\partial}{\partial s}[[\frac{p}{\rho}]]=-\frac{1}{\rho_1}\frac{\partial}{\partial s}(\kappa\sigma)-\frac{2}{\rho_1}[[\mu\left(\frac{\partial^2}{\partial s^2}(\mathbf{u}\cdot\mathbf{t})+\frac{\partial}{\partial s}(\kappa[\mathbf{u}\cdot\mathbf{n}])\right)]]+[[\frac{1}{\rho}]]\frac{\partial p_2}{\partial s}.
    \label{eq:dp_ds}
\end{equation}

\section*{Appendix B:  Equation of state in insoluble surfactants}

\begin{figure}
    \centering
\begin{tabular}{cc}
\includegraphics[width=0.48\textwidth]{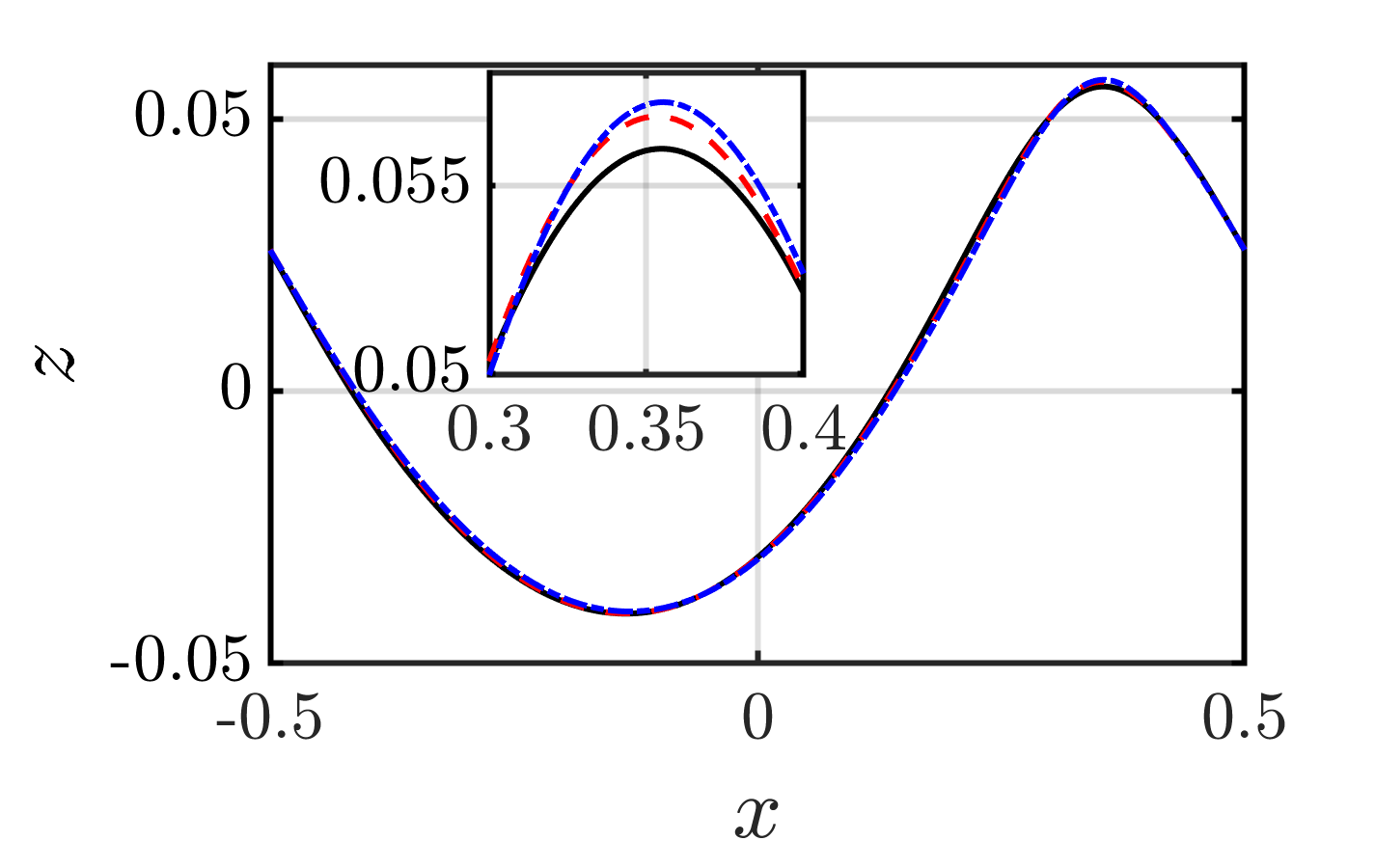}& 
\includegraphics[width=0.48\textwidth]{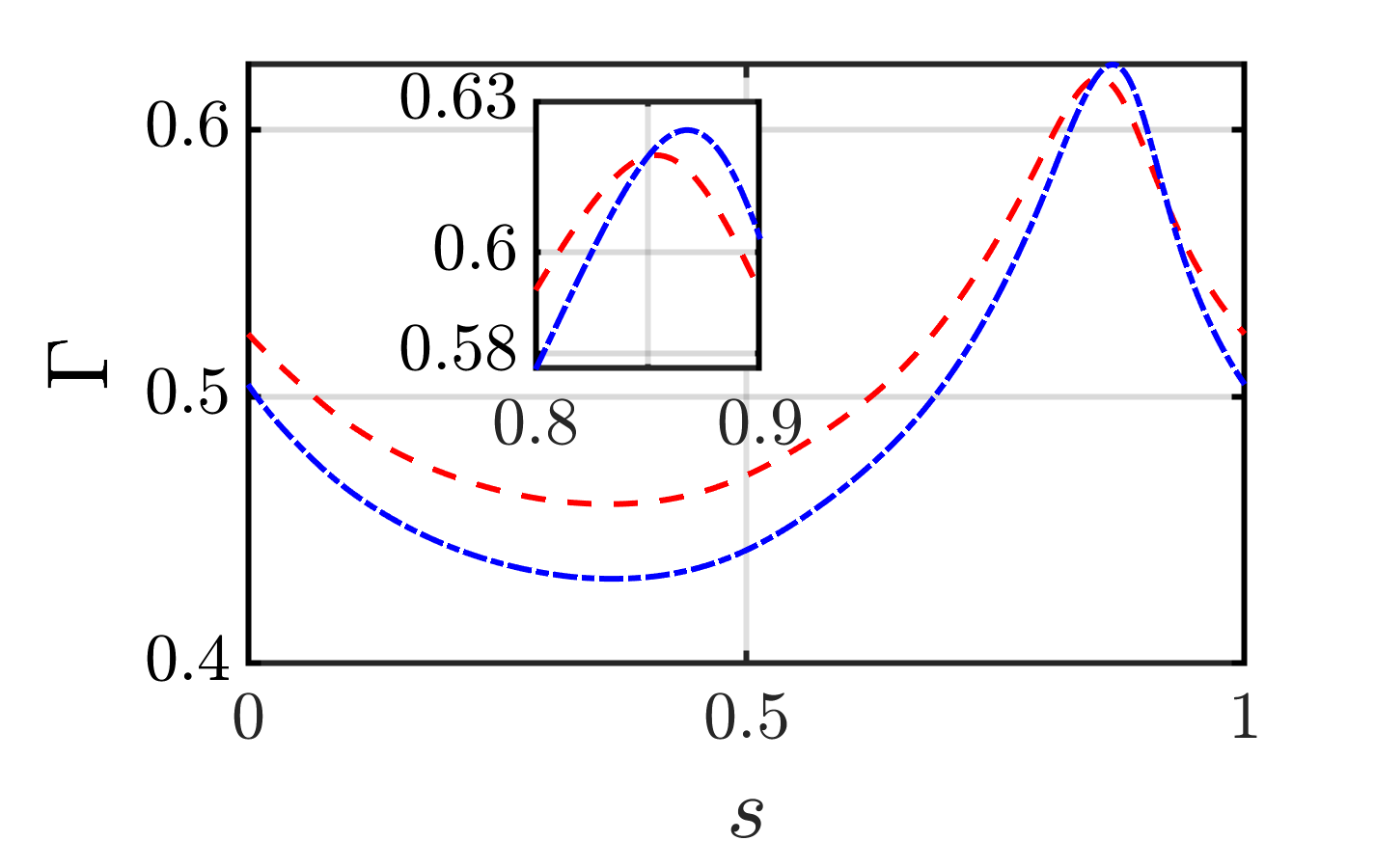}
\\
(a) & (b)\\
 \includegraphics[width=0.48\textwidth]{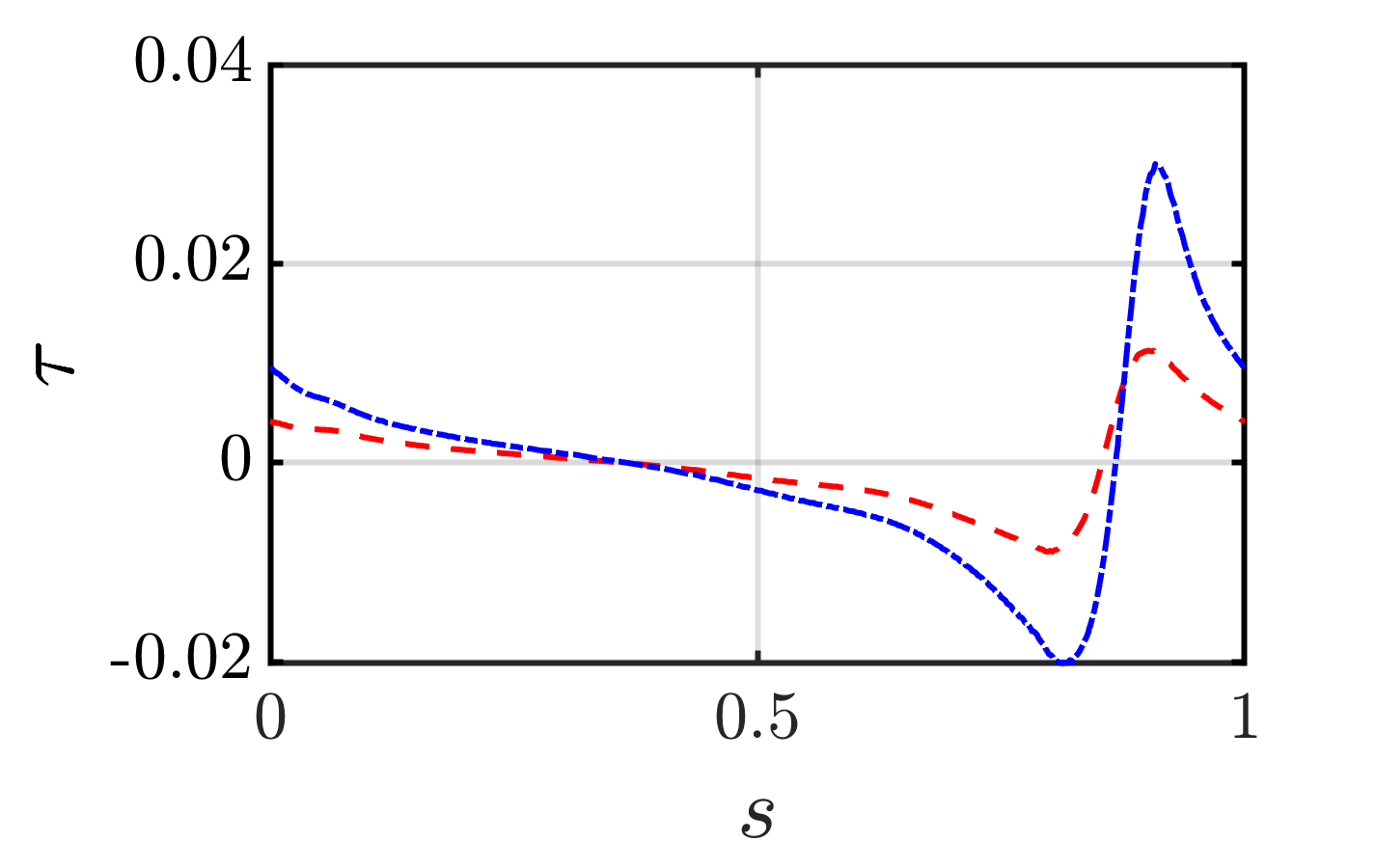}&
\includegraphics[width=0.48\textwidth]{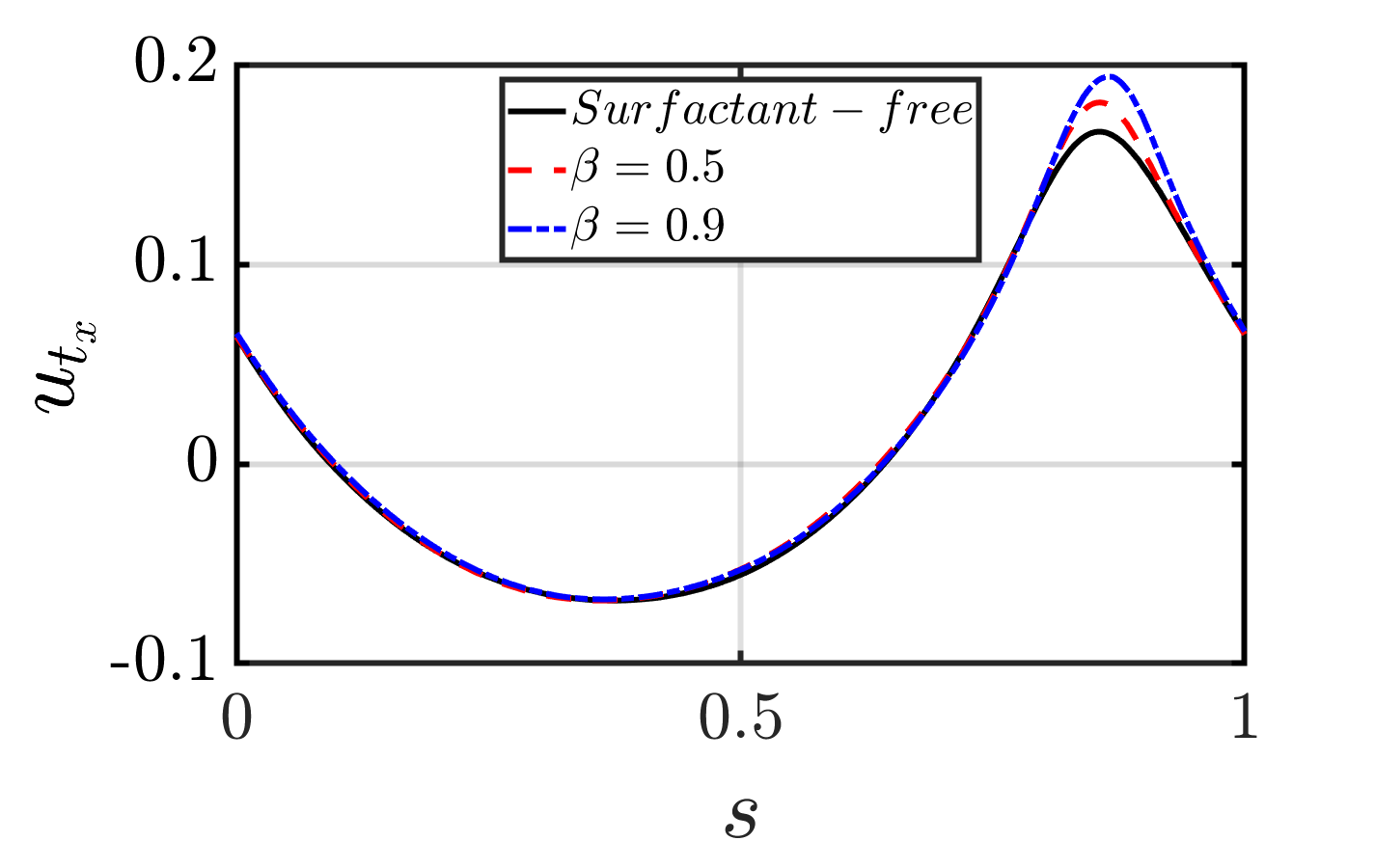}\\
(c) & (d)\\
\end{tabular}
\caption{Effect of the elasticity parameter $\beta_s$ for regular wave regime ($\epsilon=0.3$) at  $t=5.6$. Two-dimensional projections of the interface, $\Gamma$, $Ma$ and $u_{t_x}$ in the $(x–z)$ plane ($y=\lambda/8$) are shown in (a–d), respectively.  Note that the abscissa in (a) corresponds to the $x$ coordinate, and in (b–d) to the arc length, $s$.    }
    \label{ep_0d3_old}
\end{figure}

\begin{figure}
    \centering
\begin{tabular}{cc}
\includegraphics[width=0.5\textwidth]{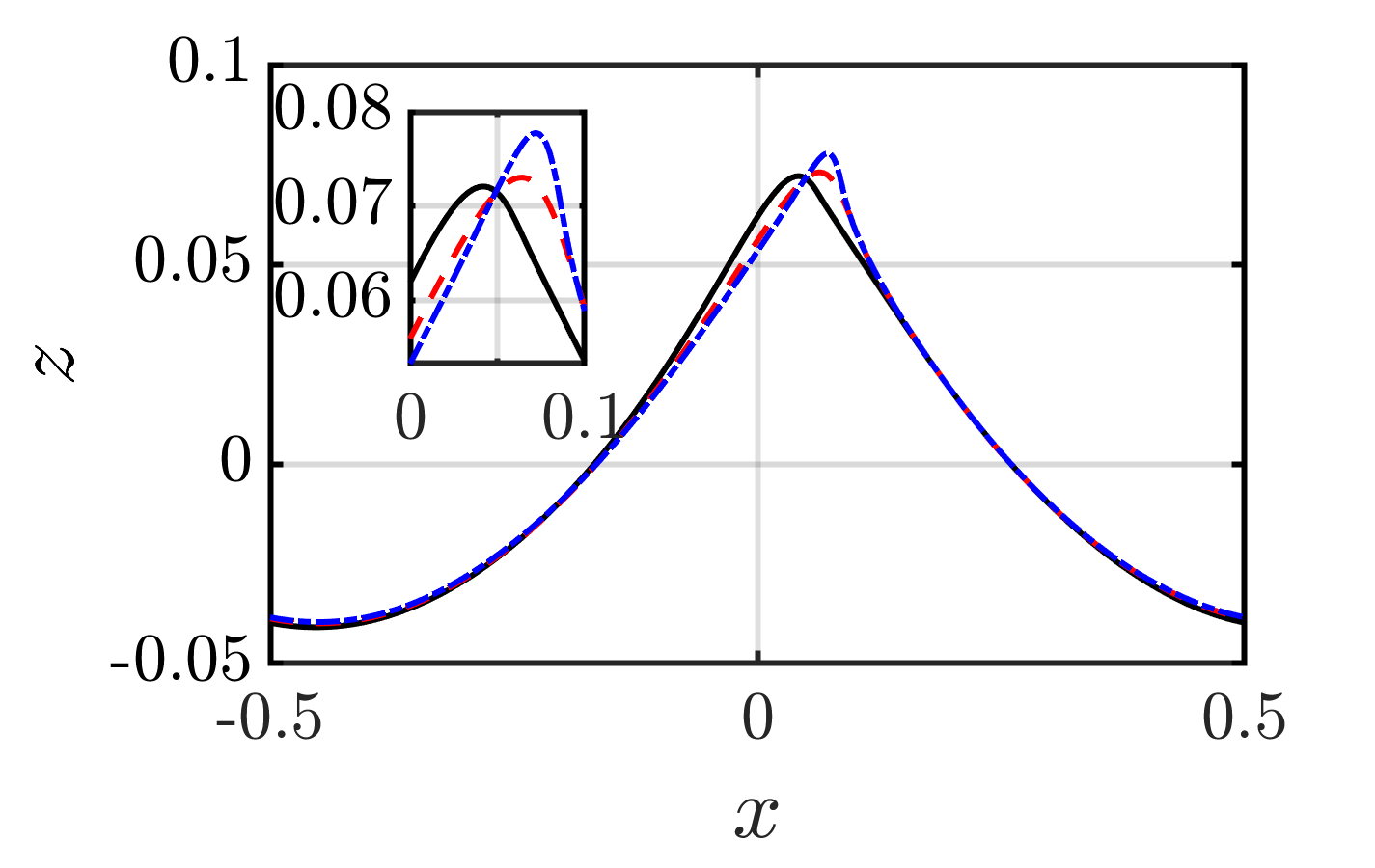} & \includegraphics[width=0.5\textwidth]{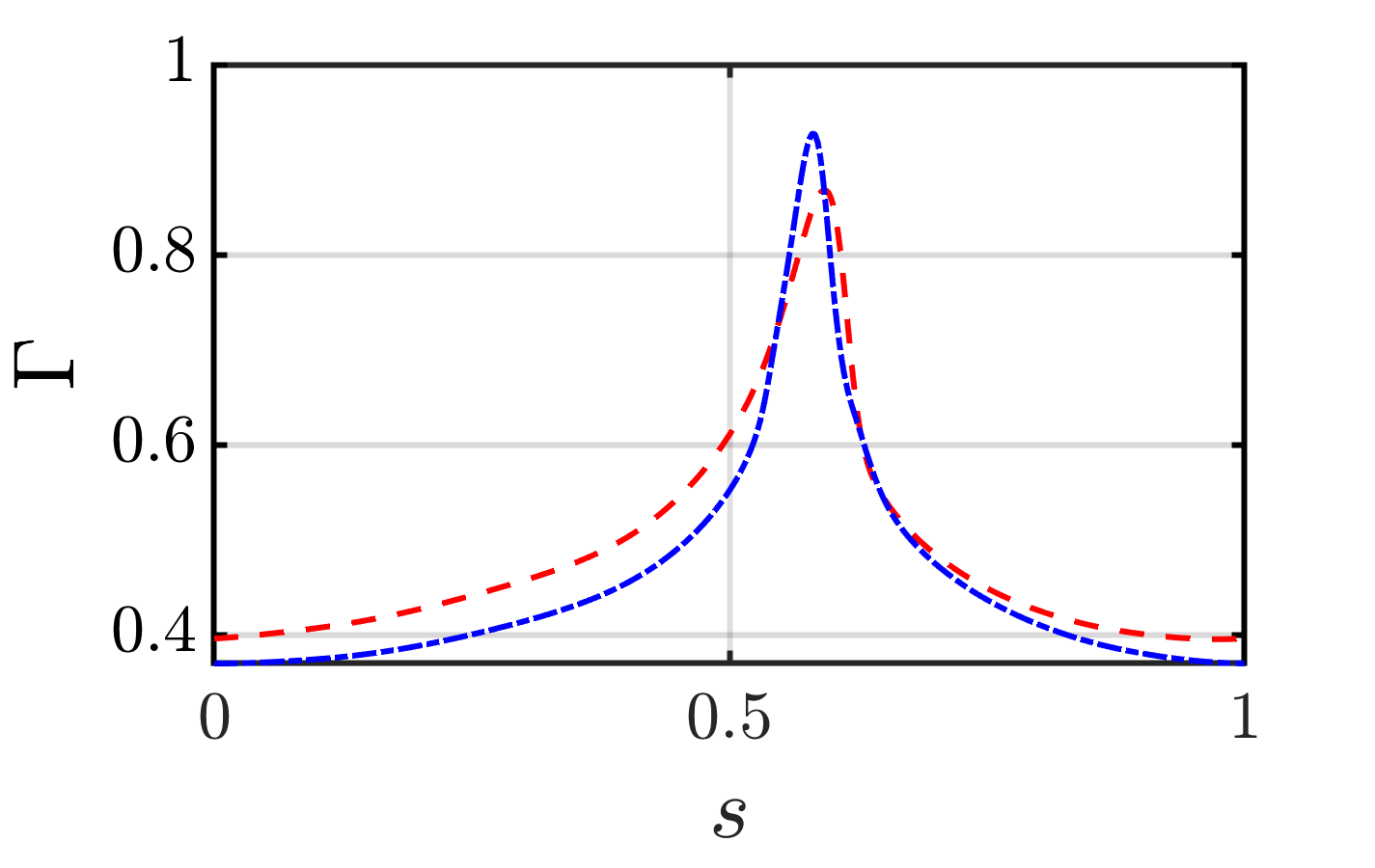} \\
(a) & (b) \\
\includegraphics[width=0.5\textwidth]{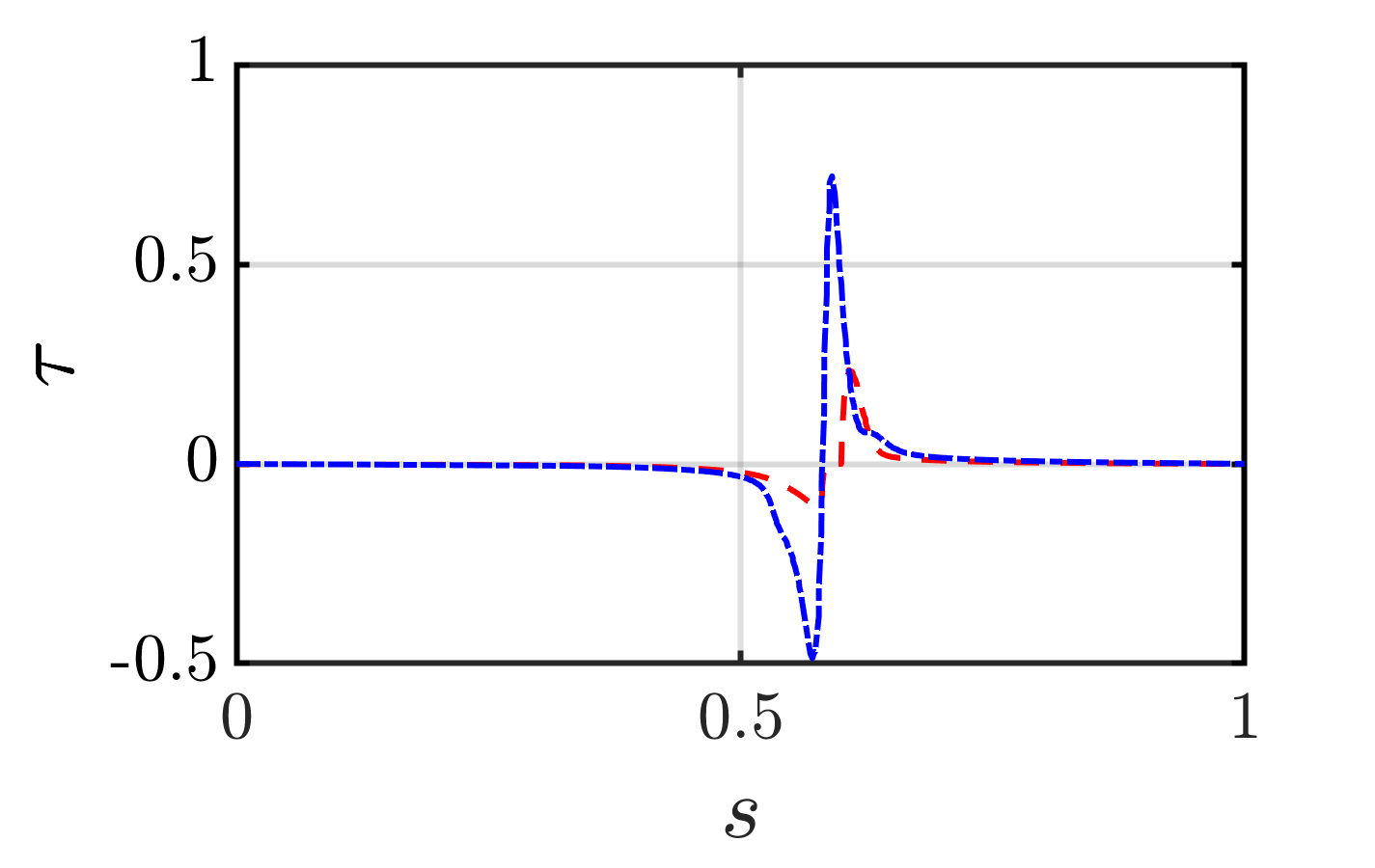} & \includegraphics[width=0.5\textwidth]{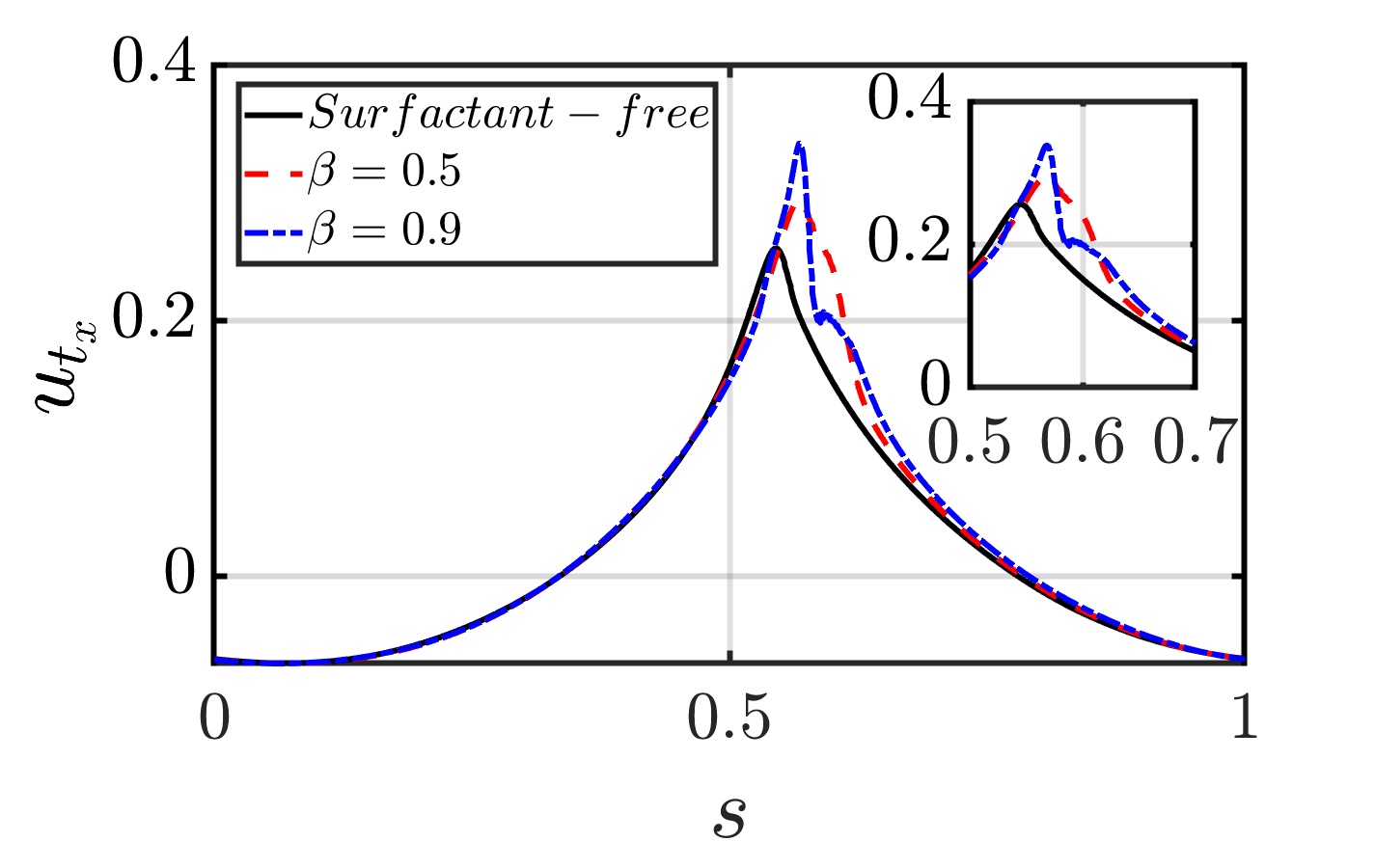} \\
(c) & (d) \\
\hline
\includegraphics[width=0.5\textwidth]{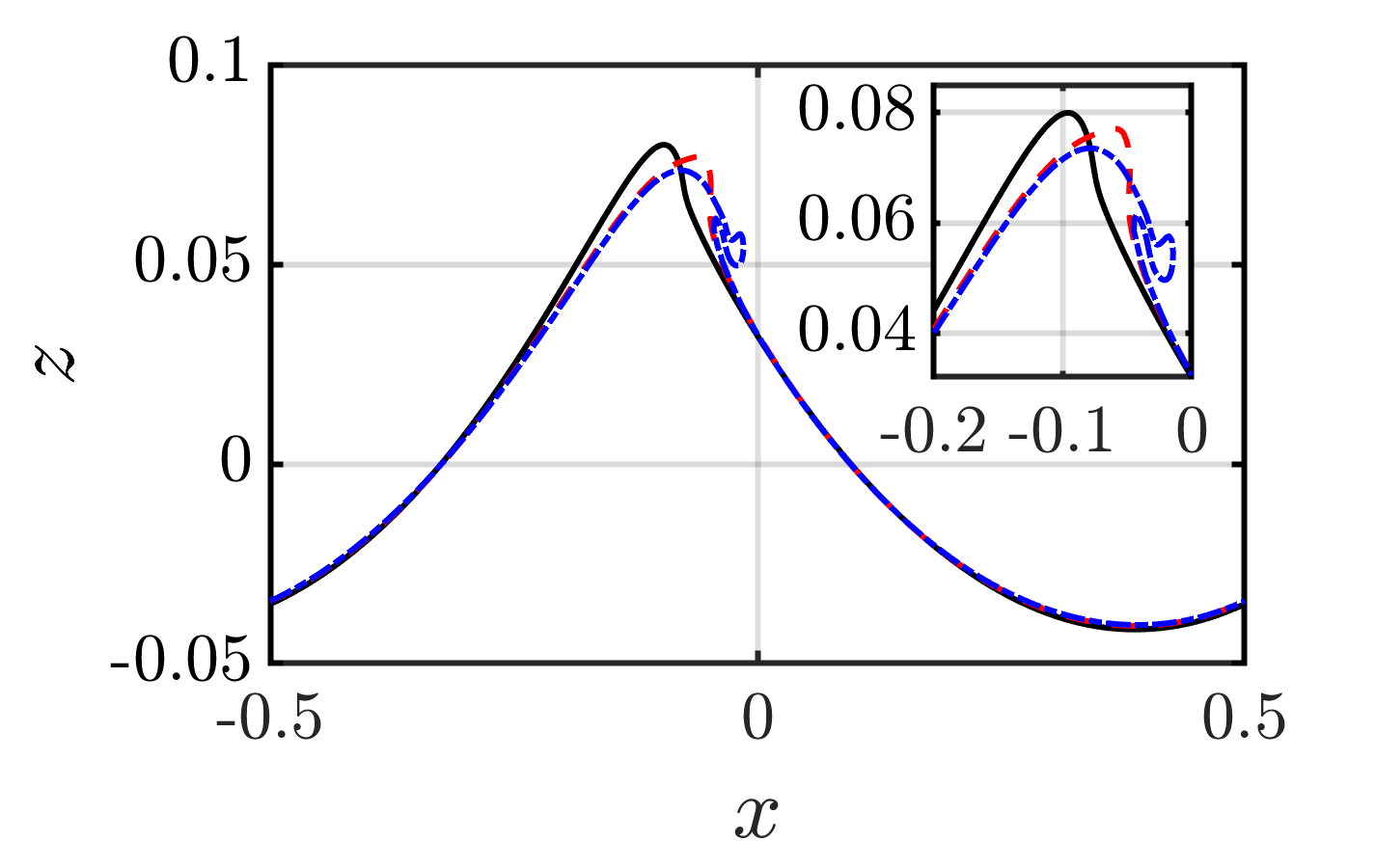} & \includegraphics[width=0.5\textwidth]{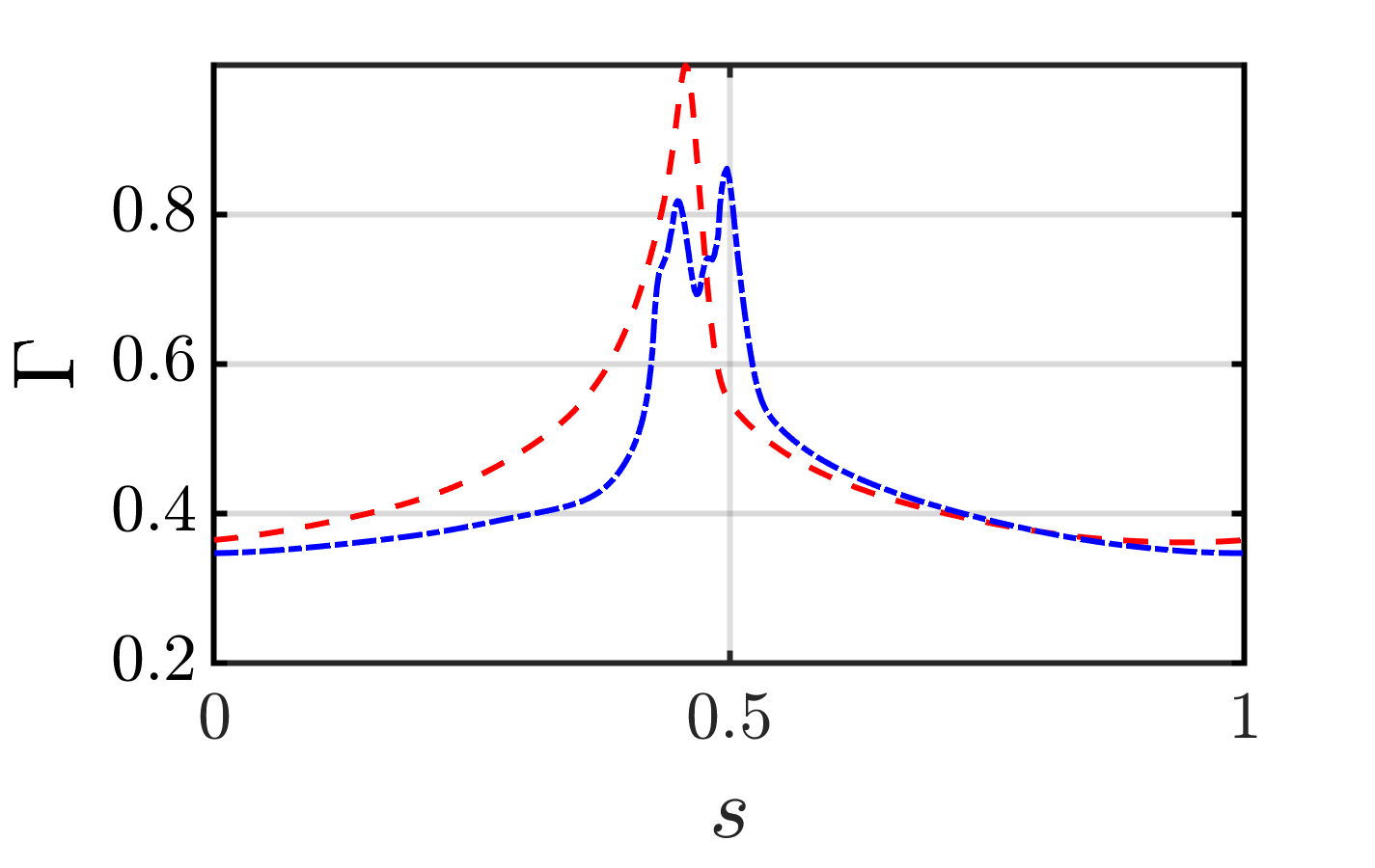} \\
(e) & (f) \\
\includegraphics[width=0.5\textwidth]{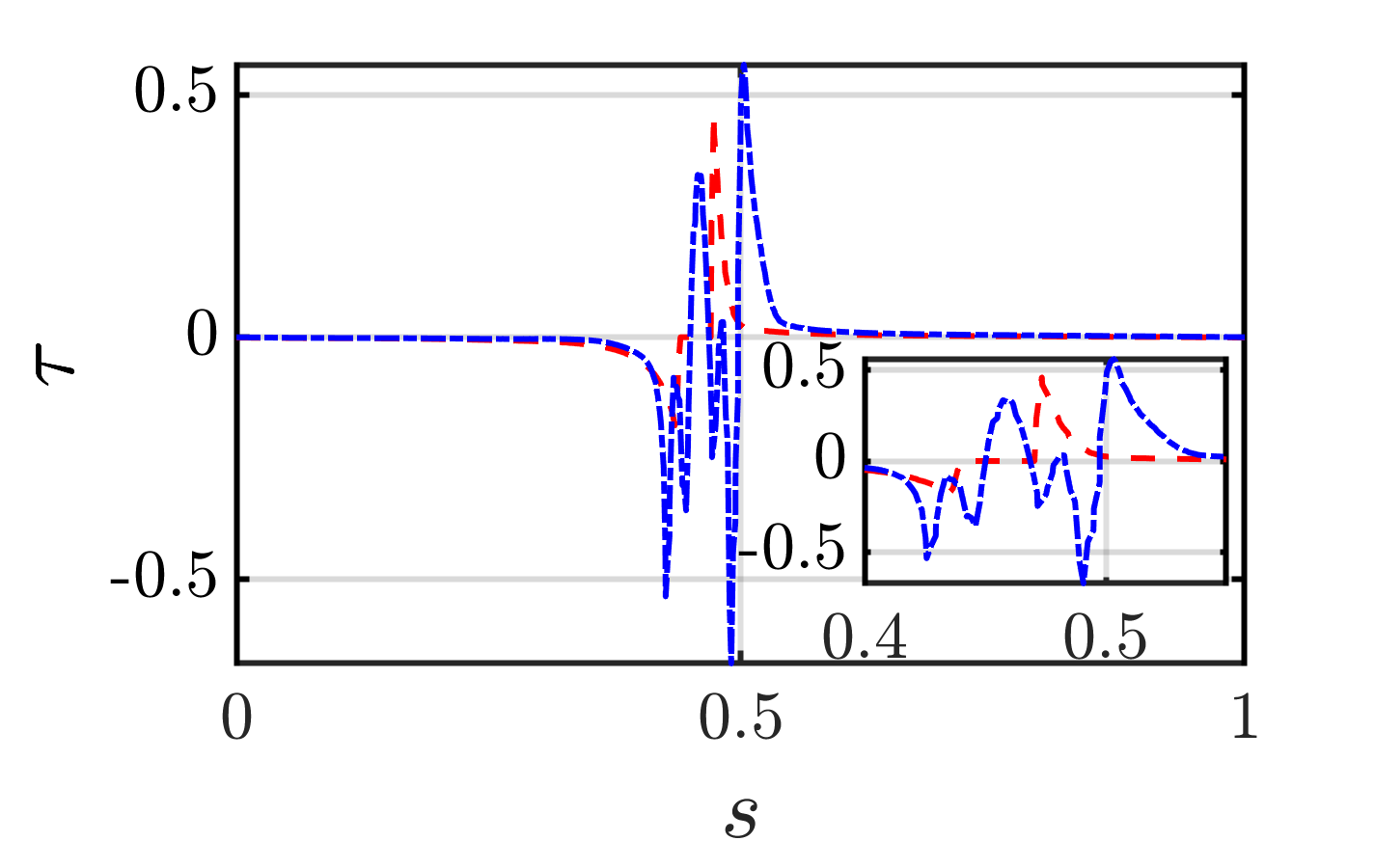} & \includegraphics[width=0.5\textwidth]{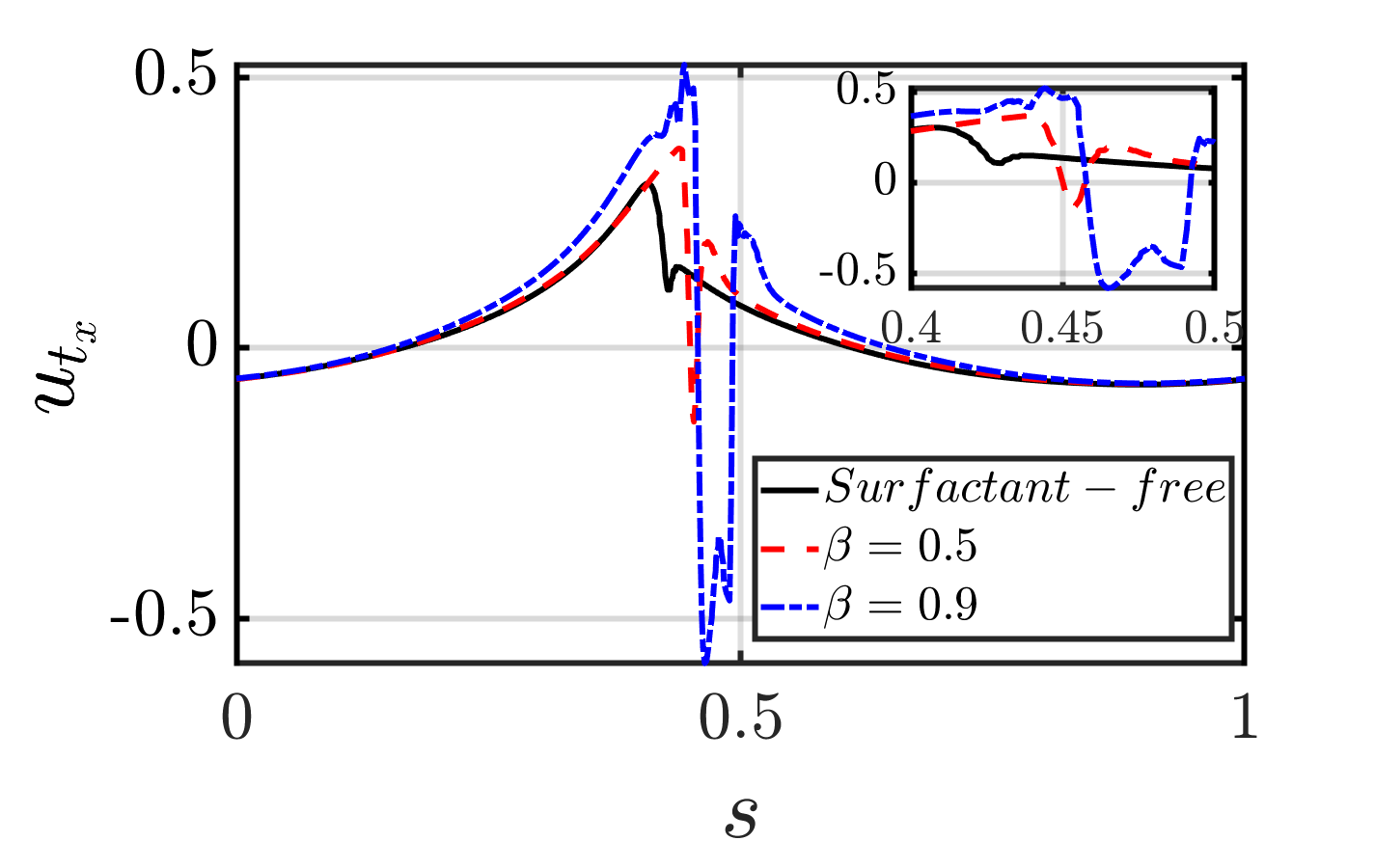} \\
(g) & (h) \\
\end{tabular}
\caption{Effect of the elasticity parameter for the spilling regime for  $\epsilon=0.324$  at $t=4.8$,  and $\epsilon=0.33$  at $t=4.4$. Two-dimensional projections of the interface, $\Gamma$, $Ma$ and $u_{t_x}$ in the $(x–z)$ plane ($y=\lambda/8$) are shown.  Panels (a–d) correspond to $\epsilon = 0.324$, and panels (e–h) to $\epsilon = 0.33$. 
} \label{fields_spilling_old}
\end{figure}

\CRCA{
In this appendix, we assess the the interfacial dynamics at elevated surfactant concentrations as it can lead to artificial numerical results.
We emphasis on the role of  the equation of state   relating  surface tension ${\sigma}$ and surfactant concentration ${\Gamma}$, by
${\sigma}=\text{max} [0.05, 1 + \beta_s \ln{ (1 -{\Gamma})]}$. While the logarithmic equation of state $ {\sigma}=1+\beta_s\ln(1-{\Gamma})$ is widely used \citep{manikantan_squires_2020,kamat_2020,constante2023impact}, and appropriate over a broad range of concentrations, the imposed lower bound ${\sigma}=0.05$, constitutes an abrupt and nonphysical modification of the system, a regularization also adopted in studies such as 
\citet{asadi_asgharzadeh_borazjani_2018,muradoglu2008front,pico2024surfactant}. This cutoff becomes active when the concentration exceeds the critical value ${\Gamma}_c=1-\exp(-0.95/\beta_s)$, beyond which the surface tension is artificially pinned. Although such a regularization would be inconsequential if ${\Gamma}<{\Gamma}_c$ everywhere or only in dynamically passive regions, \textcolor{black}{the simulations discussed in this appendix} exhibit local concentrations ${\Gamma}>{\Gamma}_c$, particularly near wave crests.
}
\textcolor{black}{We reiterate that the cutoff is active in these appendix simulations, but is not active in the simulations discussed in the main text. These results are presented to demonstrate that this numerical formulation can yield qualitatively different and non-physical dynamics, underscoring the need for caution in its use and interpretation.}

\CRCA{
Figures~\ref{ep_0d3_old} and~ \ref{fields_spilling_old} show the interfacial dynamics in the regular and spilling-like regimes for two values of the surfactant elasticity, $\beta_s=0.5$ and $0.9$, corresponding to critical concentrations $\Gamma_c \approx 0.85$ and $\Gamma_c \approx 0.65$, respectively, for an initial surfactant concentration $\Gamma_0=0.5$. In the spilling-like  regime, the dynamics exhibit an abrupt qualitative change, with a transition from plunging to spilling-type breaking as $\beta_s$ is increased. In key snapshots such as figure \ref{fields_spilling_old}f, the surfactant concentration in the vicinity of the wave crest clearly exceeds $\Gamma_c$ for both values of $\beta_s$. Since this crest region governs the dominant interfacial dynamics, the imposed surface tension cutoff becomes active precisely where  gradients in surface tension, and subsequently   Marangoni stresses are expected to play a central role. As a result, the reported relationship between surface-tension reduction, Marangoni forcing, and the transition between plunging and spilling regimes may be influenced by the numerical regularization inherent in the constitutive model.}

\CRCA{
We emphasize that these limitations are not specific to the present configuration but are inherent to inertia-dominated flows with insoluble surfactants, where strong interfacial compression can drive the concentration into regimes in which commonly used logarithmic equations of state require numerical regularization. A physically consistent route to modelling such regimes is to extract the equation of state directly from experiments and to fit a smooth constitutive relation that remains well behaved at high concentrations. This approach has recently been adopted by \citet{bb_basilisk}, leading to improved physical fidelity in simulations. In the absence of such an experimental  equation of state, artificial cutoffs may lead to nonphysical effects when high surfactant concentrations are reached. An alternative modelling approach is to allow  surfactant solubility, enabling adsorption-desorption dynamics that naturally limit surface accumulation and alleviate the need for ad hoc regularization.}

\section*{Appendix C: Mesh resolution and spanwise domain length}

\begin{figure}
    \centering
 \begin{tabular}{c}
 \includegraphics[width=0.8\textwidth]{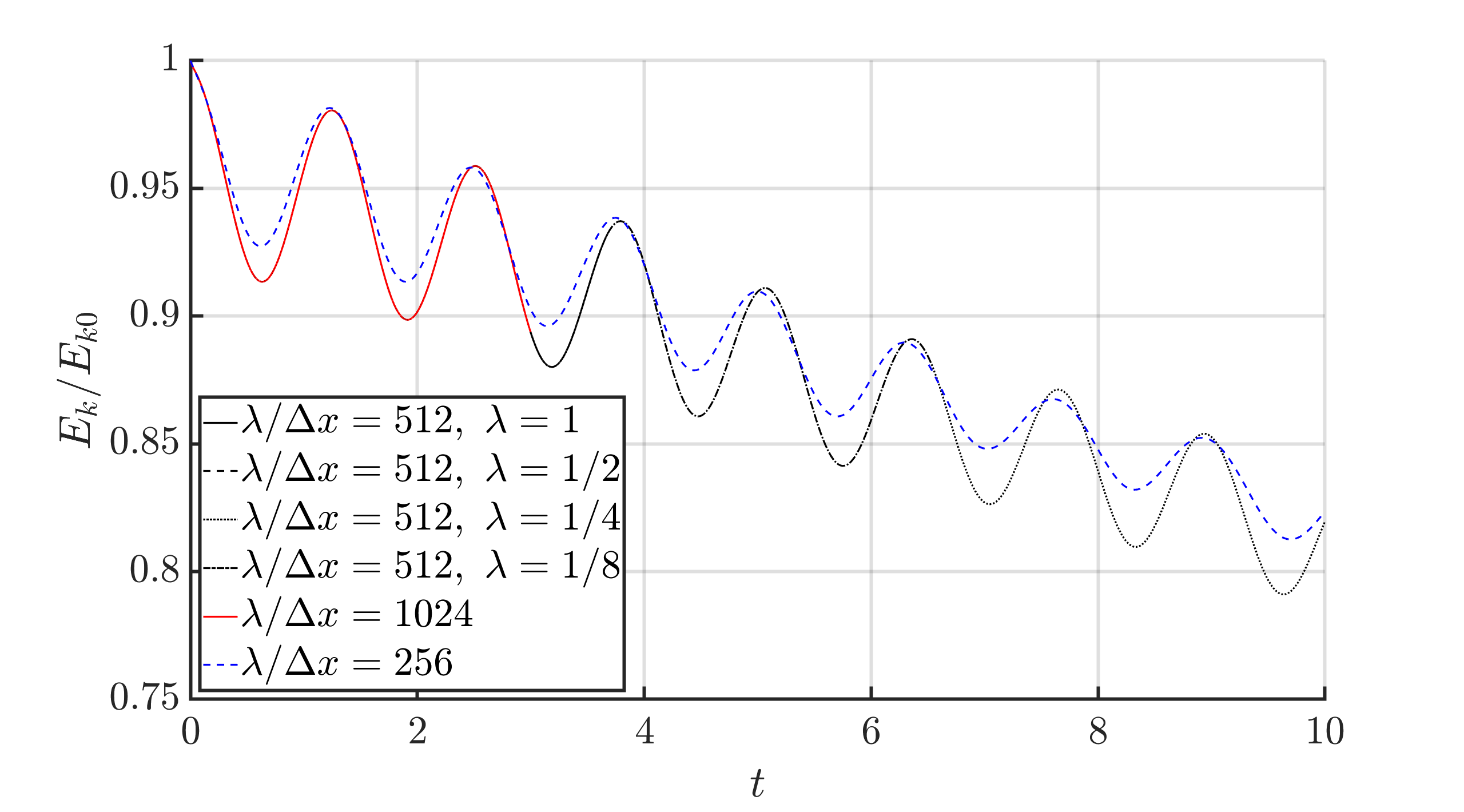} \\ 
 (a)\\
 \includegraphics[width=0.8\textwidth]{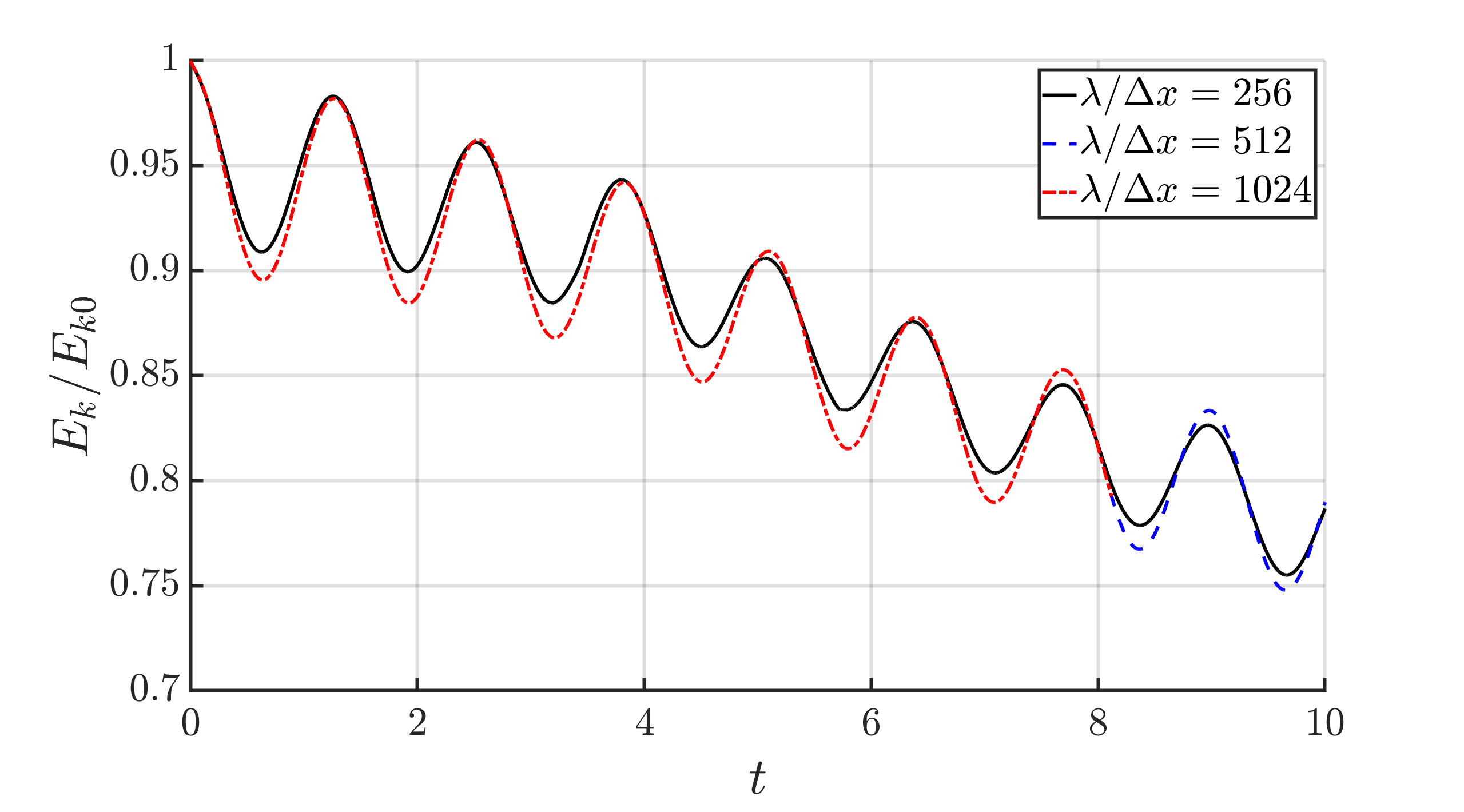} \\
 (b)
\end{tabular}
 \caption{Grid and spanwise domain-size independence of the kinetic energy evolution. The temporal evolution of the normalized kinetic energy $E_k/E_{k0}$ is shown for $\epsilon=0.30$ (regular wave) \textcolor{black}{and  $\epsilon = 0.33$ (spilling breaker) corresponding to panels (a) and (b), respectively.} In panel (a), spanwise domain widths $\lambda$, $\lambda/2$, $\lambda/4$, and $\lambda/8$ are also added   at fixed spatial resolution $\lambda/\Delta\mathbf{x}=512$.
 }\label{mesh_figure}
\end{figure}

\begin{figure}
    \centering
\begin{tabular}{cc}
\includegraphics[width=0.5\textwidth]{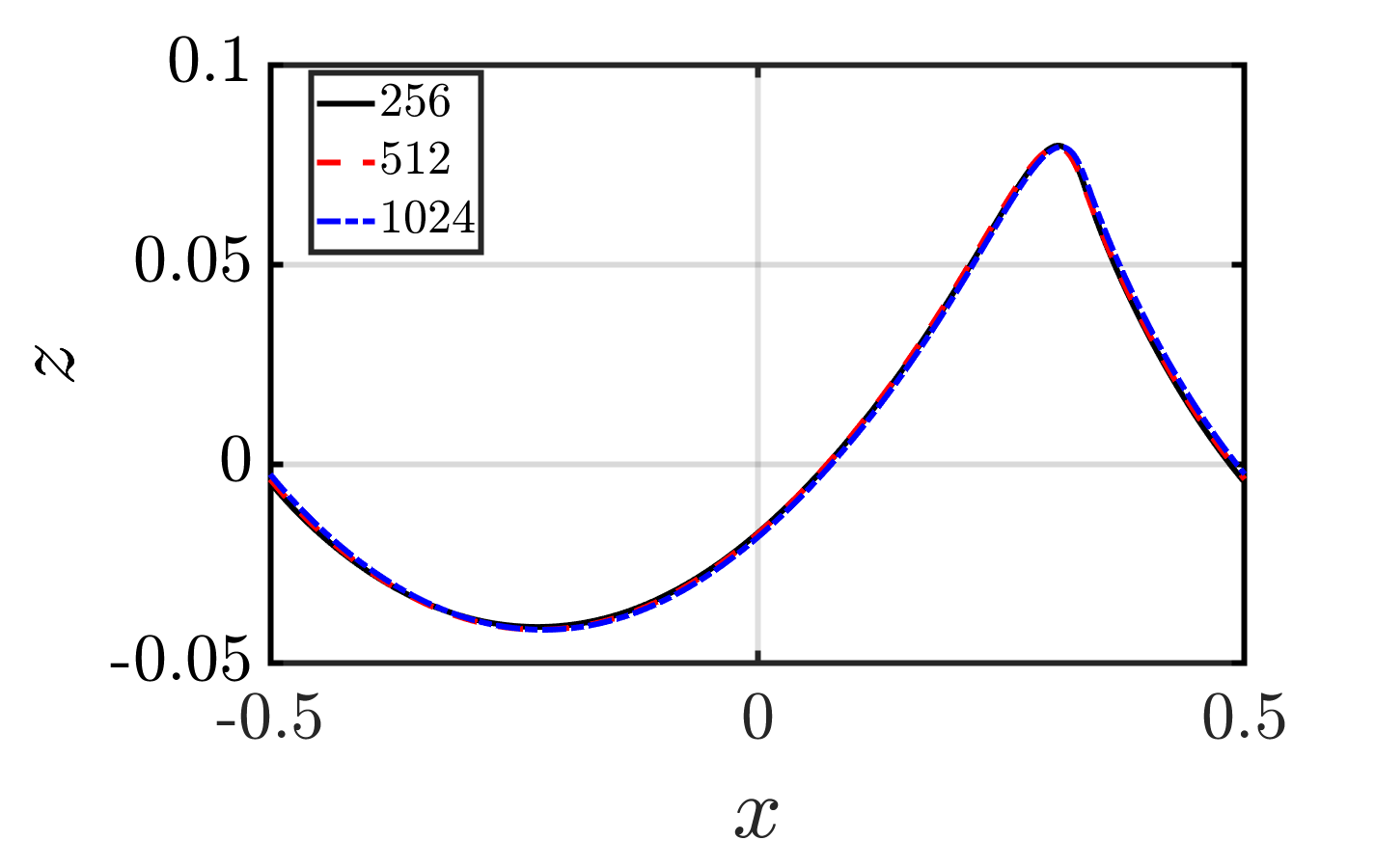} & \includegraphics[width=0.5\textwidth]{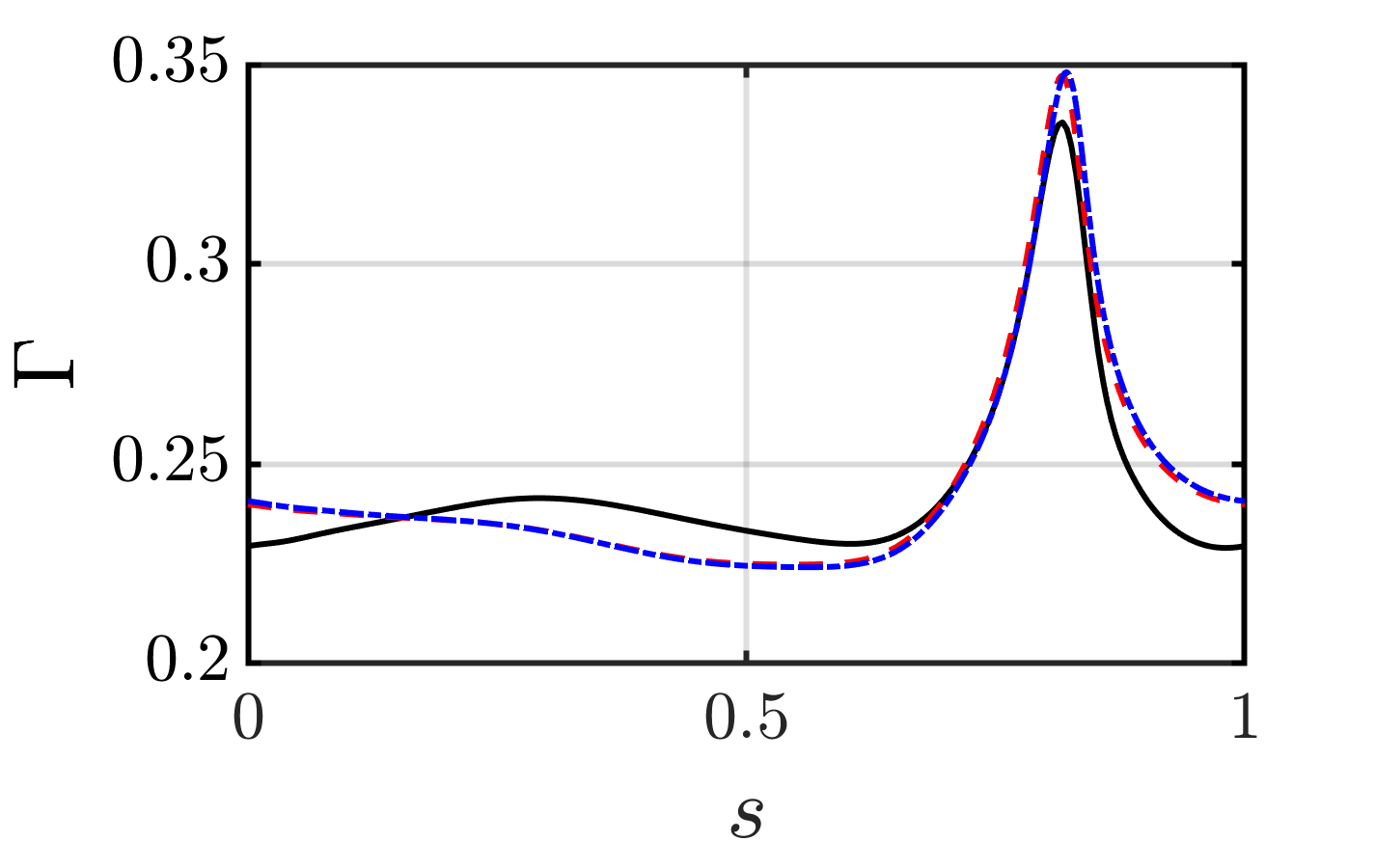} \\
(a) & (b) \\
\includegraphics[width=0.5\textwidth]{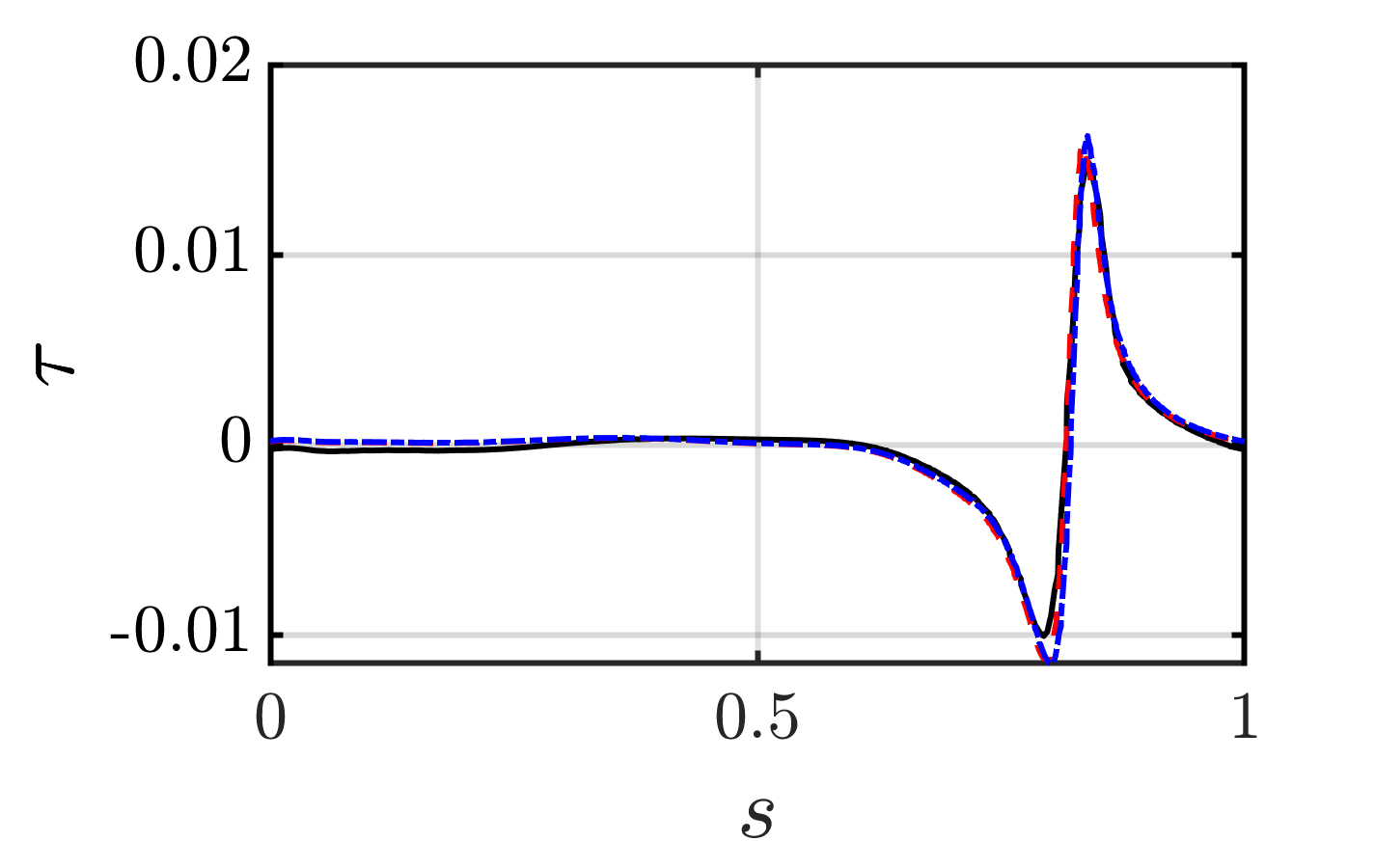} & \includegraphics[width=0.5\textwidth]{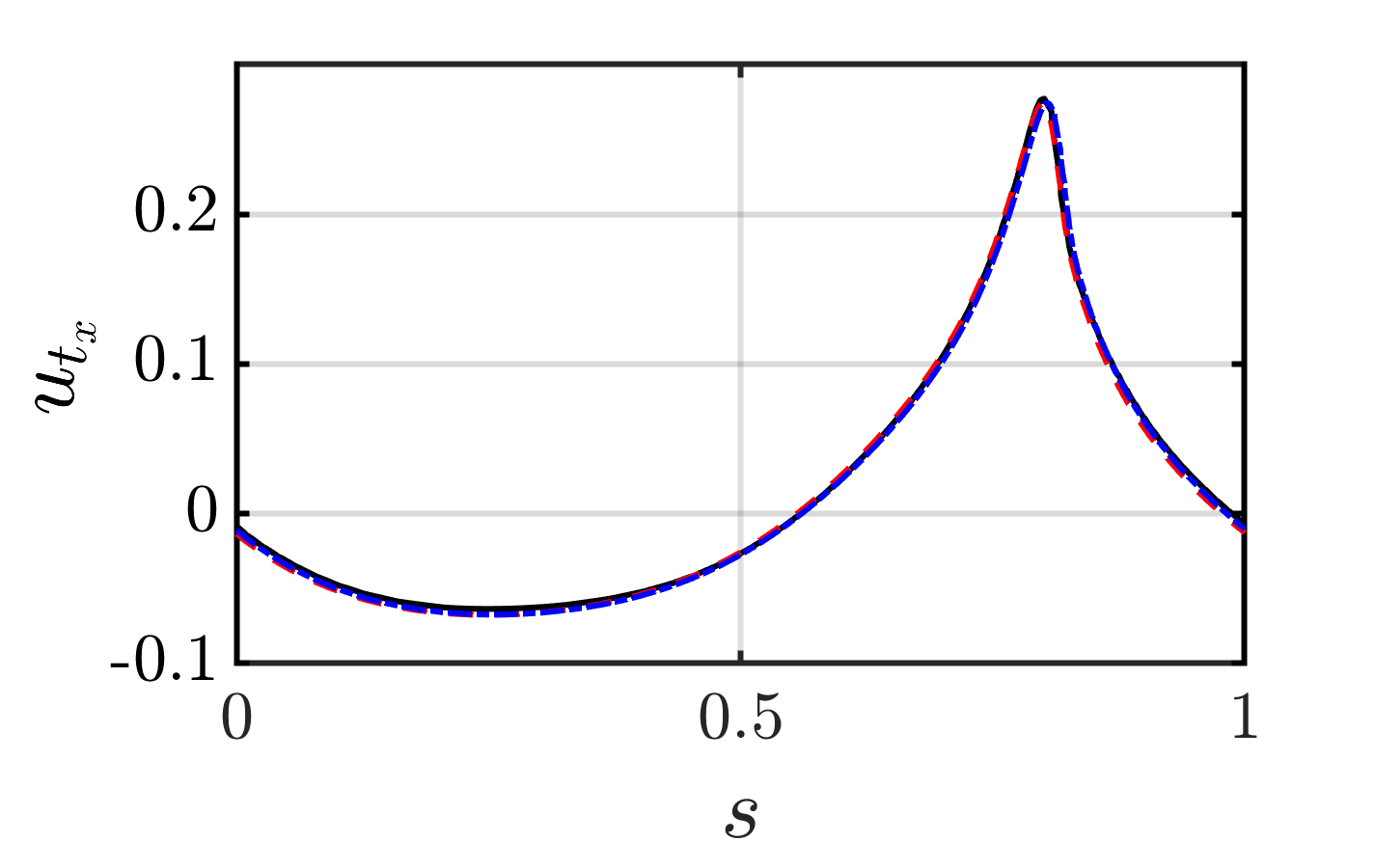} \\
(c) & (d) \\
\end{tabular}
\caption{\textcolor{black}{Grid convergence  for a representative surfactant-laden spilling-breaker  ($\epsilon = 0.33$, $\beta_s = 0.5$) at time $t=3.0$.
Comparisons of $N=256, 512,$ and $1024$ for
(a)  interface location, (b)  $\Gamma$, (c) $\tau$, and  (d)  $u_{t_{x}}$.} }
\label{grid independence}
\end{figure}

To demonstrate mesh independent results, we have monitored the temporal  variation of the kinetic energy  of the system for a resolution of $\lambda/\Delta {\bf x}= [256,512,1024]$ for waves with $\epsilon = 0.30$ \textcolor{black}{ (regular) and  $\epsilon = 0.33$ (spilling)}, displayed in figure \ref{mesh_figure}\textcolor{black}{a and \ref{mesh_figure}b, respectively. For both cases, }  
the coarsest grid, with $\lambda/\Delta \mathbf{x} = 256$, exhibits noticeable discrepancies, particularly in the valley regions of the kinetic energy evolution. In contrast, the cases with $\lambda/\Delta \mathbf{x} = 512$ and $1024$ collapse almost exactly onto one another, with discrepancies remaining below $1\%$ over the entire simulation window.   The $\lambda/\Delta \mathbf{x} =1024$ resolution was therefore run only for a short time owing to the prohibitive computational cost associated with the much smaller time step. 
The results for $\lambda/\Delta \mathbf{x} = 512$ and $\lambda/\Delta \mathbf{x} = 1024$ are essentially indistinguishable, demonstrating the mesh independence of our results \textcolor{black}{ for regular  and  spilling breaking cases  shown in this work.} All results presented in this manuscript correspond to simulations performed at $\lambda/\Delta \mathbf{x} = 512$.  
\textcolor{black}{
Finally, figure~\ref{grid independence} presents a comparison of key interfacial at $t=3.0$, including the interface  profile,  $\Gamma$, $\tau$ and $u_{tx}$ 
for different grid resolutions with  $\epsilon = 0.33$ and $\beta_s = 0.5$.
Excellent agreement is observed between the $\lambda/\Delta x = 512$ and $1024$ cases across all quantities considered, with relative differences of $\mathcal{O}(10^{-3})$. This indicates that the interfacial dynamics are well resolved at $\lambda/\Delta x = 512$,  and the results presented in the manuscript are mesh independent. 
}


\CRCA{To verify that the chosen computational width ($\lambda/4$) does not influence the dynamics, we performed a width-independence study for the clean-wave case with $\epsilon = 0.3$, measuring the kinetic energy evolution while repeating the simulation using domain widths $\lambda$, $\lambda/2$, $\lambda/4$, and $\lambda/8$, as shown   in figure \ref{mesh_figure}a. The resulting kinetic energy plot collapses onto one another, demonstrating that spanwise confinement does not affect the wave dynamics in the regime studied.}

\section*{Appendix D: Mechanical energy evolution}

 \begin{figure}
 \centering
 \begin{tabular}{c}
 \includegraphics[width=0.8\textwidth]{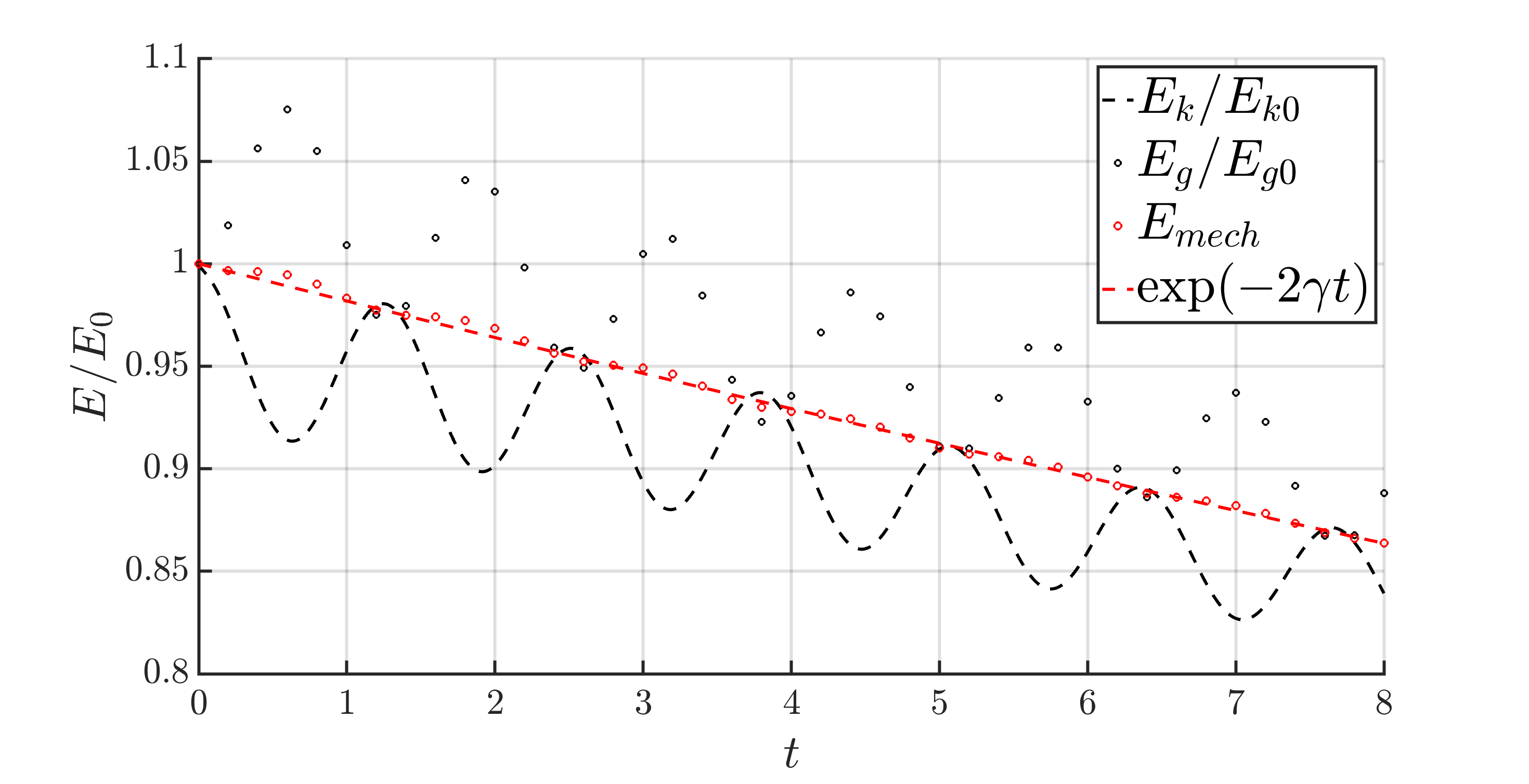}
 \end{tabular}
\caption{\CRCA{Temporal evolution of the mechanical, kinetic, and gravitational potential energies, normalised by their initial values for $\epsilon=0.3$.}  }
\label{gravity}
\end{figure}

\CRCA{ In this work,  the instantaneous kinetic and gravitational potential energies are defined as
\begin{equation}
E_K(t) = \frac{1}{2}\int_{d\ge 0} \rho \left(u^2+v^2\right)\,\mathrm{d}x\,\mathrm{d}y,
\qquad
E_P(t) = \int_{d\ge 0} \rho\, y \,\mathrm{d}x\,\mathrm{d}y,
\end{equation}
and the mechanical energy is given by
\begin{equation}
E_{\mathrm{mech}}(t)=E_K(t)+E_P(t).
\end{equation}
Following the approach of \citet{iafrati2009numerical}, we note that for weakly damped regular gravity waves, $E_K$ and $E_P$ exchange periodically over a wave period, while their sum $E_{\mathrm{mech}}$ decays monotonically due to viscous dissipation. In this regime, the decay of the mechanical energy is given by 
\[
E_{\mathrm{mech}}(t)\sim E_{\mathrm{mech}}(0)\, \mathrm{e}^{-2\gamma t},
\]
and the exponential envelope of $E_K(t)$ provides an accurate estimate of
the decay rate.}

\CRCA{Figure \ref{gravity} shows the computed potential energy $E_P(t)$ evaluated at discrete times, reflecting the fact that snapshots were saved only at selected instants. For comparison, we also include the predicted mechanical energy $E_{\mathrm{mech}}(t)$. We also find that the decay of $E_{\mathrm{mech}}$ is in excellent agreement with the exponential decay of the kinetic energy. Furthermore, the contribution of surface-tension energy is negligible in this regime, as the Weber number is large; the same assumption was adopted by \cite{iafrati2009numerical}.}

\section*{\textcolor{black}{Appendix E: Comparison of bulge shapes with experimental results}}

\begin{figure}
    \centering
\begin{tabular}{cc}
 \includegraphics[width=0.5\textwidth]{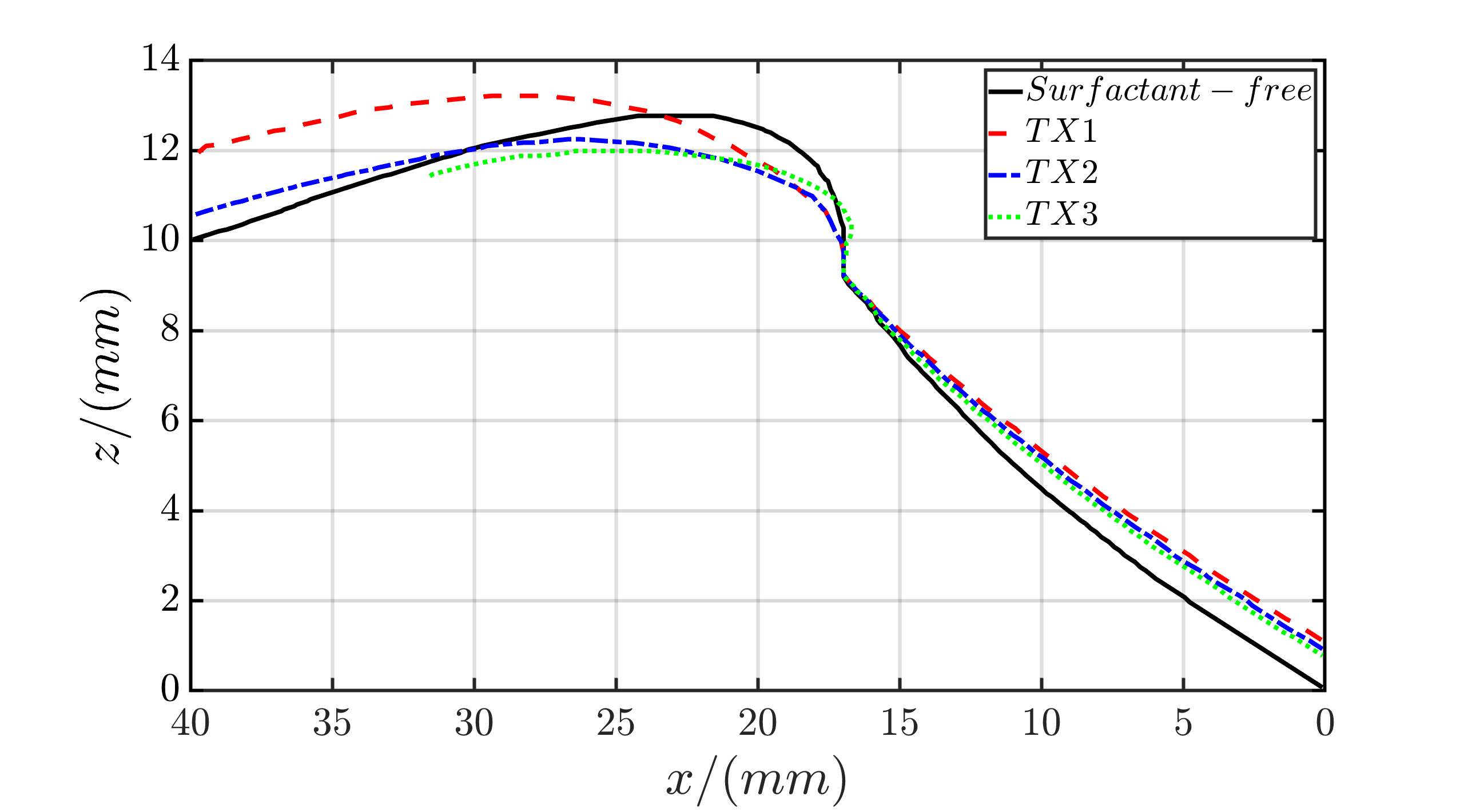} &
 \includegraphics[width=0.5\textwidth]{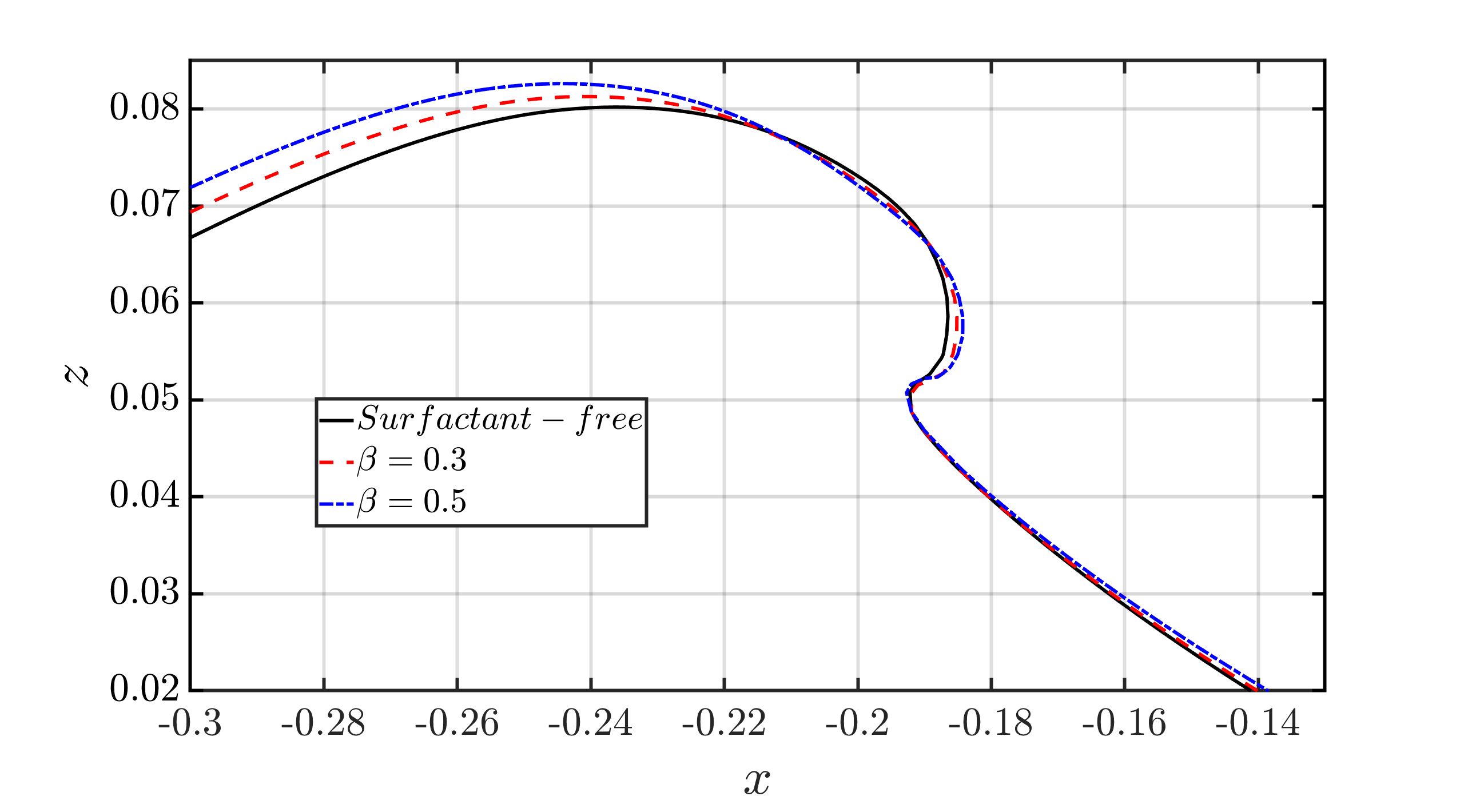}\\

(a) & (b) \\
\end{tabular}
\caption{Comparison of bulge shapes for clean and surfactant-laden conditions between the experiments of \cite{liu2006experimental} and the present simulations at $\epsilon = 0.35$, corresponding to panels (a) and (b), respectively. 
} 
\label{comparison}
\end{figure}

\textcolor{black}{Figure \ref{comparison} presents a qualitative comparison between the present simulations and the experimental profiles reported by \citet{liu2006experimental}. We emphasize that this comparison is not intended as a quantitative validation. The experiments involve soluble surfactants with adsorption--desorption kinetics, whereas the present model assumes an insoluble surfactant with no exchange with the bulk. In addition, the characteristic Reynolds and Bond numbers in the experiments are substantially larger than those considered here. Using the crest height $H_c$ as the length scale, \citet{liu2006experimental} reported $Re \approx 9.3\times10^4$, $Bo \approx 9.2\times10^2$, and wave steepness $\epsilon \approx 0.44$. When expressed in terms of the wavelength $\lambda$, these correspond to $Re \approx 1.3\times10^6$ and $Bo \approx 1.9\times10^5$. By contrast, the present simulations are performed at $Re = 10^4$ and $Bo = 253.3$ ($We = 100$), indicating differences of one to two orders of magnitude depending on the choice of length scale. Despite these differences, the overall bulge morphology, including crest elevation, forward leaning of the crest, and the curvature transition near the toe, shows similar qualitative trends across the datasets. This suggests that these large-scale interfacial features are reproduced consistently, even though the surfactant model and operating conditions differ substantially. The comparison should therefore be interpreted as evidence of qualitative consistency in interface shape, rather than as a strict quantitative agreement or a validation of the detailed dynamics.}

\bibliographystyle{jfm}
\bibliography{jfm-instructions.bib}

\end{document}